\documentclass[useAMS,usenatbib]{mnras}
\usepackage[pdftex]{graphicx} 
\usepackage{hyperref}
\usepackage{xcolor} 
\usepackage{lscape}
\usepackage{calc}
\usepackage{amsmath}
\usepackage{afterpage}
\usepackage{bm}
\usepackage{amssymb}

\definecolor{fuchsia}{rgb}{1.0, 0.0, 1.0}

\DeclareMathSymbol{\mh}{\mathord}{operators}{`\-}

\hypersetup{colorlinks,linkcolor={red!50!black}, citecolor={blue!50!black},urlcolor={blue!80!black} } 

\title[CIB-optical cross-correlations]
{Constraints on galaxy formation from the cosmic-far-infrared-background\,$-$\,optical-imaging cross-correlation using \textit{Herschel} and UNIONS}

\author[Lim et al.]
{Seunghwan Lim$^{1,2}$\thanks{E-mail: shlim@cita.utoronto.ca}\thanks{CITA National Fellow},
Ryley Hill$^{1}$, 
Douglas Scott$^{1}$,
Ludovic van Waerbeke$^{1}$, 
\newauthor
Jean-Charles Cuillandre$^{3}$,
Raymond G. Carlberg$^{4}$,
Nora Elisa Chisari$^{5}$,
Andrej Dvornik$^{6}$,
\newauthor
Thomas Erben$^{7}$,
Stephen Gwyn$^{8}$,
Alan W. McConnachie$^{8}$,
Marc-Antoine Miville-Desch\^{e}nes$^{3}$,
\newauthor
Angus H. Wright$^{6}$, 
Pierre-Alain Duc$^{9}$
\\
\vspace*{6pt} \\
$^{1}$Department of Physics and Astronomy, University of British Columbia, Vancouver, BC, Canada V6T 1Z1 \\
$^{2}$Canadian Institute for Theoretical Astrophysics, University of Toronto, 60 St. George Street, Toronto, ON, Canada M5S 3H8 \\
$^{3}$AIM, CEA, CNRS, Universit\'{e} Paris-Saclay, Universit\'{e} de Paris, F-91191 Gif-sur-Yvette, France \\
$^{4}$Department of Astronomy \& Astrophysics, University of Toronto, Toronto, ON M5S 3H4, Canada \\
$^{5}$Institute for Theoretical Physics, Utrecht University, Princetonplein 5, NL-3584 CC Utrecht, the Netherlands \\
$^{6}$Ruhr-University Bochum, Astronomical Institute, German Centre for Cosmological Lensing, Universitätsstr 150, D-44801 Bochum, Germany \\
$^{7}$Argelander-Institut f\"ur Astronomie, University of Bonn, Auf dem H\"ugel 71, 53121 Bonn, Germany \\
$^{8}$NRC Herzberg Astronomy and Astrophysics, 5071 West Saanich Road, Victoria, BC V9E 2E7, Canada \\
$^{9}$Universit\'{e} de Strasbourg, CNRS, Observatoire astronomique de Strasbourg (ObAS), UMR 7550, 67000 Strasbourg, France
}

\begin{document} 

\pagerange{\pageref{firstpage}--\pageref{lastpage}}

\date{\today}
\pubyear{2021}

\maketitle

\label{firstpage}

\begin{abstract} 
Using {\it Herschel}-SPIRE imaging and the Canada-France Imaging Survey (CFIS) Low Surface Brightness data products from the Ultraviolet Near-Infrared Optical Northern Survey (UNIONS), we present a cross-correlation between the  cosmic far-infrared background and cosmic optical background fluctuations. The cross-spectrum is measured for two cases: all galaxies are kept in the images; or all individually-detected galaxies are masked to produce `background' maps. We report the detection of the cross-correlation signal at $\gtrsim 18\,\sigma$ ($\gtrsim 14\,\sigma$ for the background map). The part of the optical brightness variations that are correlated with the submm emission translates to an rms brightness of $\simeq 32.5\,{\rm mag}\,{\rm arcsec}^{-2}$ in the $r$ band, a level normally unreachable for individual sources. A critical issue is determining what fraction of the cross-power spectrum might be caused by emission from Galactic cirrus.  For one of the fields, the Galactic contamination is 10 times higher than the extragalactic signal; however, for the other fields, the contamination is around 20~per cent. An additional discriminant is that the cross-power spectrum is of the approximate form $P(k)\propto 1/k$, much shallower than that of Galactic cirrus. We interpret the results in a halo-model framework, which shows good agreement with independent measurements for the scalings of star-formation rates in galaxies. The approach presented in this study holds great promise for future surveys such as FYST/CCAT-prime combined with {\it Euclid} or the Vera Rubin Observatory (LSST), which will enable a detailed exploration of the evolution of star formation in galaxies. 
\end{abstract} 


\section[intro]{Introduction}
\label{sec_intro}

The cosmic infrared background (CIB) is the relic of UV/optical starlight reprocessed (absorbed and re-radiated) by dust grains \citep[e.g.][]{Savage1979, Heinis2014}. With its intensity containing about half of the combined starlight ever emitted \citep[e.g.][]{Hauser2001, Dole2006, HillMasuiScott2018}, it traces the star-formation history of the Universe over a wide redshift range, making its measurement crucial to understanding the formation and evolution of galaxies. The far-infrared part of the CIB, in particular, peaks at around $150\,\micron$ and can be observationally investigated over a wide range of wavelengths. It probes obscured star formation \citep{CharyElbaz2001,Lagache2005}, in contrast to the near-IR, optical and near-UV backgrounds (that we collectively refer to in this paper as the `cosmic optical background', COB), which mostly tracks direct starlight \citep[e.g.][]{Conselice2016}. This means that while the COB traces stellar mass, the CIB traces star-formation rate, and hence combining them offers the opportunity to learn important new information about the evolving populations of galaxies that cannot be discerned from optical data alone. Since there may be some confusion with terminology, we stress that when we say ``CIB'' in the rest of this paper, we are focusing on the part of the CIB peaking at far-IR wavelengths. 

Unfortunately, because of the limited angular resolution of current single-dish telescopes operating at far-infrared and submillimetre (submm) wavelengths \citep[e.g.][]{Dole2003, Dole2004}, sources are `confused', i.e. blended within the same instrumental beam \citep[e.g.][]{Nguyen2010}. For example, $\lesssim 15$ per cent of the total flux density (coming from the brightest 1~per cent of sources) is resolved into individual galaxies \citep{Oliver2010} at $250\,\micron$ by data from the {\it Herschel\/} satellite (without the use of priors or deconvolution techniques). If the analysis is solely based on detecting sources and creating catalogues, then the source confusion makes it challenging to carry out unbiased analyses of the star-formation history, as well as to identify their counterparts at other wavelengths. On the other hand, the complete maps contain information from all sources (both resolved and unresolved), integrated over luminosity and redshift into large-scale fluctuations. This would be true even if the galaxies were Poisson distributed, but also included in the CIB anisotropies is the clustering of star-forming galaxies \citep[e.g.][]{Scott1999}. In fact, dusty star-forming galaxies (DSFGs) at high redshifts, which are responsible for the CIB, are found to be strongly correlated with each other \citep[e.g.][]{Farrah2006, Wilkinson2017}. Measurements of the statistical properties of the maps, such as cross-correlations and power spectra, are excellent tools for placing constraints on galaxy evolution, and for probing the large-scale structure of the Universe. Such measurements also enable a joint analysis across various wavelengths, free from the difficulties of finding multi-wavelength counterparts in the band-merging process. 

Following the discovery of the far-IR CIB \citep{Puget1996, Fixsen1998, Hauser1998, Schlegel1998}, the power spectra and anisotropies of the CIB have been particularly well measured, with the detection of clustering signals from {\it Spitzer} \citep{Grossan2007, Lagache2007}, the Balloon-borne Large Aperture Submillimeter Telescope \citep[BLAST;][]{Viero2009, Hajian2012}, the Atacama Cosmology Telescope \citep[ACT;][]{Dunkley2011}, and the South Pole Telescope \citep[SPT;][]{Hall2010}. Later, the anisotropies from far-infrared (far-IR) to microwave wavelengths were more precisely determined by {\it Herschel\/} \citep{Amblard2011, Viero2013} and {\it Planck\/} \citep{Planck2011, Planck2014}, which constrain the signals from star formation within large-scale structures. Measurements of the CIB anisotropies have then been interpreted within halo-based modelling frameworks \citep[e.g.][]{Lagache2007, Viero2009, Viero2013, Amblard2011, Planck2011, Planck2014, Penin2012, Shang2012, Xia2012, Bethermin2013, Maniyar2021}, which use halo occupation distributions to link galaxies with their dark matter haloes and to predict clustering properties \citep[e.g.][]{Seljak2000, Peacock2000, Scoccimarro2001, Cooray2002}. 

The COB is another probe of the evolution of galaxies and large-scale structure, tracing direct starlight observed at optical wavelengths.\footnote{This includes near-IR light, which has been extensively studied through auto-correlation functions \citep{Kashlinsky2005a, Kashlinsky2018}, with some debate over the origin of the clustered signals \citep[see e.g.][]{Kashlinsky2005b,Kashlinsky2007,Thompson2007,Cooray2007, Cooray2012,Donnerstein2015}.} Accounting for roughly the other half of the light emitted from star-forming galaxies, as well as the starlight emitted from galaxies with old stellar populations, the COB is an important observable that provides constraints on galaxy evolution complementary to those from the CIB \citep[e.g.][]{BernsteinIII2002, Lauer2021}. However, measuring this background is complicated in many cases (particularly for ground-based telescopes) by local foregrounds, such as the Earth's atmosphere and instrumental artefacts \citep[e.g.][]{BernsteinI2002}. This can be bypassed by cross-correlating optical maps with space-based surveys at longer wavelengths. Measuring cross-correlations and cross-power spectra between optical and submm wavelengths has great potential to probe the build-up of stellar components over cosmic time. Such cross-correlation measurements can also be used to estimate the part of the optical (or submm) fluctuations (not the average background level) that correlates with the submm (or optical) fluctuations. 

In this paper, we investigate the signal contained in the cross-correlation between Canada-France Imaging Survey (CFIS; see \citealt{Ibata2017} for the first results from the $u$-band) $r$-band images and {\it Herschel}-Spectral and Photometric Imaging Receiver (SPIRE; \citealt{Griffin2010}) submm images, covering a total area of $91\,{\rm deg}^2$ of the sky. Most of the minor systematics normally affecting auto-correlations, such as noise and instrumental artefacts, are reduced to the level of the statistical uncertainties in the cross-correlation, and do not bias the results. However, to ensure the robustness of the measurements, careful attention still needs to be paid to other systematics that may potentially bias the measurements, such as spatial filtering and masking, and particularly the impact of Galactic cirrus. We can also perform tests by extracting the signals in diffuse light from unresolved sources separately from resolved sources. We interpret our measurements through a halo model framework, putting constraints on the star formation in galaxies and dark matter haloes across a wide range of masses and redshifts. With upcoming survey missions and telescopes such as {\it Euclid\/} \citep{Laureijs2011}, the Vera Rubin Observatory (previously referred to as LSST) \citep{LSST}, and the FYST/CCAT-prime \citep{CCATp}, which promise dramatically improved data in the near future, as well as currently available wide surveys including the Dark Energy Survey \citep[DES;][]{DES} and the UNIONS, we expect that the methodology presented in this paper will become important to our understanding of the cosmic star-formation history. 

The structure of this paper is as follows. In Sect.~\ref{sec_data}, we describe the surveys and fields selected for our analysis and the construction of the CFIS mosaic maps from the raw data, as well as the detailed prescription of how we measure and correct the cross-power spectra from the images. In Sect.~\ref{sec_results}, we present our measurements and our method to estimate and correct for the impact of our Galaxy on these measurements. Section~\ref{sec_test} explains how we test the robustness of our method against potential systematic effects using simulated light cones. Finally, Sect.~\ref{sec_model} describes our modelling and fitting of the data, and presents the resulting constraints on galaxy formation. Some discussion is contained in Sect.~\ref{sec_disc} and conclusions are given in Sect.~\ref{sec_sum}. Additionally, Appendix~\ref{sec_appA} presents our main results in flux unit, while Appendices~\ref{sec_appB} and \ref{sec_appC} present the results of some null tests, and  some filtered images that show the cross-correlation visually, respectively.  
Throughout this paper, we assume a {\it Planck\/} cosmology \citep{Planck2018VI} and a \citet{Chabrier2003} initial mass function (IMF).

\section[data]{Measuring the CFIS--SPIRE cross-correlation}
\label{sec_data}

\subsection{Surveys}
\label{ssec_survey}

The Ultraviolet Near Infrared Optical Northern Survey (UNIONS) is a scientific collaboration of wide-field imaging surveys of the Northern hemisphere, which is also part of the ground-based support for the {\it Euclid\/} space mission. For this study, we use the CFIS $r$-band data, which is one component of UNIONS, along with the Panoramic Survey Telescope and Rapid Response System (Pan-STARRS), the Wide Imaging with Subaru HyperSuprimeCam of the Euclid Sky (WISHES), and Waterloo-IfA G-band Survey (WIGS). CFIS is a high-resolution, deep survey carried out in the $u$ and $r$ bands, conducted with the 3.6-metre Canada-France-Hawaii Telescope (CFHT) on Mauna Kea, using the MegaCam wide-field optical imager with a field-of-view (FOV) of 1\,${\rm deg}^2$. The survey makes three single-exposure visits, with the offset being one third of an FOV, optimizing astrometric and photometric calibration. The data reduction is performed using {\sc MegaPipe} \citep{Gwyn2008}. The images are astrometrically calibrated against {\it Gaia\/} \citep{Gaia2016,Gaia2018}, and are photometrically calibrated using Pan-STARRS $3\pi$ $r$-band photometry \citep{Chambers2016}. With a median seeing of 0.65\,arcsec and a 5$\,\sigma$ point-source depth of 24.85 AB mag, the CFIS $r$-band survey will eventually cover about $5{,}000\,{\rm deg}^2$ of the sky above a declination of $30^\circ$. For the data products that we use here, the survey had completed $r$-band imaging over approximately $3{,}000\,{\rm deg}^2$.  We use the CFIS $r$-band Low Surface Brightness (LSB) version of the images \citep[][in preparation]{Cuillandre2022}, which preserves large-scale brightness variations. On the contrary, we find that the cross-power spectra measured using the regular (non-LSB) CFIS data, where the images have been flattened by removing all background fluctuations as is routinely executed to conduct compact source science, have essentially zero signal. This is true even after applying the transfer function computed between the LSB and non-LSB images, in order to account for loss of signal due to the large-scale filtering applied in the non-LSB production and thus to recover the underlying signal. This means that the filtering applied in the non-LSB data, at $\gtrsim 20$--$30$\,arcsec, is so effective that signals at much larger scales as probed in our analysis are almost completely removed. 

The individually-processed CFIS images are stacked into LSB tiles of $0.5\,{\rm deg} \times0.5\,{\rm deg}$ each (0.187\,arcsec per pixel) through {\sc Swarp} \citep{Bertin2002,Bertin2010}, by resampling the images according to the astrometric calibration, by scaling them in accordance with the photometric calibration, and by combining them with weights. {\sc Swarp} is applied to the images without its background removal, in order to retain large-scale signals. The weights are used to handle bad columns and cosmic rays. 

We compare CFIS images with maps from the SPIRE instrument \citep{Griffin2010} aboard the {\it Herschel\/} Space Observatory \citep{Pilbratt2010} as a tracer of star formation through dust emission. The SPIRE instrument provides separate data in bands centred at 250, 350 and 500\,$\micron$. In particular, we use SPIRE data from the HerMES \citep{Oliver2012} survey, specifically the fourth data release (DR4).\footnote{\url{https://hedam.lam.fr/HerMES/}}  HerMES maps were created by processing the raw images from the {\it Herschel\/} Science Archive using standard European Space Agency  software and the software package {\sc SMAP} \citep[known earlier as SHIM;][]{Levenson2010, Viero2013}. {\sc SMAP} is a code designed to minimize large-scale noise artefacts by iteratively removing a low-order polynomial baseline from each scan, while trying to preserve real large-scale structure as much as possible. The full-width at half-maximum (FWHM) values of the SPIRE maps are 18.1, 25.5 and 36.6\,arcsec at 250, 350 and 500\,$\micron$, respectively, with the pixel sizes of each map being a third of the FWHM values. For the {\it Herschel\/} fields for which the HerMES maps are not available, we make use of SPIRE maps from the {\it Herschel\/} Extragalactic Legacy Project (HELP\footnote{\url{https://herschel.sussex.ac.uk/}}; \citealt{Vaccari2016,Shirley2019}). HELP is a European-funded project aimed at providing homogeneously-calibrated multi-wavelength data covering roughly 1,300\,${\rm deg}^2$ of the {\it Herschel\/} survey fields, including fields not contained in HerMES, such as the {\it Herschel}-Astrophysical Terahertz Large Area Survey \citep[H-ATLAS;][]{Eales2010} and the {\it Herschel\/} Stripe~82 Survey \citep[HerS;][]{Viero2014}. We specifically use version~1.0 of the SPIRE maps from HELP for the HATLAS-NGP field, which is the only field contained in our CFIS overlap without available HerMES data.

\subsection{Field selection}
\label{ssec_field}

\begin{figure*}
\includegraphics[width=0.88\linewidth]{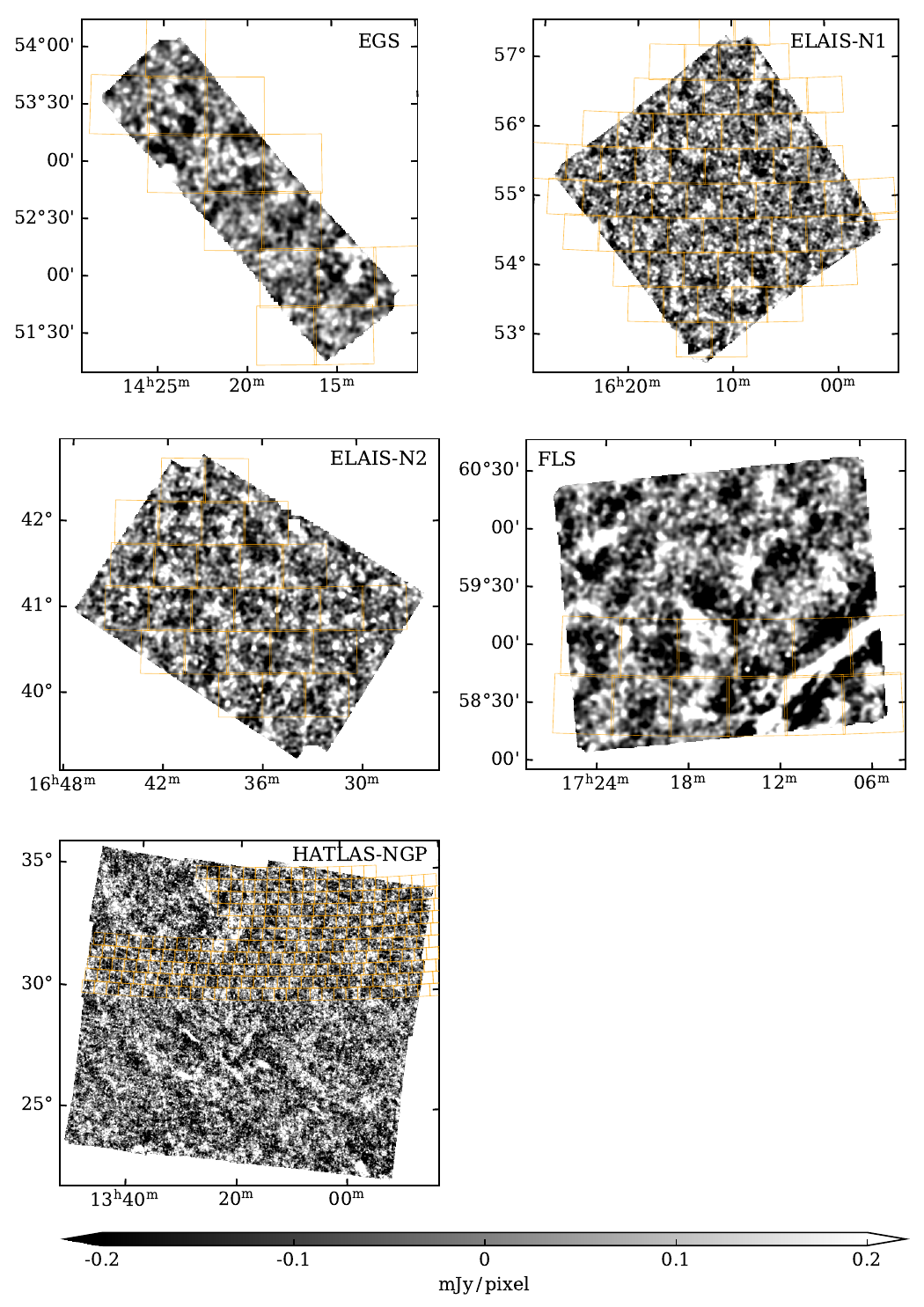}
\caption{The five SPIRE maps at $250\,\micron$ used in our analysis (smoothed with a Gaussian kernel of 3\,arcmin to illustrate the CIB fluctuations), namely the EGS, ELAIS-N1, ELAIS-N2, FLS, Lockman and HATLAS-NGP fields, shown together with the CFIS $r$-band tiles overlaid (orange squares). The linear intensity scales used are identical for all panels. Note that the large feature in the lower-right of the FLS image is of Galactic origin. }
\label{fig_fields}
\end{figure*}

\begin{table*}
 \renewcommand{\arraystretch}{1.2} 
 \centering
 \begin{minipage}{102mm}
  \caption{List of basic information on the fields used in this study.}
  \begin{tabular}{ccccccc}
\hline
Field & SPIRE area & Overlap\textsuperscript{\footnote{This is the area of overlap between the CFIS $r$-band and SPIRE maps for each field used in our analysis.}} & RA &  Dec & $l$ & $b$ \\
 & $[{\rm deg}^2]$ & $[{\rm deg}^2]$ & $[{\rm deg}]$ & $[{\rm deg}]$ & $[{\rm deg}]$ & $[{\rm deg}]$ \\
\hline
\hline
EGS & 3.6 & 3.5 & 215.0 & 52.7 & 96.1 & 59.6 \\  
ELAIS-N1 & 13.5 & 13.5 & 242.9 & 55.1 & 85.0 & 44.5 \\  
ELAIS-N2 & 9.2 & 6.5 & 249.2 & 41.1 & 65.1 & 42.2 \\  
FLS & 7.4 & 3.0 & 259.0 & 59.4 & 88.2 & 35.2  \\  
HATLAS-NGP & 177.7 & 64.3 & 199.5 & 29.0 & 52.2 & 83.8 \\  
\hline
Total & 211.4 & 90.8 & & & & \\
\hline
\hline
\\
\vspace{-8mm}
\end{tabular}
\textbf{Notes.}
\vspace{-5mm}
\label{tab_fields}
\end{minipage}
\vspace{0mm}
\end{table*}

To measure the CIB-optical cross-correlation, we select essentially all substantial (i.e.\ large and contiguous) high Galactic latitude {\it Herschel\/}-SPIRE patches that overlap with the CFIS $r$-band coverage, regardless of whether the overlap is complete. In the current data release of CFIS $r$-band data, the LSB tiles are available only for the survey regions that have received all three planned visits from CFHT. The selected SPIRE fields are the Extended Groth Strip (EGS), the European Large-Area ISO Survey-North 1 (ELAIS-N1), the European Large-Area ISO Survey-North 2 (ELAIS-N2), the First Look Survey (FLS), and the  HATLAS North Galactic Pole (HATLAS-NGP) field, overlapping with a total of 371 CFIS $r$-band LSB tiles, resulting in a combined area of 91\,${\rm deg}^2$.  Figure~\ref{fig_fields} shows each of the five SPIRE patches (at $250\,\micron$), together with outlines of the overlapping CFIS tiles. Note that the FLS field contains a filamentary structure in the southwestern quadrant, which is a prominent part of the Galatic cirrus. While this can potentially (and indeed does) contaminate the extragalactic signal that is our focus, including the FLS field in our analysis is rather useful as a reference to be compared with the other fields where the fluctuations are mostly extragalactic, as will be seen in Sect.~\ref{ssec_res}. The fields used for this study, and a summary of their properties, are given in Table~\ref{tab_fields} (and see the HELP site for more details). 

\subsection{CFIS \textit{r}-band map mosaics}
\label{ssec_mosaic}

The CFIS images that we use are the LSB versions described in the previous section. The tiles are then cut exactly along RA and Dec limits so that two neighbouring tiles do not overlap. For each of the five larger SPIRE fields, all CFIS-$r$ LSB images available over the same footprint are positioned on the celestial sphere, and a larger mosaic is constructed using {\sc Swarp}. {\sc Swarp} runs on the tiles without its internal background subtraction, so that the large-scale fluctuations across the tiles are preserved in the mosaics. Stars, satellite trails and large gaps between CCDs are masked. In addition, we also generate a version of the mosaic where all galaxies individually detected are masked, using an elliptical mask that extends to 10 times the half-light-radius of the galaxies. The regular UNIONS galaxy catalogue produced through {\sc Megapipe} was used for this process. The mosaicked, galaxy-masked CFIS-$r$ LSB images constitute our `background'-only versions of the mosaics. We present our results primarily based on these two versions of mosaics, considering them as fiducial CFIS LSB data for our analysis. While residual (not masked) outer part of haloes around bright stars may display a brightness greater than the extragalactic signals, we find that more aggressive masking on stars and artefacts do not change our results or reduce the associated noise noticeably. This is because of the forgiving nature of cross-correlation with another survey at submm wavelengths; the bright stellar haloes do not appear in the SPIRE images so that there is no contribution from them in the cross-correlation. As a result, the total fraction of masked pixels in the CFIS maps is 47 and 53 per cent for the less and more aggressively masked maps, respectively, when combined for the five fields, with some variations among the fields of about 5 per cent. From the tests with varying amounts of masking, as well as using simulations, we verify that our measurements are consistent regarding masking. 

\begin{figure*}
\includegraphics[width=0.95\linewidth]{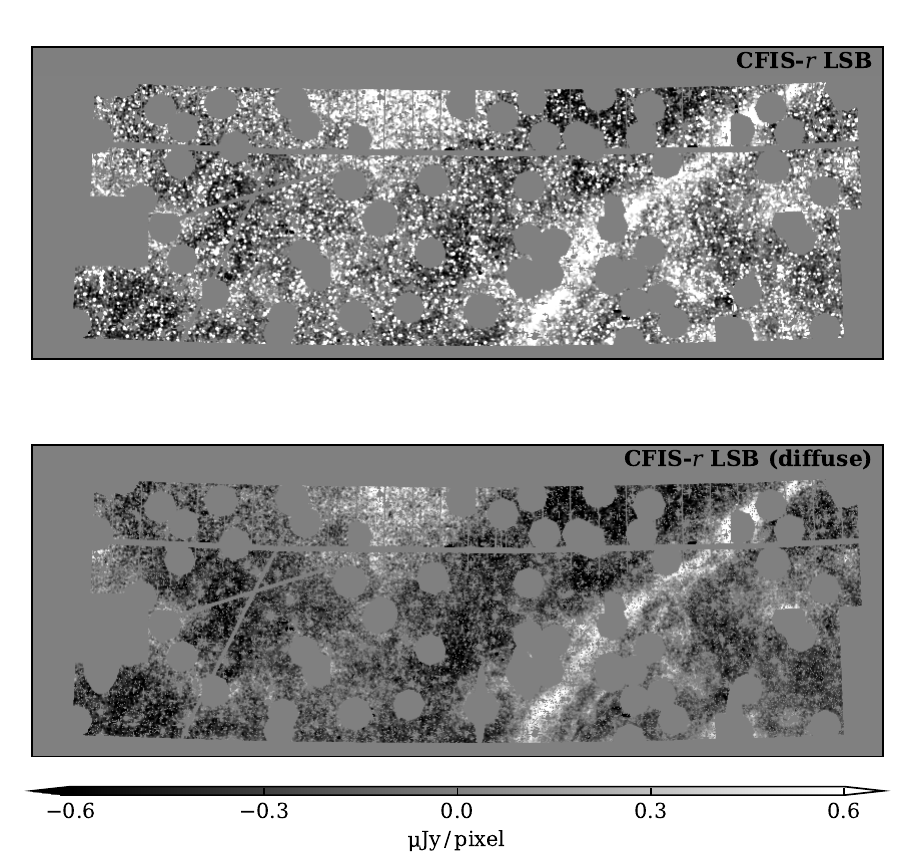}
\caption{Example of CFIS $r$-band LSB mosaic maps, in this case for the FLS field. The maps shown were smoothed with a Gaussian kernel of 10\,arcsec, i.e. more than 10 times lower resolution than the raw data. The intensity scales used are identical for both panels, and are linear. A filamentary structure of Galactic origin is obvious on the right-hand side of the images. {\it Top:} Map after only masking stars and artefacts. {\it Bottom:} Same as above, but with additional masks for all identified compact objects, including galaxies.}
\label{fig_FLSmosaic}
\end{figure*}

Figure~\ref{fig_FLSmosaic} shows two versions of the CFIS-$r$ LSB mosaic for the FLS field: in the top panel, only stars, satellite trails and large CCD gaps are masked; and in the bottom panel, detected galaxies are also masked. The flux density in janskies, $f_\nu$, is obtained from the AB magnitude, $m_{\rm AB}$, by $m_{\rm AB} = -2.5\,\log_{10}{f_\nu} + 8.90$. Only nine CFIS tiles overlap with the FLS field. As mentioned earlier, the LSB tiles are assembled with a single pedestal adjustment, unlike the regular non-LSB CFIS tiles, for which large-scale variations are substantially diminished due to subtraction of local backgrounds through {\sc Swarp}, hence preserving more of the large-scale variations; this is nicely demonstrated with a prominent large filament of Galactic light that is captured in the figure. Even in the lower panel, where all detected sources are masked, we can still see filamentary structure of Galactic origin. 

To make sure that our results are robust against potential systematics and uncertainties arising from the pipelines used to create the mosaics, we use another version of the mosaic maps, which are constructed in a different way based on the latest development of CFIS LSB images, with stacking processing beyond that adopted for the current data release. For this version of the mosaics, the pixel size of stacks is 3 times larger than the native MegaCam resolution, namely 0.561 arcsec per pixel, while the size of each stack is also larger, being $1.2\,{\rm deg} \times 1.2\,{\rm deg}$. One issue with the CFIS LSB stacks in the data release so far has been signatures of the CCDs in the stacks in areas around bright stars, which arise from a skewness in median backgrounds due to bright features. The stacks are mosaicked using {\sc Montage}, a software package for assembling FITS images into mosaics. {\sc Montage} is optimized for preserving large-scale modes by using overlapping areas between the images. The masking scheme used for this version of the mosaics is the same as that for the fiducial version. The `background'-only mosaic map of this version for the FLS field is shown in Fig.~\ref{fig_Galactic_map_FLS}. We refer to this set of mosaics as `B3+{\sc Montage}' hereafter. We also tried various combinations of slightly different methods for stacking images, and found that the statistical properties of the CFIS LSB maps relevant for our study are fairly insensitive to the choice between reasonable methods for creating the stacks and mosaics that preserve the LSB signal carefully.

\subsection{Cross-power spectra}
\label{ssec_Xcorr}

We follow methods that are essentially the same as those in other studies of correlations within either the CIB or COB \citep[e.g.][]{Kashlinsky2007,Thompson2007, Viero2009,Arendt2010,Amblard2011,Matsumoto2011,Planck2011,Kashlinsky2012,Hajian2012,Viero2013,Planck2014,Seo2015,Matsumoto2019}, although the notation varies among previous papers.  While most of the earlier studies are concerned with the auto-correlation function or power spectrum, some do consider cross-correlations \citep[e.g.][]{Cappelluti2013,Thacker2015,MitchellWynne2016,Cappelluti2017,Li2017}, but we are not aware of any published papers specifically looking at the cross-correlation between far-IR/submm and optical images. 

We calculate the cross-power spectra between the CFIS $r$-band and SPIRE maps using the fast Fourier transform (FFT) method. Both maps have their global means subtracted, so that the average of the unmasked pixels for each map is zero. For the SPIRE maps, we exclude the edges of the fields (out to about 5\,arcmin) from our analysis because these regions contain boundary effects. To be explicit, the Fourier transform of each map, $F(\boldsymbol{k})$, is defined through 
\begin{align} \label{eq_FFT}
S_{\rm C}(\boldsymbol{n}) &= \sum_{k_x=-N_x/2}^{N_x/2} \sum_{k_y=-N_y/2}^{N_y/2}F_{\rm C}(\boldsymbol{k}) e^{i2\pi k_x n_x/N_x} e^{i2\pi k_y n_y/N_y},  \nonumber \\
S_{\rm S}(\boldsymbol{n}) &= \sum_{k_x=-N_x/2}^{N_x/2} \sum_{k_y=-N_y/2}^{N_y/2}F_{\rm S}(\boldsymbol{k}) e^{i2\pi k_x n_x/N_x} e^{i2\pi k_y n_y/N_y}, 
\end{align}
where $S(\boldsymbol{n})$ is the pixel value of each map in the direction $\boldsymbol{n}$ of the sky, the subscripts `C' and `S' are used to stand for CFIS and SPIRE, respectively, and $N_x$ and $N_y$ are the total number of pixels along each of the two-dimensional axes. The Fourier transforms can be further expressed as 
\begin{align}
F_{\rm C}(\boldsymbol{k}) &= C_{\rm C}(\boldsymbol{k})e^{i\theta_{\rm C}(\boldsymbol{k})}, \nonumber \\
F_{\rm S}(\boldsymbol{k}) &= C_{\rm S}(\boldsymbol{k})e^{i\theta_{\rm S}(\boldsymbol{k})},
\end{align}
where the real number $C$ (with the subscripts defined as above) is the amplitude (modulus) of the Fourier mode $\boldsymbol{k}$, and $\theta$ is the argument (phase). 

\begin{figure*}
\includegraphics[width=0.9\linewidth]{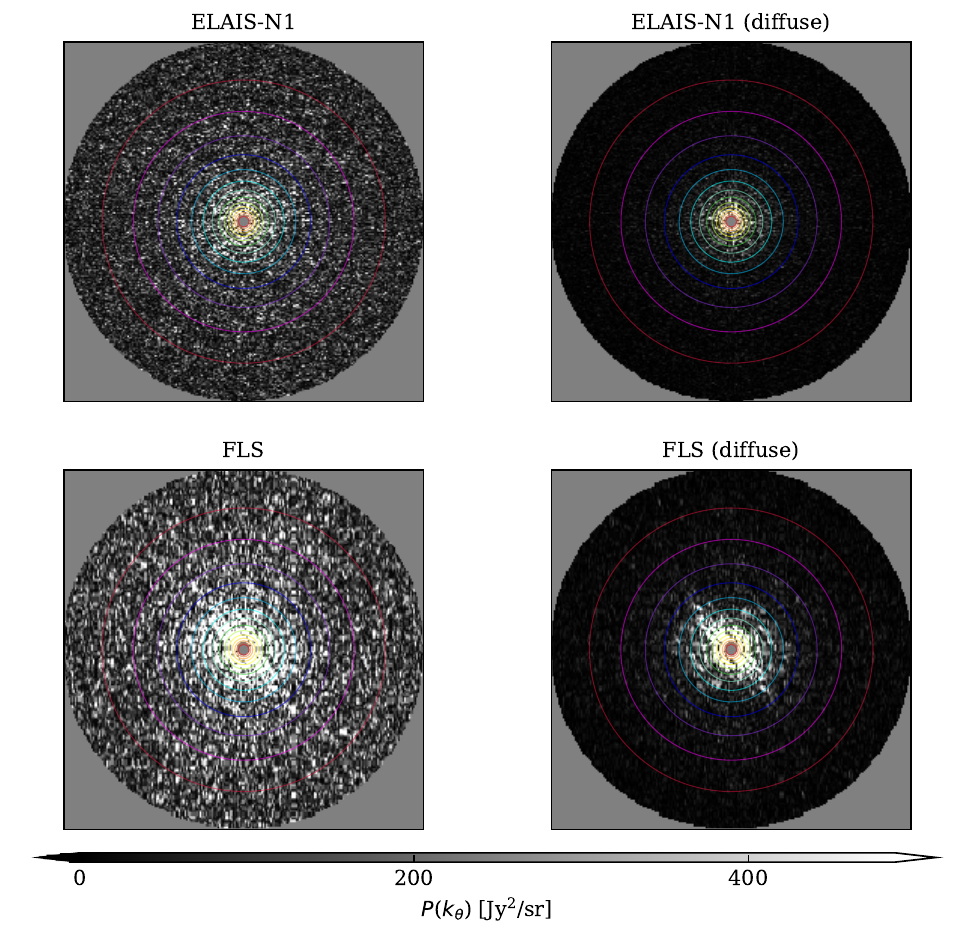}
\caption{Two-dimensional Fourier transforms of the cross-correlation between the SPIRE map at 250\,$\micron$ and the CFIS maps without (left column) and with (right column) masking on individually-detected galaxies. The ELAIS-N1 and FLS fields are presented as examples. As can be seen, the ELAIS-N1 field shows essentially isotropic Fourier transforms, implying that the maps are free of obviously anisotropic features; this is true for all the other fields, except for the FLS field. On the other hand, the FLS field, which clearly has anisotropic imprints of Galactic cirrus, shows an enhancement perpendicular to the filament in the Fourier transforms. The circles of colors represent the radial bins that we define for the analysis (see text for details).}
\label{fig_FFT_2D}
\end{figure*}
\begin{figure*}
\includegraphics[width=0.85\linewidth]{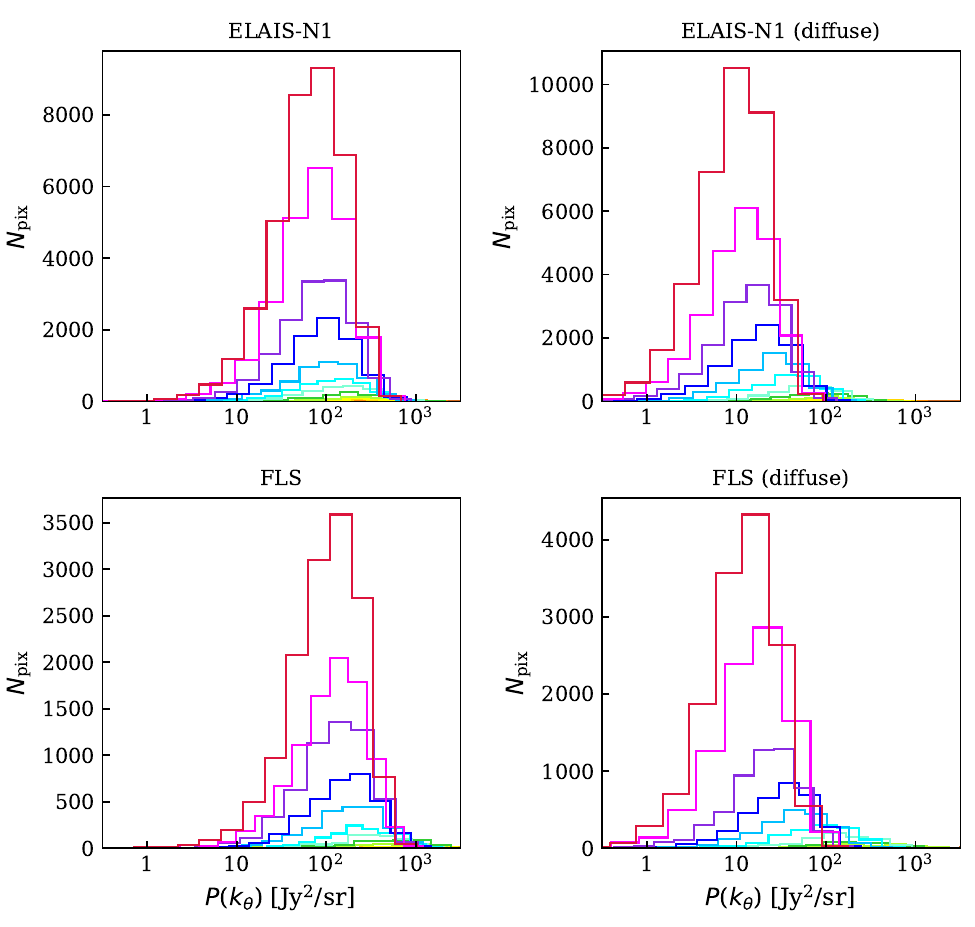}
\caption{Histograms for the Fourier rings of the cross-correlation between the SPIRE map at 250\,$\micron$ and the CFIS maps for the ELAIS-N1 and FLS fields as examples, as defined in Fig.~\ref{fig_FFT_2D} and the text, from which we calculate the cross-power spectra. The colors of the histograms match those of the radial bins in Fig.~\ref{fig_FFT_2D}. }
\label{fig_FFT_2D_hist}
\end{figure*}

The two-dimensional Fourier transforms of the cross-correlation between the SPIRE and CFIS maps, with and without the masks on galaxies, are shown in Fig.~\ref{fig_FFT_2D} for the ELAIS-N1 and FLS fields, as examples. The result for ELAIS-N1 is representative of the other fields (other than the FLS field), and is fairly isotropic, as can be seen in the figure. The FLS field, on the contrary, which has a clearly non-isotropic structure in the image, i.e.\ the strong Galactic filament (see Fig.~\ref{fig_FLSmosaic}), is shown to have a diagonal feature in Fourier space. For analysis, we compute the azimuthally averaged one-dimensional power spectra for all fields, including the FLS field. To compute one-dimensional power spectra, we define radial bins, $k_i$, equally spaced in log-space with $\delta \log k_i \simeq 0.11$. Next, we define annuli in two-dimensional Fourier space, such that each annulus is a set of pixels belonging to a given radial bin $k_i$. The annuli are represented in Fourier space in Fig.~\ref{fig_FFT_2D} for the ELAIS-N1 and FLS fields, as examples. Note that our choice of bin size is larger than the widths of `rings' that can be discerned in the Fourier image, which arise from the masking of the maps. We further tested increasing and decreasing the bin size, and confirm that the results are robust to the choice of bin size, provided that bins of our chosen size or larger are used. The cross-power spectrum is then calculated for each annulus defined by $k_i$, the $i$-th bin in $k$-space, by taking the average over all pixels within the annulus, 
\begin{align} \label{eq_Xpower}
P_{\rm C\times S}(k_i)&=\frac{\sum_{\boldsymbol{k}\in k_i}{F_{\rm C}(\boldsymbol{k})  F_{\rm S}^*(\boldsymbol{k})}} {N_{\boldsymbol{k}\in k_i}}\,, \nonumber \\
&=\Big<{C_{\rm C}(\boldsymbol{k})C_{\rm S}(\boldsymbol{k})e^{i\Theta(\boldsymbol{k})}}\Big>_{\boldsymbol{k}\in k_i}\,,  
\end{align}
where $N_{\boldsymbol{k}\in k_i}$ is the total number of pixels in bin $k_i$, and $\Theta(\boldsymbol{k})\equiv \left[\theta_{\rm C}(\boldsymbol{k})-\theta_{\rm S}(\boldsymbol{k})\right]$. Any one-dimensional quantity averaged from two-dimensional Fourier space is obtained in the same way throughout this paper. For example, in Fig.~\ref{fig_FFT_2D_hist} we present histograms of the Fourier transforms for each of the annuli, from which we compute the average cross-power spectra, corresponding to Fig.~\ref{fig_FFT_2D}. The histograms for the FLS field are shown to have higher average values and larger dispersions, relative to the ELAIS-N1 and other fields; that is because of the strong Galactic emission present in this field. Finally, since the quantities and maps that we are dealing with are all real values only, we only take the real parts of $P_{\rm C\times S}$ to present the cross-power spectra. 

The true underlying power spectra are related to the measured spectra through the effects of masking, the map-making process, and the instrumental beam as
\begin{equation} \label{eq_Pk}
P^{\rm meas}(k) = \sum_{k'} M_{kk'} T(k') B^2(k') P^{\rm true}(k') + N(k), 
\end{equation}
where $T(k)$ is the transfer function for SPIRE maps, which is a convolution in real space and thus a multiplication in $k$-space, representing the suppression of modes from the map-making process, $B^2(k)$ is the beam function, describing beam-smoothing effects on the power spectrum (the square of the beam that affects the map), and $N(k)$ is the noise power spectrum, which we assume to be zero on average for cross-correlations. The quantity $M_{kk'}$ is the mode-mode coupling matrix \citep{Hivon2002}, which describes the impacts of masking, and is approximated for a flat sky as
\begin{equation} \label{eq_Mk}
M_{kk'} = \sum_{\boldsymbol{k}\in k} \sum_{\boldsymbol{k}\in k'} \langle w^{\rm C}_{kk'}w^{\rm *S}_{kk'} \rangle / N_{\boldsymbol{k}\in k},
\end{equation}
where $\langle w^{\rm C}_{kk'}w^{\rm *S}_{kk'} \rangle$ is the cross-power spectrum of the masks for a pair of SPIRE and CFIS maps. 

Writing Eq.~\ref{eq_Pk} in vector and matrix form makes it convenient to describe the next steps in our analysis process. We have 
\begin{equation} \label{eq_Pk_vec}
{\boldsymbol P^{\rm meas}} = \mathbfss{M} {\boldsymbol P^{\rm decoup}} + {\boldsymbol N}, 
\end{equation}
where ${\boldsymbol P}^{\rm meas}$ is a vector containing $P^{\rm meas}(k_i)$ as its elements, $\mathbfss{M}$ is the mode-coupling matrix, and ${\boldsymbol P^{\rm decoup}}$ is a vector containing the mode-decoupled spectrum at each $k_i$ as its elements, which is the Hadamard product (for which we use the symbol $\odot$) between ${\boldsymbol P^{\rm true}}$, the transfer function, and the beam function, namely ${\boldsymbol P^{\rm decoup}} = {\boldsymbol T}\odot {\boldsymbol B}^2 \odot {\boldsymbol P^{\rm true}}$ ($P^{{\rm decoup},i} = T_iB^2_iP^{{\rm true}, i}$, if expressed element-wise). Ignoring the noise term, the true cross-power spectrum is then recovered by inverting the mode-coupling matrix, 
\begin{equation} \label{eq_Pdecoupled}
{\boldsymbol P^{\rm decoup}} = {\boldsymbol M}^{-1} {\boldsymbol P^{\rm meas}}, 
\end{equation}
and by dividing ${\boldsymbol P^{\rm decoup}}$ with ${\boldsymbol T}$ and ${\boldsymbol B}^2$ element-wise, namely via Hadamard division (for which we use the symbol $\oslash$).  Thus we recover the true power spectrum, 
\begin{equation} \label{eq_Ptrue}
{\boldsymbol P^{\rm true}} = ({\boldsymbol P^{\rm decoup}}\oslash {\boldsymbol B}^2) \oslash {\boldsymbol T}, 
\end{equation}
or equivalently $P^{{\rm true}, i} = P^{{\rm decoup}, i}/(T_i B^2_i)$ considering the terms element by element.

\subsubsection{Masking}

We correct for the impact of masking on the measured power spectra by computing the mode-coupling matrix and then inverting it (Eq.~\ref{eq_Mk} and ~\ref{eq_Pdecoupled}). We calculate the power spectra of the masks from maps whose values equal unity in the masked regions and taper to zero with a Gaussian profile that has a FWHM of 90\,arcsec. From tests using simulated maps with similar power spectra to the data, we find that the recovered power spectra using the mode-coupling matrix obtained in this way are unbiased relative to the true input power spectra.

\subsubsection{Filtering and transfer functions}

Measurements of power spectra derived from observations can also be affected by map-making processes when generating final data products. A typical example is spatial filtering. Many surveys implement some filtering that suppresses or retains certain Fourier modes, in order to minimize noise and remove unwanted artefacts. However, this process will bias the signal that we are trying to measure. To account for this effect, one has to either apply the same map-making process to the models, or directly undo the filtering on the data. We choose the latter approach, since it is the unbiased spectra that we are interested in. The filtering leading to a suppression of modes is called the `transfer function'. It is corrected for by dividing the mode-decoupled spectrum $P^{\rm decoup}$ with this transfer function (see Eq.~\ref{eq_Ptrue}).

The {\sc SMAP} mapmaker \citep{Levenson2010,Viero2013} used to produce the HerMES maps performs a mild high-pass filter on the SPIRE maps, which is a convolution process in real space; this suppresses not only large-scale correlated noise, but also some of the large-scale physical signal that we are looking for. To correct for this suppression, we adopt the average transfer function from \citet{Viero2013}, which was measured from Monte Carlo simulations where the {\sc SMAP} pipeline was run on simulated maps with identical masking and filtering identical as in the data. We divide the decoupled spectra by the transfer function to compute the true underlying power spectra. The variation in the transfer function between the different fields is negligible relative to the uncertainty of the estimate, particularly\footnote{In our notation, $k$ is related to multipole, $\ell$, by $\ell=2\pi k$, and thus approximately related to the angular scale, $\theta$, by $\theta\simeq1/2k$.} at $k\gtrsim0.02\,{\rm arcmin}^{-1}$. Because the estimation of the transfer function is less reliable at $k\lesssim0.02\,{\rm arcmin}^{-1}$ or at $k\gtrsim(0.6\times 250\,\micron / \lambda )\,{\rm arcmin}^{-1}$, we limit our analysis to $k=[0.02, 0.6\times 250\,\micron/\lambda]\, {\rm arcmin}^{-1}$. 

Unlike the SPIRE maps, the mosaicking processes used to create the CFIS $r$-band maps are designed to preserve the modes on scales beyond the size of tiles. One step in the whole process that can potentially lead to loss of signal is modulation of the median background in each CFIS frame, which is mainly to remove the large scale tilt caused by extended straylight at
the camera field-of-view scale. While this can produce a `dip' in the amplitude of some modes, the scales impacted by this are expected to be the size of each frame, namely about 1\,deg, thus outside the range of scales concerned in this study. We therefore apply no transfer function correction for the CFIS images, since the signals are expected to be preserved on all relevant scales.

\subsubsection{Instrumental beams}

Finally, we correct for both the \textit{Herschel} and CFHT instrumental beams, which attenuate power on small scales. 
The CFIS maps are prepared to have the same resolution as the SPIRE maps. To do this, the CFIS maps are convolved with the same beam and re-sampled with the same grid as in the SPIRE maps. They thus have the same beam functions as the SPIRE maps to correct for in the extraction of the power-spectrum signal.  We approximate the CFIS/SPIRE beam as Gaussian, which closely matches the SPIRE maps of Neptune contained in the HerMES DR4, following the procedure described in \citet{Viero2013}. Alternatively, we apply a Gaussian kernel with the FWHM of the SPIRE beams (namely, 18.1, 25.5 and 36.6\,arcsec at 250, 350 and 500\,$\micron$, respectively) to our simulated maps described in Sect.~\ref{sec_test} to measure the impact of the beam on the power spectra. In this case, the correction for the beam is measured as the ratio between the power spectra before and after the smoothing. We find that the beam functions computed in these two ways are consistent, with negligible difference over our range of interest in $k$-space. In order to obtain $P^{\rm true}$, we divide the decoupled spectra by the beam function (Eq.~\ref{eq_Ptrue}). 

\subsubsection{Estimating uncertainties}

We estimate the uncertainties for the cross-power spectra in a similar manner as jackknife resampling. Each SPIRE field, overlapping with a CFIS area, is divided into 200 sub-regions of roughly equal size. The cross-power spectra is computed after eliminating one of the 200 sub-regions (which we denote as $P^j(k)$), and then we replace it and remove the next sub-region, thus obtaining a total of 200 measurements. From the set of $P^j(k)$, the error on the mean for each field, $\sigma_{P^{\rm field}(k)}$, is obtained as 
\begin{equation} \label{eq_err}
\sigma_{P^{\rm field}(k)} = \sqrt{\frac{N_{\rm J}-1}{N_{\rm J}}\sum_{j=1}^{N_{\rm J}}{(P^j(k)-P^{\rm field}(k))^2}},
\end{equation}
where $N_{\rm J}$ is the total number of sub-regions, and $P^{\rm field}(k) = \sum_{j=1}^{N_{\rm J}}{P^j(k)} / N_{\rm J}$. 
Later in Sect.~\ref{ssec_modelfit} where we interpret the observational measurements in a halo-model framework via fitting to a model, we account for a bias in the inverse of the covariance matrix by multiplying the covariance with the so-called Hartlap factor \citep{Hartlap2007} of $(N_{\rm J} - 1)\,/\,(N_{\rm J} - N_k - 2)$, where $N_k$ is the number of bins in $k$-space.

\section[results]{Results}
\label{sec_results}

\subsection{Accounting for Galactic cirrus}
\label{ssec_cirrus_res}

\begin{figure*}
\includegraphics[width=1.01\linewidth]{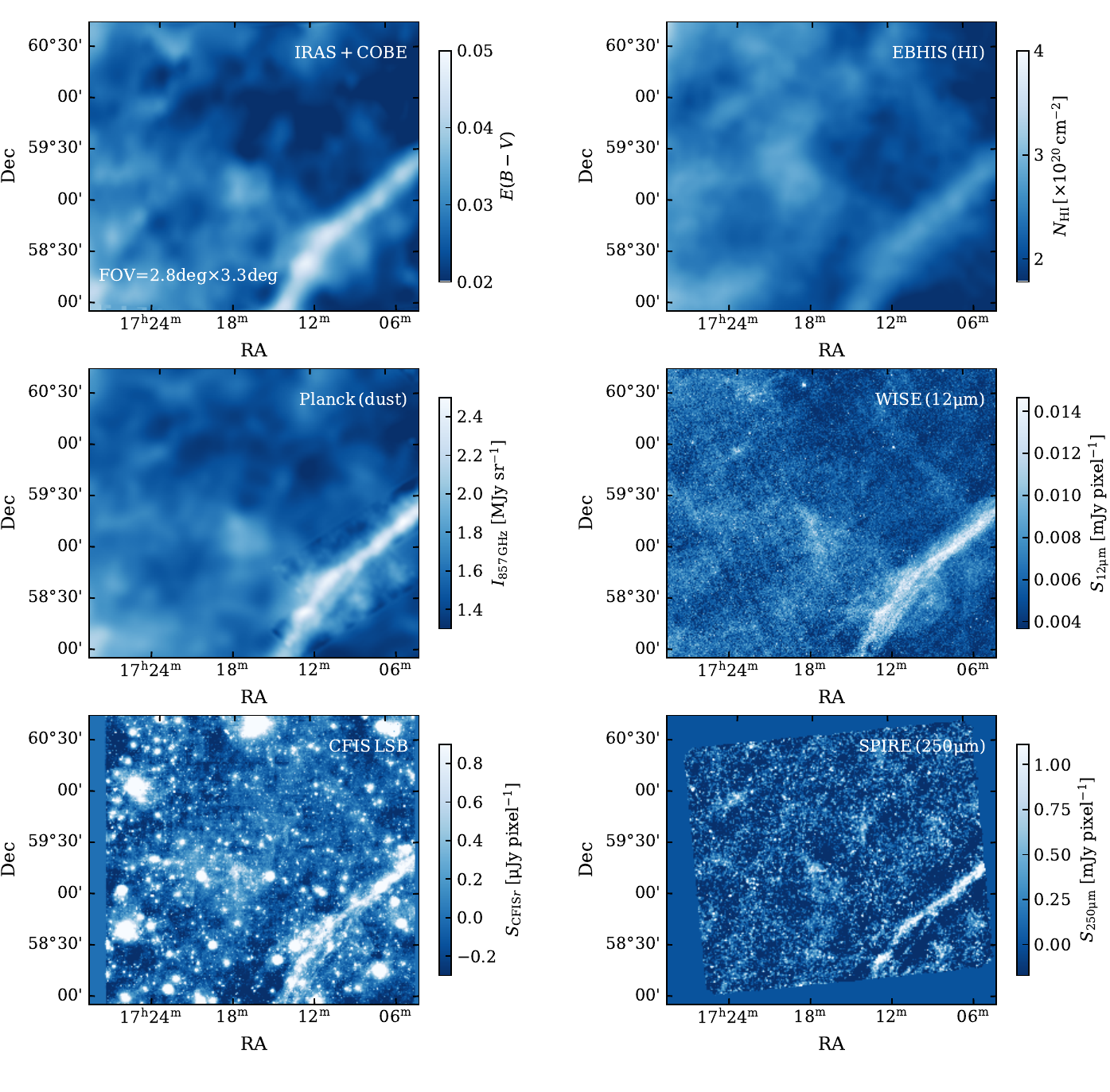}
\caption{Various maps of the FLS field. The intensity scales used are linear for all panels. {\it Top left:} Galactic reddening map from \citet{Schlegel1998}. {\it Top right:} Effelsberg-Bonn H\,{\sc i} Survey \citep[EBHIS;][]{Winkel2016} map, showing the H\,{\sc i} column density, obtained by integrating over $|v_{\rm LSR}| < 600\,{\rm km\,s^{-1}}$. {\it Middle left:} {\it Planck\/} GNILC map \citep{Planck2016b} at $857\,{\rm GHz}$. {\it Middle right:} {\it WISE\/} 12-$\micron$ map \citep{Meisner2014}. {\it Bottom left:} CFIS $r$-band LSB `B3+{\sc Montage}' version. {\it Bottom right:} {\it Herschel}-SPIRE map at $250\,\micron$. The correlation visually between the various maps is clearly present. Note, however, the scales probed by our cross-power spectra analysis are too small in this figure to be identified by eyes. }
\label{fig_Galactic_map_FLS}
\end{figure*}

\begin{figure*}
\includegraphics[width=0.9\linewidth]{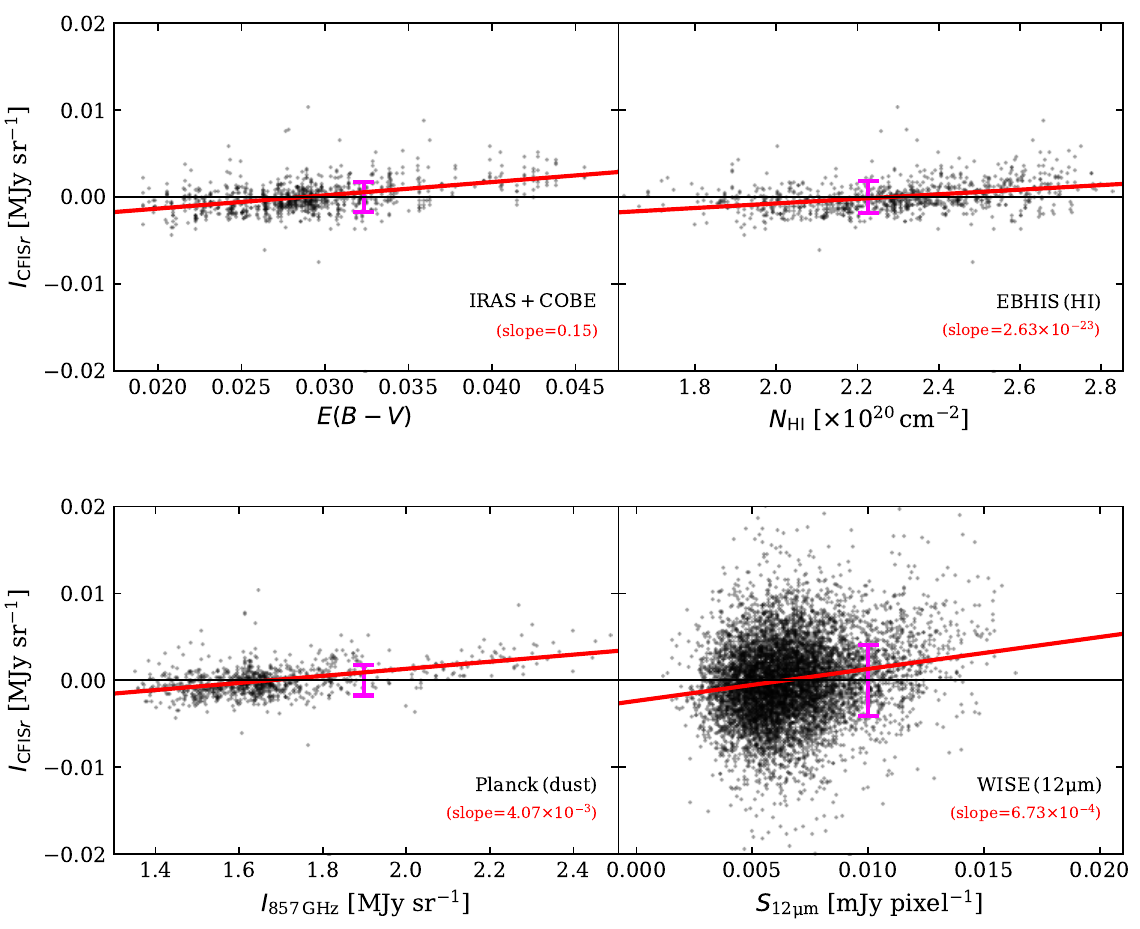}
\caption{Scatter plot of pixel values (dots) from the original CFIS LSB map versus the four independent `Galactic' maps for the FLS field. The red solid lines show the linear regression between the maps, namely $f(D)$ in Eq.~\ref{eq_cirrus_fit}. The fit $f(D)$ is subtracted from the data such that the resulting map has an average of zero (black horizontal line), to minimize the contribution from Galactic cirrus. The error bars in the middle of the panels show the scatter in the CFIS map. The size of error bars higher than or comparable to the trends seen in the linear relation for most of the pixels indicate that the Galactic contribution is only moderate relative to the CIB, even in the original maps before subtraction. This is particularly true when the same relation is used to subtract the Galactic cirrus in the other fields, in which the values and range of horizontal axis, i.e. the Galactic contamination, are significantly smaller. } 
\label{fig_regression_CFIS}
\end{figure*}

\begin{figure*}
\includegraphics[width=0.9\linewidth]{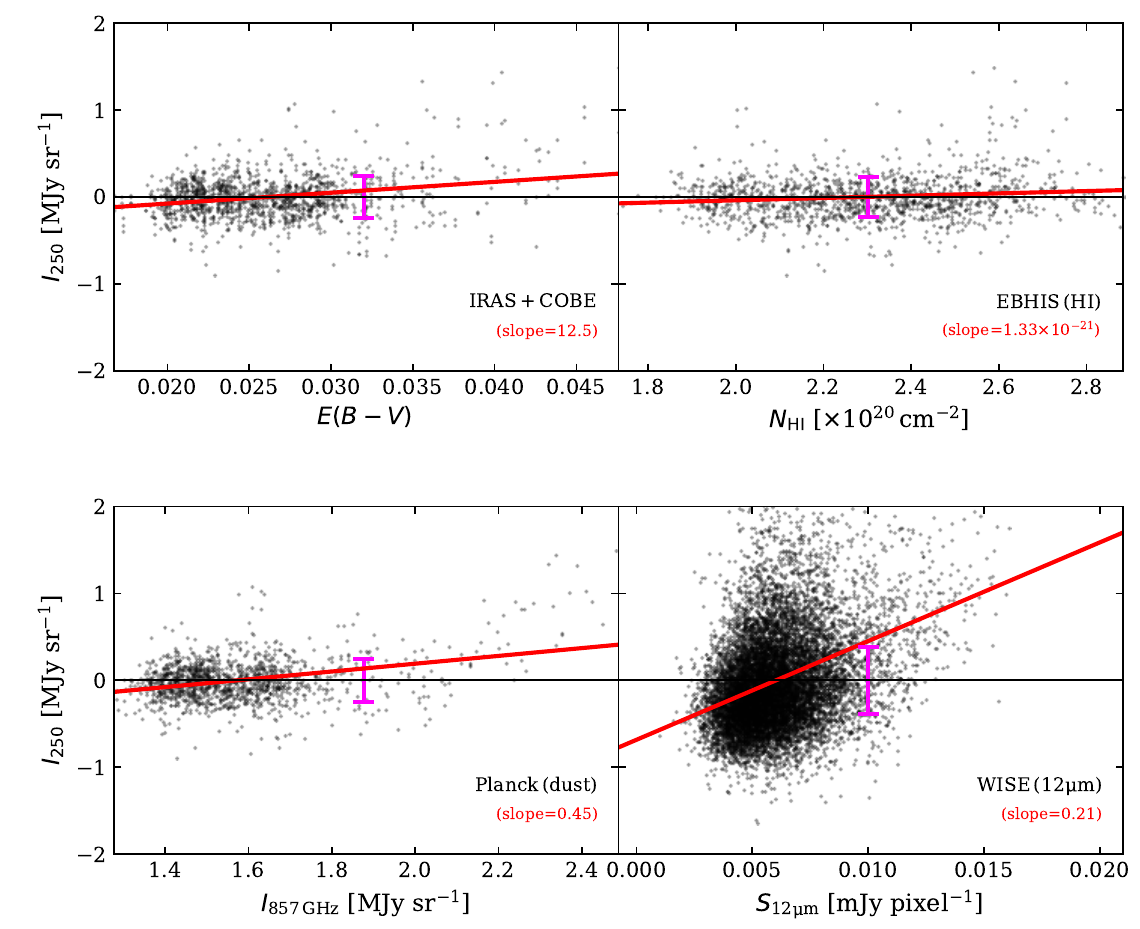}
\caption{Scatter plot of pixel values (dots) from the original SPIRE $250\,\micron$ map versus the four independent `Galactic' maps for the FLS field. The red solid lines show the linear regression between the maps, namely $f(D)$ in Eq.~\ref{eq_cirrus_fit_SPIRE}. The fit $f(D)$ is subtracted from the data such that the resulting map has an average of zero (black horizontal line), to minimize the contribution from Galactic cirrus. The size of error bars higher than or comparable to the trends seen in the linear relation for most of the pixels indicate that the Galactic contribution is only moderate relative to the CIB, even in the original maps before subtraction. This is particularly true when the same relation is used to subtract the Galactic cirrus in the other fields, in which the values and range of horizontal axis, i.e. the Galactic contamination, are significantly smaller. }
\label{fig_regression_SPIRE}
\end{figure*}

The most important source of residual systematics in the cross-correlation measurement is the Milky Way's cirrus, i.e. diffuse dust clouds, which emit in the far-IR and either absorbs or scatter light in the optical. \citet{MivilleDeschenes2016}, for example, demonstrated strong correlations of Galactic dust emission existing between the {\it WISE}, {\it Planck\/} and MegaCam images (also seen in Fig.~\ref{fig_Galactic_map_FLS}). Even though none of the fields chosen in our analysis are close to the Galactic plane (all fields have $b\gtrsim35^{\circ}$; see Table~\ref{tab_fields}), the cirrus contamination in our measurements may still be significant because of the faint signals we are looking for. In order to assess the impact of Galactic contamination, we use four independent maps as indirect tracers of Galactic dust: the {\it IRAS-} and {\it COBE}-based Galactic dust extinction map \citep[][SFD]{Schlegel1998}; the H\,{\sc i} column density data from the Effelsberg-Bonn H\,{\sc i} Survey \citep[EBHIS;][]{Winkel2016}; the {\it Planck\/} GNILC dust map \citep{Planck2016b}; and the {\it WISE\/} 12-$\micron$ map \citep{Meisner2014}, which traces line emission of PAH molecules and is known to correlate to a certain degree with dust emission. Each of these maps has its own advantages and disadvantages, and thus a unique set of systematics, as will be described below. By using this set of maps, instead of relying on one particular map alone, we ensure that our conclusions are robust against residual Galactic contamination. \newline

{\bf 1---The Schlegel, Finkbeiner \& Davis (SFD) Galactic extinction map:}
\citet{Schlegel1998} combined the {\it IRAS\/} and {\it COBE}-DIRBE data to produce an all-sky map of Galactic reddening, $E(B-V)$ (in magnitude units), at a resolution of a few arcminutes. They first photometrically calibrated the {\it IRAS\/} 100-$\micron$ intensity images using {\it COBE}-DIRBE, then used the ratio of the intensities at 100\,$\micron$ to 240\,$\micron$ to derive the temperature and column density of the dust, assuming a single temperature through a given line of sight. The column-density map was then converted to a reddening map, $E(B-V)$, using Mg\,{\sc ii} index measurements of early-type galaxies, which are known to tightly correlate with intrinsic $B-V$ colour. The SFD map for the FLS field used for subtracting the Galactic cirrus is shown in Fig.~\ref{fig_Galactic_map_FLS} as an example. We use the updated calibrations and conversions provided in table~6 of \citet{Schlafly2011} to estimate bandpass-specific amounts of extinction for the CFIS $r$-band, defined as $A_r$, from the $E(B-V)$ map. Galactic cirrus is expected to not only attenuate the extragalactic light, but also to scatter light from within the Galaxy. To correct for the overall impact of Galactic cirrus in the CFIS images, we fit the following function to the maps: 
\begin{equation} \label{eq_cirrus_fit}
\overline{S}_{\rm C, map} = \overline{S}_{\rm C, exgal} \times 10^{-A_r/2.5} + f(D).
\end{equation}
Here $\overline{S}_{\rm C, map}$ is the expected average flux density in the CFIS image for pixels of a given value $D$ of an external map ($D=E(B-V)$ for the SFD map), $\overline{S}_{\rm C, exgal}$ is a constant representing the average flux density in the CFIS image of extragalactic origin along a line of sight before the effect of Galactic extinction, and $f(D)$ is a linear relation to describe the Galactic emission. Note that the first term on the right-hand side of Eq.~\ref{eq_cirrus_fit} describes the average extragalactic emission after attenuation by Galactic cirrus for the CFIS pixels that have a corresponding value of $D$. We simultaneously fit $\overline{S}_{\rm C, exgal}$ and the coefficients of the polynomial. While the fitting can be performed jointly or individually for each of the five fields, we find that our results do not change depending on the choice. Throughout the analysis, we choose to use the $f(D)$ obtained from the FLS field to subtract the cirrus for all fields. This is because the other fields have much less cirrus, the fluctuations in the maps being dominated by the CIB, and thus their results are significantly noisier compared to the FLS field.  Furthermore, the relation is expected to be more or less uniform over the sky in the density regime probed in our analysis. Figure~\ref{fig_regression_CFIS} shows a scatter plot of pixel values between the CFIS map and the SFD map for the FLS field, together with the best fit $f(D)$. The scatter of pixel values from the CFIS map, shown by the error bar in the middle, is higher than the change due to the slope of $f(D)$ for most of the pixels, indicating that the Galactic contribution is in most cases dominated by or at most comparable to the CIB anisotropies even before the subtraction. This is particularly true for the fields other than the FLS where $E(B-V)$ and its range are smaller. The true flux density of extragalactic origin for {\it each\/} pixel in the CFIS image, $S_{\rm C, exgal}$, is then estimated by, 
\begin{equation} \label{eq_cirrus}
S_{\rm C, exgal} = [S_{\rm C, map} - f(D)] \times 10^{0.4A_r},
\end{equation}
where $S_{\rm C, map}$ is the flux density value of a given pixel from the original CFIS map. 

Similarly, to estimate the contamination of Galactic cirrus in the SPIRE bands, we perform another linear fit to pixel intensities from the dust map versus those from the SPIRE maps. Unlike the correction for the CFIS data, only emission is considered for the impact of Galactic cirrus at submm wavelengths,
\begin{equation} \label{eq_cirrus_fit_SPIRE}
\overline{S}_{\rm C, map} = \overline{S}_{\rm C, exgal} + f(D) \,.
\end{equation}
From the resulting fits, we determine the emission from Galactic cirrus in each of the fields, from which we estimate the extragalactic emission, 
\begin{equation}
\label{eq_cirrus_SPIRE}
S_{\rm C, exgal} = S_{\rm C, map} - f(D) \,.
\end{equation}
Figure~\ref{fig_regression_SPIRE} shows the scatter plot for the SPIRE 250-$\micron$ map versus the SFD map. As seen for the case of CFIS map, the SPIRE map also shows fluctations due to the CIB that are mostly higher than or comparable to that from Galactic cirrus even in the FLS field. The dominance by the CIB is found to be much greater in the other fields, due to the much weaker presence of Galactic contamination in those fields. \newline

{\bf 2---The H\,{\sc i} column-density map:}
The EBHIS \citep{Winkel2016} is a 21-cm survey conducted at the 100-m Effelsberg telescope (with approximately 10\,arcmin resolution), covering the entire northern sky out to $z\simeq 0.07$. The specific product used in our analysis is an H\,{\sc i} column-density map constructed by integrating all velocity components (relative to the local standard of rest) with $|v_{\rm LSR}| < 600\,{\rm km\,s^{-1}}$, accounting for most of the gas in the Milky Way. One advantage of using the H\,{\sc i} maps to account for Galactic cirrus is that, unlike the dust maps that we test, the H\,{\sc i} maps do not contain extragalactic contamination \citep[see e.g.][]{Chiang2019}. Using a CIB-contaminated dust map to subtract Galactic cirrus will result in a loss of the CIB signal. Similar to the analysis with the reddening map, we correct for the impact of Galactic cirrus in the CFIS images by jointly fitting for $S_{\rm C, exgal}$ and a linear relation representing the Galactic emission. This time, $D$ in Eq.~\ref{eq_cirrus_fit} is the H\,{\sc i} column density, and we estimate $A_r$ from the empirical relation of $N_{\rm H}\,/ \, A_V = 1.8\times10^{21}\, {\rm [mag^{-1}\, cm^{-2} ]}$ \citep{Predehl1995}, together with $R_{V}=3.1$ and the conversion from \citet{Schlafly2011}. The amount of cirrus contamination in the SPIRE maps can be estimated by performing a linear fitting between the EBHIS and SPIRE maps. The EBHIS map for the FLS field is presented in Fig.~\ref{fig_Galactic_map_FLS}, showing a case with stronger Galactic contamination in contrast to the other fields. Figures~\ref{fig_regression_CFIS} and \ref{fig_regression_SPIRE} show the scatter plot of pixel values from the CFIS map and the SPIRE map at 250\,$\micron$ (before the subtraction), respectively, versus the EBHIS map, together with the best-fitting $f(D)$. \newline

{\bf 3---The Galactic cirrus from the \textit{Planck} GNILC map \citep{Planck2016b}:}
Compared to the other dust maps, the GNILC map was built by implementing an explicit separation of the Galactic dust and the CIB, with the goal of minimizing the amount of extragalactic dust emission in the resulting Galactic map. While the map contains less extragalactic  emission compared to the SFD and {\it WISE\/} maps, a non-negligible amount of the CIB is still present, unlike in the H\,{\sc i} emission maps \citep{Chiang2019}. The component-separation process is carried out via a method called a generalized needlet internal linear combination (GNILC) by exploiting a spatial prior, namely the predominance of each component at different scales; away from the Galactic plane, the CIB anisotropies are dominant over the Galactic dust emission on small scales, while the Galactic cirrus dominates on large scales. To generate the map, the small-scale fluctuations, most of which are assumed to be CIB anisotropies, have been smoothed out by the GNILC processing. The GNILC map also has the advantage of having slightly better angular resolution compared with the EBHIS map, ranging from a few and up about 10 arcmin, varying with Galactic latitude. Similar to the analysis using the EBHIS and SFD maps, we follow Eq.~\ref{eq_cirrus_fit} to perform a linear fit and subtract that from the CFIS and SPIRE maps, to limit the contamination of Galactic cirrus in the maps. For the GNILC map, $D$ in Eq.~\ref{eq_cirrus_fit} is the specific intensity from the map. Following \citet{Chiang2019}, we derive a linear conversion from the specific intensity of the GNILC map to $E(B-V)$, such that the resulting average of $E(B-V)$ in a given field matches that from the SFD map. $E(B-V)$ thus obtained is then converted to $A_r$ in Eq.~\ref{eq_cirrus_fit} by using the conversion table from \cite{Schlafly2011}. Figure~\ref{fig_Galactic_map_FLS} shows part of the GNILC map corresponding to the FLS field. Figures~\ref{fig_regression_CFIS} and \ref{fig_regression_SPIRE} show the scatter plot of pixel values from the CFIS map and the SPIRE map at 250\,$\micron$ (before the subtraction), respectively, versus the GNILC, together with the best-fitting $f(D)$. \newline

{\bf 4---The full-sky \textit{WISE} 12-$\micron$ map of \citet{Meisner2014}:}
These data measure PAH emission, assumed to be an indirect tracer of the dust. The isolation of the Galactic component from the CIB in the {\it WISE\/} map is relatively poor compared to the GNILC \citep{Chiang2019} map. However, a great advantage of the {\it WISE\/} map is that it has much higher angular resolution (about 15\,arcsec) than the other maps we test here, roughly matching that of the SPIRE maps. The high resolution of the {\it WISE\/} map allows us to probe the impact of Galactic cirrus in the cross-correlation measurements on small angular scales. To minimize the contamination by Galatic cirrus, we perform the same analysis as for the GNILC map, using a polynomial fit, with a linear conversion of the {\it WISE\/} map to $E(B-V)$, and then a conversion to $A_r$, with $D$ in Eq.~\ref{eq_cirrus_fit} being the brightness per pixel. Figure~\ref{fig_Galactic_map_FLS} shows the {\it WISE\/} map for the FLS field as an example. The raw data were converted from data number (DN) to mJy per pixel using the conversion factor of $1.8326\times10^{-6}\,{\rm Jy}\,{\rm DN}^{-1}$ \citep{Wright2010, Cutri2012}. Figures~\ref{fig_regression_CFIS} and \ref{fig_regression_SPIRE} show the scatter plot of pixel values from the CFIS map and the SPIRE map at 250\,$\micron$ (before the subtraction), respectively, versus {\it WISE\/}, together with the best-fitting $f(D)$.

As clearly seen in Fig.~\ref{fig_Galactic_map_FLS} particularly with the Galactic features nicely represented in the CFIS LSB images, a strong correlation of Galactic origin is present in the data, which must be subtracted to extract the extragalactic signal. This is particularly true for the FLS field, with a weaker contamination in the other fields. 

\begin{figure*}
\includegraphics[width=0.99\linewidth]{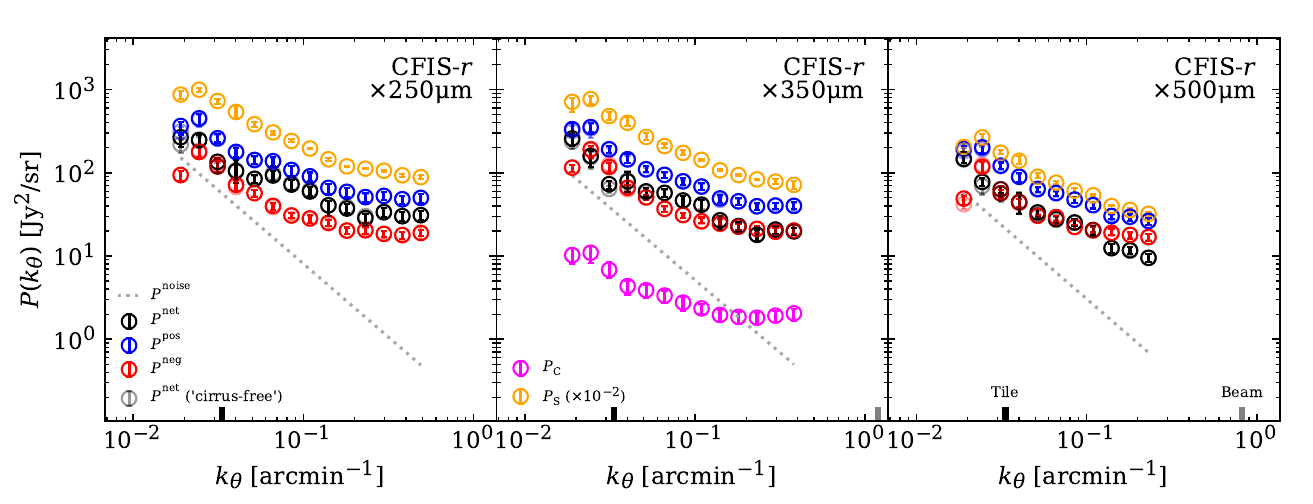}
\caption{Power spectra measured from the CFIS $r$-band and SPIRE maps at 250, 350 and 500\,$\micron$. The black data points show the net cross-power spectra as calculated using Eq.~\ref{eq_Xpower}. The blue and red data points show the positive and negative cross-power spectra, i.e. the power spectra calculated only from the pixels with $|\Theta(\boldsymbol{k})|$ (the phase difference of the Fourier transform between the two maps) smaller and greater than $\pi/2$, respectively (see Eq.~\ref{eq_Xpower_pos_neg} and the text). The positive power spectra are larger than the negative power spectra on all scales and in all wavebands, indicating that the net power spectra measures a physical, correlated signal rather than a statistical deviation from zero. As another check, we present the expected noise level in the case of no correlation (dotted line), obtained by shuffling the modulus and randomizing the phases of the Fourier transforms among pixels, and by measuring the cross-power spectra. The absolute values of the noise were taken to present only its magnitude, regardless of the sign. As seen, the noise level is much lower than the the net cross-power spectra, meaning that the measurement is a detection of correlation rather than noise. The faint symbols, which are almost indistinguishable from the dark symbols, are the measurements from the cirrus-free maps, indicating that the impact of Galactic cirrus is negligible. The impact of Galactic cirrus was estimated and subtracted here using the EBHIS map \citep{Winkel2016} with the conversion table of \citet{Schlafly2011}. A polynomial fit was used to compute and subtract the contribution of Galactic cirrus from the data (as described in detail in the text). We find the same conclusion when using other Galactic maps, such as the reddening map of \citet{Schlegel1998}. To guide the eye, the thick vertical tickmarks indicate the scales of a CFIS tile (black; 0.5\,deg) and the SPIRE beam (grey; 18.1, 25.5, and 36.6\,arcsec at 250, 350 and 500\,$\micron$, respectively); note that we cut off our estimates of the power spectrum at scales 2--3 times larger than the beam out of concern about controlling uncertainties in the transfer function corrections. For reference, the auto-power spectra of the CFIS (magenta; only shown in the middle panel) and SPIRE (orange) maps estimated using the same approach are also presented; note that in these units the CIB fluctuations are much higher than the COB ones, and so the SPIRE auto-power spectra have been divided by 100.}
\label{fig_Xpower_nomask}
\end{figure*}

\begin{figure*}
\includegraphics[width=0.99\linewidth]{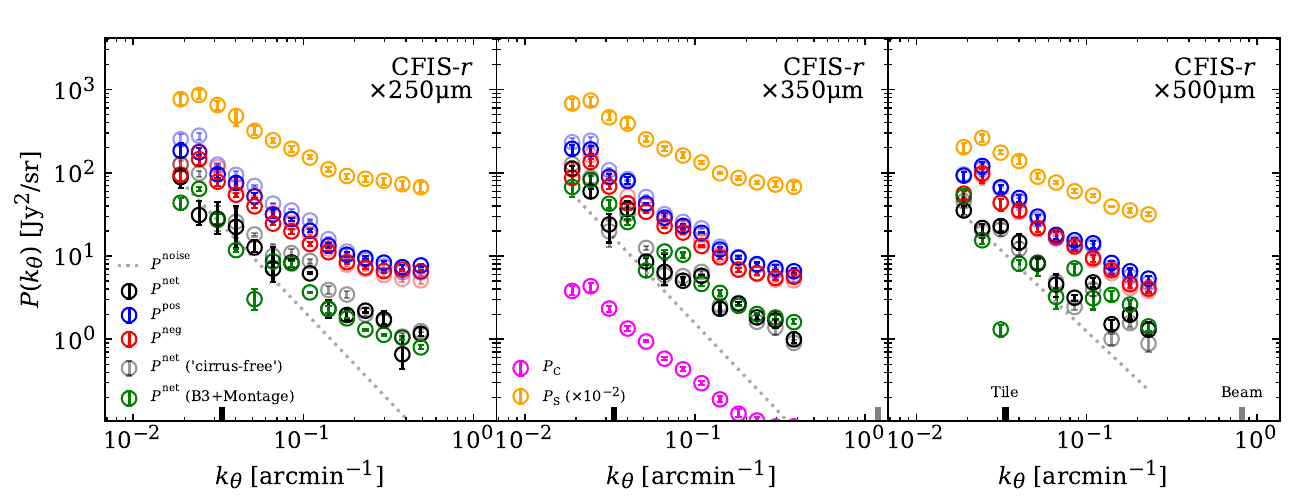}
\caption{Same power spectra measurements as in Fig.~\ref{fig_Xpower_nomask}, but after masking all sources detected in the CFIS map, including galaxies (see Sect.~\ref{ssec_mosaic}). The positive cross-power spectra (blue) are higher than the negative cross-power spectra (red), and the net cross-power spectra are higher than the noise levels expected in the case of no correlation (dotted) on all scales, meaning that there is a detection of a positive correlation from the diffuse components. As in Fig.~\ref{fig_Xpower_nomask}, the faint symbols represent the results from the `cirrus-free' maps. From the fractional difference between the black and faint symbols (i.e. dividing the latter with the former and subtracting it from 1 for each $k$ bin), we find that the contribution from Galactic cirrus in the signal is $\simeq 30$ per cent at maximum and typically a few per cent across the $k$ bins and wavebands. Also, the green symbols show the results based on another version of mosaics, constructed using improved stacks from the latest development for stable backgrounds, as well as using {\sc Montage} software optimized for preserving large-scale modes (see Sect.~\ref{ssec_mosaic}). No significant discrepancy between the results from two mosaics, as shown in this figure, thus mean that our measurements are not sensitive to the uncertainties in the two carefully crafted versions of the mosaics. The vertical tickmarks indicate the scales of a CFIS tile (black) and the SPIRE beam (grey).}
\label{fig_Xpower_mask}
\end{figure*}

\begin{figure*}
\includegraphics[width=0.99\linewidth]{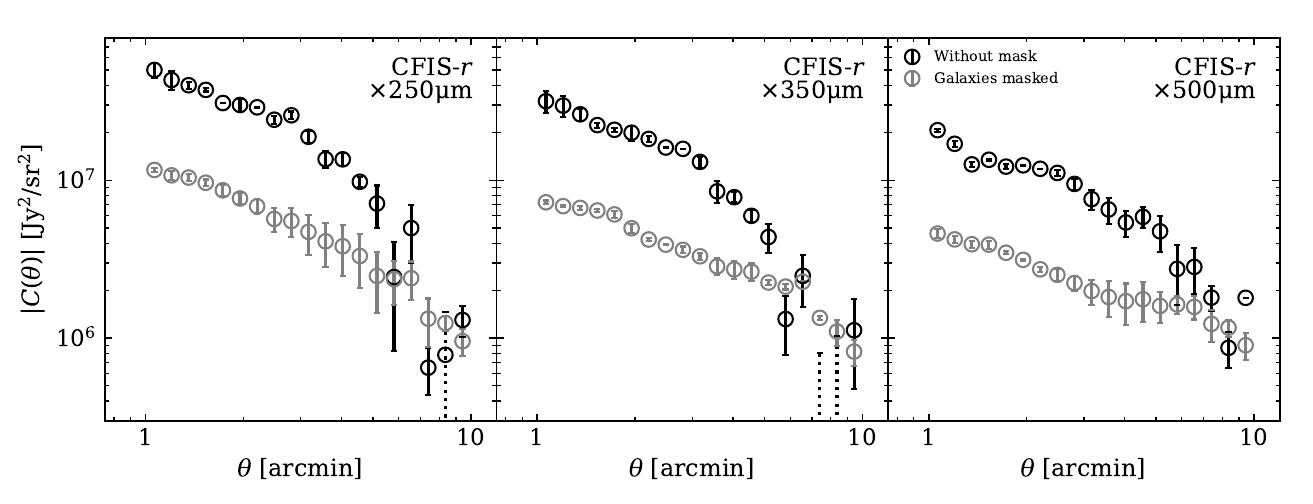}
\caption{Cross-correlation in real space, measured out to about 10\,arcmin, between the CFIS $r$-band and SPIRE maps at 250, 350 and 500\,$\micron$. The black data points represent the cross-correlation from the maps where individually detected galaxies are not masked, while the measurements from the galaxy-masked maps are shown in grey. Note that because some of the data points are negative, the absolute value has been taken on the vertical axis of this plot; dotted error bars are used to indicate negative values.}
\label{fig_Xreal}
\end{figure*}

\subsection{CFIS-SPIRE cross-power spectra measurements}
\label{ssec_res}
The cross-power spectra estimated from the `cirrus-free' (in an ideal case) map pairs are shown in Figs.~\ref{fig_Xpower_nomask} and \ref{fig_Xpower_mask}, without and with masks for detected galaxies, respectively. The same measurements, but presented in flux units, are also provided in Appendix~\ref{sec_appA} for those who are more familiar with these units. The true underlying power spectra are measured following the method described in Sect.~\ref{ssec_Xcorr}. For reference, the auto-power spectra of both the SPIRE and CFIS maps estimated the same way are also presented in these two figures. As mentioned in Sect.~\ref{ssec_Xcorr}, due to potential uncertainties included in the transfer functions for SPIRE maps from \cite{Viero2013}, we measure the power spectra only on scales larger than the SPIRE beam at least by a factor of $\simeq 3$. The average of the EGS, ELAIS-N1, ELAIS-N2, and HATLAS-NGP fields are shown in these figures. The power spectra from the FLS field are excluded because of the significant contamination from Galactic cirrus, particularly coming from the western side. This will be further discussed in this section.   

\begin{figure}
\includegraphics[width=0.93\linewidth]{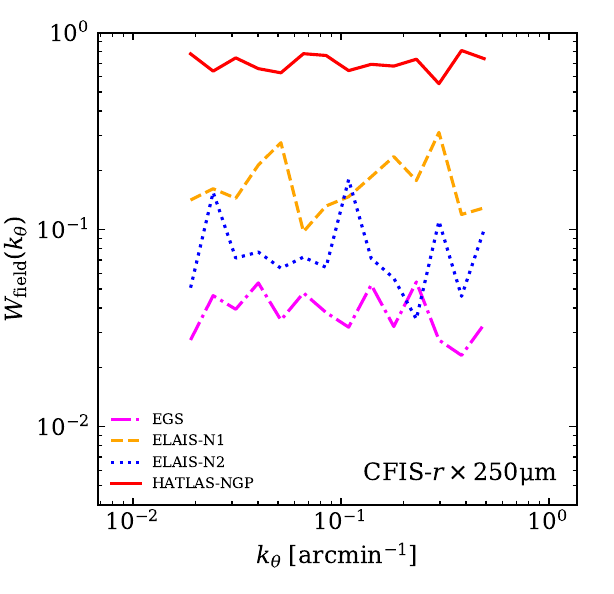}
\caption{Weight functions for the individual fields, $W_{\rm field}(k)$ in Eq.~\ref{eq_weight}, with the SPIRE maps at 250\,$\micron$ are shown. The functions for different SPIRE bands are similar qualitatively. }
\label{fig_weight}
\end{figure}

The combined cross-spectrum $P^{\rm comb}(k)$, averaged over all fields, is defined as
\begin{equation} \label{eq_Pcombined}
P^{\rm comb}(k) = \sum_{\rm field} W_{\rm field}(k) P^{\rm field}(k), 
\end{equation}
where $P^{\rm field}(k)$ is the power spectrum for a given field and $W_{\rm field}(k)$ its weight, defined as
\begin{equation} \label{eq_weight}
W_{\rm field}(k) = \frac{\sigma_{P^{\rm field}(k)}^{-2}}{\sum_{\rm field} \sigma_{P^{\rm field}(k)}^{-2}},
\end{equation}
where $\sigma_{P^{\rm field}(k)}$ is defined in Eq. \ref{eq_err}. $W_{\rm field}(k)$ for the individual fields with the SPIRE maps at 250\,$\micron$ are shown in Fig.~\ref{fig_weight}. The results for the other SPIRE bands are similar qualitatively. 

We measure the cross-power spectra using their absolute magnitudes (Eq.~\ref{eq_Xpower}), therefore the values are always non-zero and positive, even in the case of negative or no correlation. Finding positive values, therefore, does not necessarily mean that there is a correlation. One way to check whether the correlation is positive, negative, or zero, is to measure the power spectra from the Fourier transform separately for pixels `in phase', $P^{\rm pos}$, and `out of phase', $P^{\rm neg}$. To this end, we define the `positive' and `negative' parts of the cross-power spectra as
\begin{align} \label{eq_Xpower_pos_neg}
P^{\rm pos}(k_i)&=\frac{\sum_{\boldsymbol{k}\in k_i,\ |\Theta|\le\pi/2}{C_{\rm C}(\boldsymbol{k})C_{\rm S}(\boldsymbol{k})e^{i\Theta(\boldsymbol{k})}}} {N_{\boldsymbol{k}\in k_i}}, \nonumber \\
P^{\rm neg}(k_i)&=\frac{\sum_{\boldsymbol{k}\in k_i,\ |\Theta|\ge\pi/2}{C_{\rm C}(\boldsymbol{k})C_{\rm S}(\boldsymbol{k})e^{i\Theta(\boldsymbol{k})}}} {N_{\boldsymbol{k}\in k_i}}, 
\end{align}
where $\Theta$, which is defined in Eq.~\ref{eq_Xpower}, corresponds to the phase difference, at a given mode $\boldsymbol{k}$, between the Fourier transforms of the two maps. Similar to the net spectra, only the real parts of the left-hand sides of the equations are taken as the cross-power spectra. Furthermore, the sign is also reversed to show $P^{\rm neg}(k_i)$ in the figures, for ease of presentation. These are useful quantities to examine because they indicate whether we are looking at a correlation, an anti-correlation, or no correlation, by comparing the amplitudes of $P^{\rm pos}$ and $P^{\rm neg}$. In the case of no correlation, $P^{\rm pos}$ and $P^{\rm neg}$ are expected to be similar.\footnote{While they are expected to be exactly the same theoretically, resulting in a completely zero net spectra, there is still a residual noise that is not perfectly zero because of the finite realisations on the finite number of pixels.} In Appendix~\ref{sec_appB}, we present null tests that demonstrate this point, in which $P^{\rm pos}$ is seen to be similar to $P^{\rm neg}$. In Fig.~\ref{fig_Xpower_nomask}, $P^{\rm pos}$ (blue points) is higher than $P^{\rm neg}$ (red points) at all scales and in all SPIRE wavebands, showing that there is a net positive correlation between the SPIRE and CFIS maps. In Fig.~\ref{fig_Xpower_nomask}, we also plot the noise spectrum in the case of no correlation (dotted line). The noise spectrum is estimated by shuffling the modulus and randomizing the phases (for the $\Theta$s to be uniformly distributed between $-\pi$ and $\pi$) of the Fourier transforms among pixels that belong to each given radial bin of $k$, and by measuring the cross-power spectra, as defined by Eq.~\ref{eq_Xpower}. The shuffling and randomization of the modulus and phases are performed 1000 times to compute the average noise spectra. The absolute value of the resulting noise spectra are taken to show only the magnitudes of the noise, regardless of its sign. Note that since our choice of radial $k$ bins are equally spaced in log-space, the number of pixels in two-dimensional Fourier space belonging to each radial bin scales as $k^2$, making the average noise spectrum scale as $\{P_{\rm C}(k)\, P_{\rm S}(k)\}^{0.5}/k$. The net cross-power spectra are higher than the estimated noise spectrum, as shown in Fig.~\ref{fig_Xpower_nomask}, which confirms that SPIRE and CFIS maps are positively correlated, with a correlation signal significantly exceeding the noise level. From tests varying the bin sizes, the significance of the detection for each of the SPIRE bands converges to be no less than $18\,\sigma$. As seen in Appendix~\ref{sec_appB}, on the other hand, the net spectra agree with the noise spectra in the case of no correlation (i.e.\ null tests). 

Figure~\ref{fig_Xpower_mask} shows the cross-power spectra computed in the same way as above, but using the CFIS `background' map, where all detected sources in the CFIS images, including galaxies, are masked, as described in Sect.~\ref{ssec_mosaic}.  As in Fig.~\ref{fig_Xpower_nomask}, we present the net, `positive' and `negative' cross-power spectra separately. It is also found that $P^{\rm pos}$ is higher than $P^{\rm neg}$ by roughly a factor of 2, and that the spectra are higher than the noise level on almost all scales, indicating that there is a physical correlation from the diffuse component detected in excess of noise. The detection of the net cross-correlation signal for the background maps, combined over all $k$-space bins and investigated by changing the bin sizes, is greater than $14\,\sigma$. The statistical correlation that we detect from the power spectra, when divided by the auto-spectra of SPIRE maps, converts to an rms brightness of about $32.5\,{\rm mag/arcsec^2}$ in the CFIS $r$-band (as we discuss further in Sect.~\ref{sec_disc}). This is more than ten orders of magnitude fainter than the typical night sky brightness in the CFIS $r$-band data, which is $21.4\,{\rm mag/arcsec^2}$. We also find that the results from the two different versions of the mosaics (see Sect.~\ref{ssec_mosaic}) are in close agreement, meaning that our measurements are not sensitive to uncertainties potentially included in the construction of the mosaics. 

While we have tested and validated our method in Sect.~\ref{sec_test}, there may still be a concern about our treatment of masking effects and recovering the true power spectra using the mode-coupling matrix, as presented in Sect.~\ref{ssec_Xcorr}. As a consistency check, we also show the real-space cross-correlation in Fig.~\ref{fig_Xreal}, which should not be subject to the uncertainties introduced by masking. The detection significance is found to broadly agree with that from the power spectrum analysis, being greater than $18\,\sigma$.

\begin{figure*}
\includegraphics[width=0.99\linewidth]{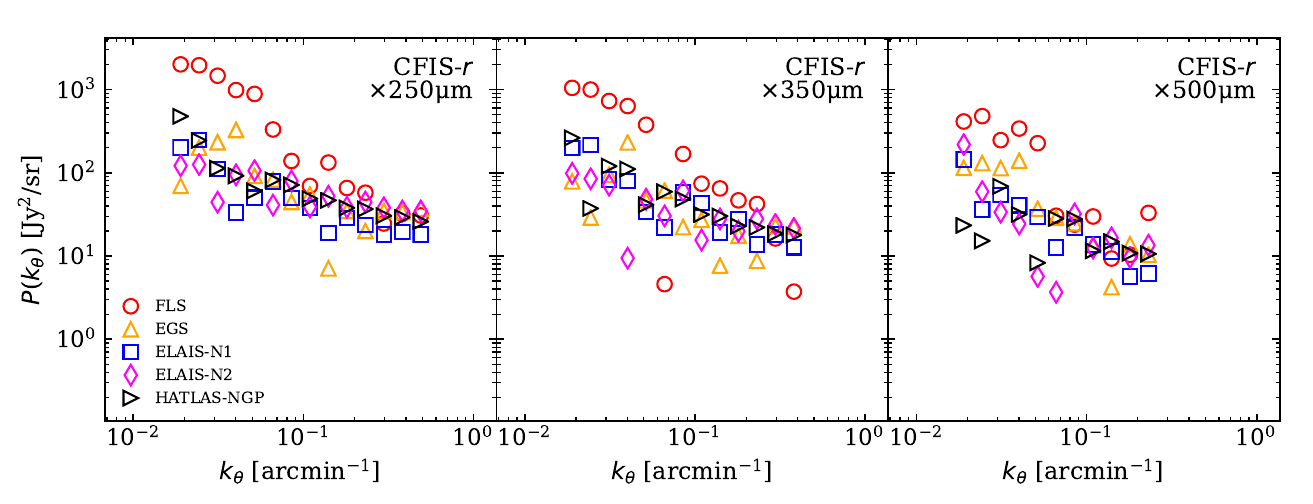}
\caption{Cross-power spectra measured for individual fields before the correction for the Galactic cirrus, {\it without\/} masking the galaxies detected in the CFIS images. Note that here we only show the net power spectra, without separating it into the positive and negative correlations. The results from the different fields are similar, given the relatively large measurement errors, except for that from the FLS field, which is highly contaminated by the Galactic foreground. }
\label{fig_Xpower_nomask_fields}
\end{figure*}

\begin{figure*}
\includegraphics[width=0.99\linewidth]{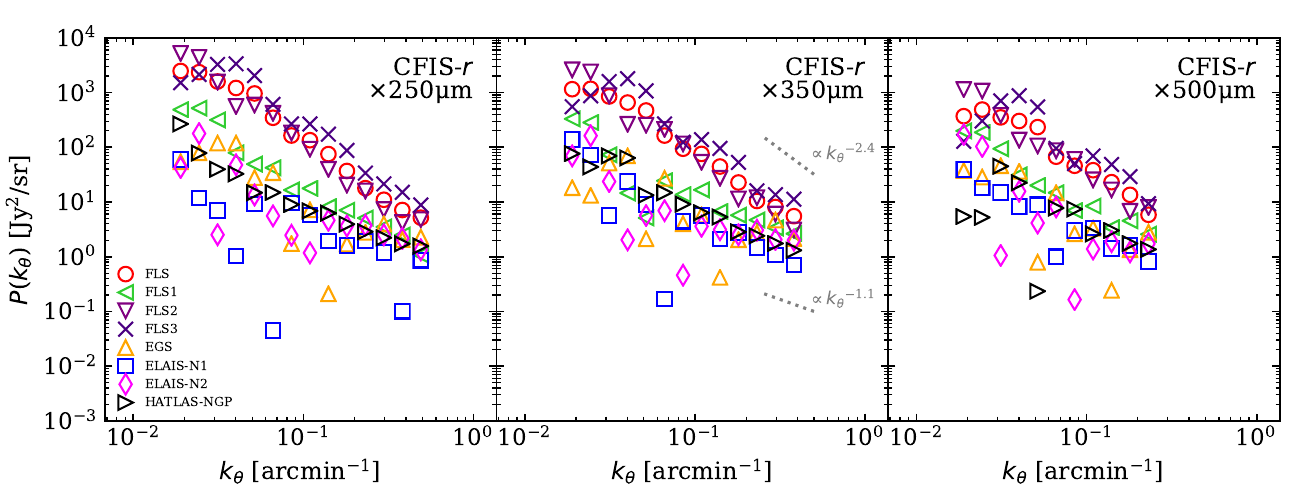}
\caption{The same cross-power spectra measured for individual fields before the correction for the Galactic cirrus as in Fig.~\ref{fig_Xpower_nomask_fields}, except that here the galaxy-masked CFIS images are used. The FLS field is further divided into three regions: the eastern third of the field, with the least contamination by Galactic cirrus (FLS1); the central third (FLS2); and the western third, with the most Galactic contamination, including a strong filament (FLS3).}
\label{fig_Xpower_mask_fields}
\end{figure*}

In Figs.~\ref{fig_Xpower_nomask} and \ref{fig_Xpower_mask}, the measurements before the correction for Galactic cirrus are also shown for ease of comparison; one can see that Galactic cirrus is not a significant source of contamination, even for the CFIS diffuse map where bright galaxies are masked. It is possible to quantify the contamination by dividing the power spectrum after the correction with that before the correction, and by subtracting the ratio from 1. We find that the contamination obtained for each $k$ bin and waveband is $\simeq 30$ per cent at maximum, and typically only a few per cent, for the CFIS diffuse maps. For our primary measurements, we present only the results obtained using the EBHIS data. This is because the other three maps tested are not confined to Galactic emission, but to a varying degree include some extragalactic signals along a given line of sight, as is the case for almost any kind of dust map in general \citep[e.g.][]{Chiang2019,Planck2014a,Planck2016a}. 

The method used for subtracting the cirrus in Sect.~\ref{ssec_cirrus_res} is complicated, with one potential bias being from the transfer function of SPIRE, which is used to recover the true power spectra only after the linear regression and subtraction steps are already carried out. This may lead to underestimation of the cirrus, by underestimating the linear regression. To check this, we used another version of the SPIRE maps, namely the `Level-3' products from the {\it Herschel\/} archive\footnote{\url{http://archives.esac.esa.int/hsa/whsa/}}, in which weaker large-scale filtering was applied through the map-making process compared to the HerMES or HELP maps. Using these Level-3 data, we repeated the same analysis, including the linear regression and subtraction of the Galactic cirrus, and found the same basic result that the correlation signal has a similar amount of Galactic contamination in it, not exceeding $\simeq 30$ per cent. This similarity in results is due to the fact that our analysis is based on cross-correlation, and thus any potential uncertainty or bias that is only included in one data set (the SPIRE map, in this case) does not propagate to the signal unless the other map (the CFIS data) is also affected by the same uncertainty. We stick with our use of the HerMES products here, since the transfer
function has been well characterized for these maps.  Since we do not extend to the largest angular scales beyond $\simeq30$\,arcmin where the transfer function becomes unreliable, in this sense the use of these SPIRE maps is also conservative. It is worth pointing out that the impact of the uncertainty in the transfer function and the map-making process for {\it both\/} data sets is something that could be pursued further in future studies. 

In Fig.~\ref{fig_Xpower_nomask_fields} we show the power spectra obtained for each of the four fields (namely EGS, ELAIS-N1, ELAIS-N2, and HATLAS-NGP) individually before the correction for the Galaxy. Figure~\ref{fig_Xpower_mask_fields} shows the same results as in Fig.~\ref{fig_Xpower_nomask_fields} but with the CFIS `background' maps. As can be seen, there is no obvious field-to-field variation, although the measurement errors and the fluctuations across scales are large. Here, for clarity, we only show the net power spectra, without showing the positive and negative spectra. We find similar conclusions for the positive and negative spectra, namely that there is no significant variation among the fields beyond the uncertainties in the measurements. 

\begin{figure*}
\includegraphics[width=0.99\linewidth]{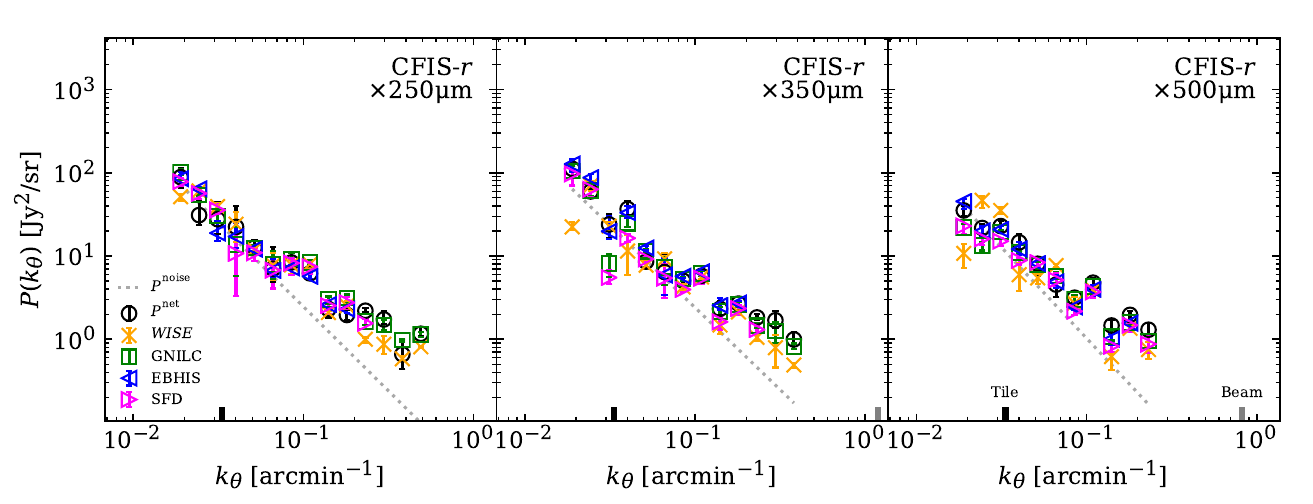}
\caption{Estimation of the cross-correlation signal between CFIS (galaxies masked) and SPIRE after accounting for Galactic dust contamination using {\it WISE\/} (cross), {\it Planck\/} GNILC (square), EBHIS (left triangle) and SFD (right triangle) maps. The subtraction of Galactic cirrus is performed in the same way as in Figs.~\ref{fig_Xpower_nomask} and \ref{fig_Xpower_mask}, i.e.\ by using a linear relation between each external map and the CFIS/SPIRE maps. By dividing each of the results after the subtraction of Galactic cirrus with those before the subtraction (the latter shown by the solid circles), we find that the median Galactic contribution to the signal among the four external maps for each $k$ bin and waveband is typically about 20 per cent (around 50 per cent at maximum). The scatter between the results from using the four maps (which reflects systematic uncertainties in using each of the maps) is also typically around 40~per cent (relative to the mean).}
\label{fig_Gal_sub}
\end{figure*}

\begin{figure*}
\includegraphics[width=0.99\linewidth]{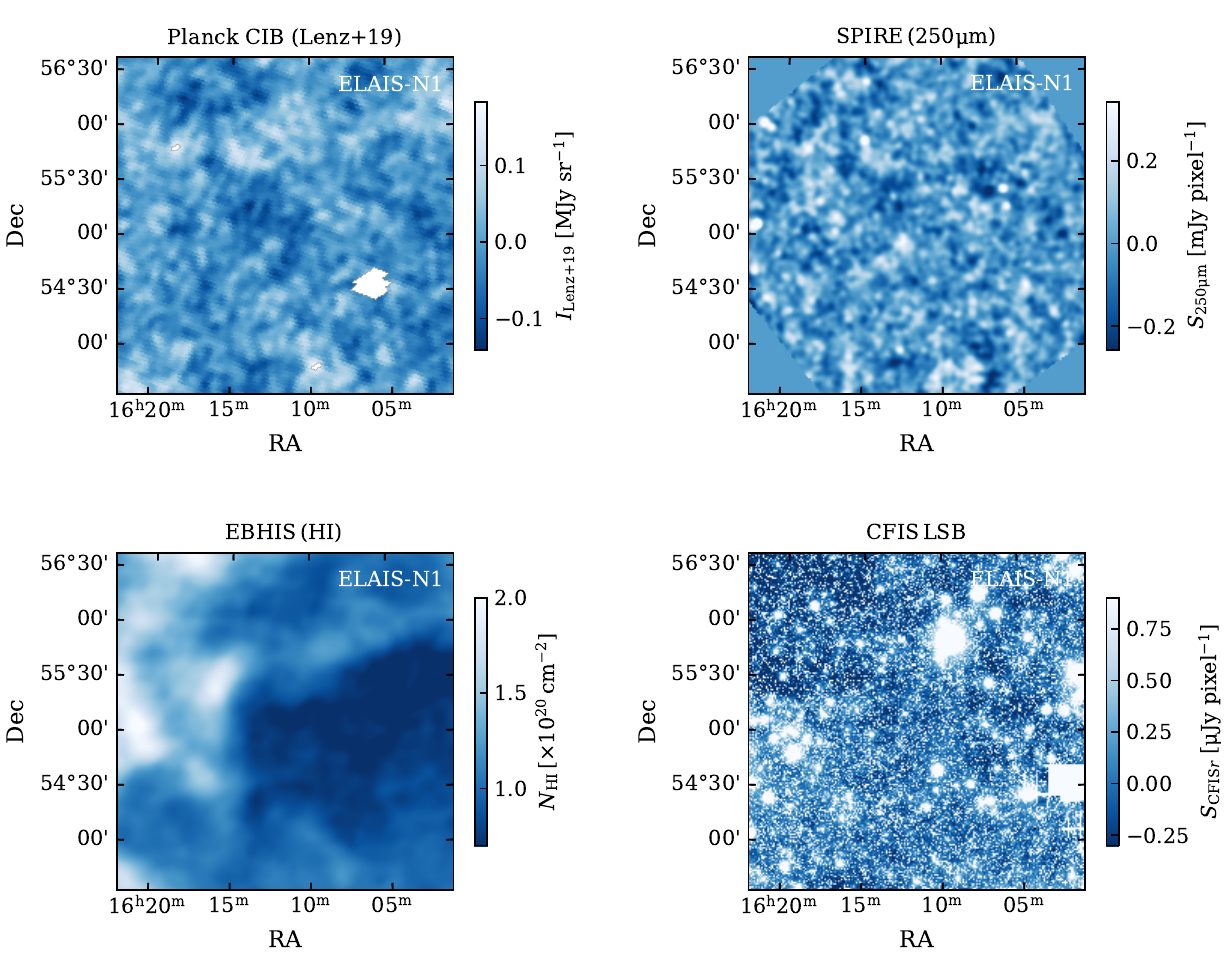}
\caption{Visual comparison in the ELAIS-N1 field (as an example) between the SPIRE $250\,\mu$m map, the {\it Planck\/} CIB map from \citet{Lenz2019}, which is intended to show only the CIB component derived from {\it Planck\/} using a regression with H\,{\sc i} surveys, EBHIS survey, and CFIS $r$-band LSB `B3+{\sc Montage}' version (galaxies are not masked). The strong visual agreement between the SPIRE and {\it Planck\/} map, in contrast to the much weaker correlation with the EBHIS, implies that fluctuations are dominated by the CIB rather than Galactic emission. In the bottom panels, the CFIS LSB map is also compared to the EBHIS map to present its correlation with the Galactic emission, although it is disturbed significantly by the bright individual sources.} 
\label{fig_SPIRE_Lenz}
\end{figure*}

In Fig.~\ref{fig_Gal_sub} we show the resulting cross-correlation obtained using each of the four Galactic cirrus estimators described above. We see that the maps are roughly consistent, with a typical Galactic contamination obtained as above being $\simeq 20$ per cent ($\simeq 50$ per cent at maximum) across the $k$ bins and wavebands. The numbers quoted here are the averages over the four estimators. This is also consistent with the results from tests using simulated maps in Sect.~\ref{ssec_cirrus}, where we mimic and apply the impact of Galactic cirrus to simulated maps of extragalactic signals. Note that all of the `Galactic' maps we use (except for the {\it WISE\/} map) have much lower resolution (a few to 10\,arcmin) compared to the SPIRE and CFIS maps. This, in principle, means that the fluctuations of Galactic origin in the SPIRE and CFIS maps on small scales are not accounted for by the subtraction method. Including the {\it WISE\/} map is important in this regard, since it confirms that accounting for the Galactic cirrus on small scales does not significantly change our results, as can be seen in Fig.~\ref{fig_Gal_sub}. This is because the power of Galactic cirrus drops with decreasing scale much more rapidly ($P(k)\propto 1/k^{2.5-3.5}$) than that of the CIB, which is known to be approximated by $P(k)\propto 1/k^{1-1.5}$, \citep[e.g.][]{MivilleDeschenes2007,MivilleDeschenes2010,MivilleDeschenes2016,Martin2010,Viero2013,Blagrave2017}. As can be seen, the slopes of the cross-power spectra in our measurements are also consistent with the published range of the exponent for the CIB. We further tested other `Galactic' maps, such as those by  \citet{Schlafly2014} and \citet{Green2019}, as well as the Green Bank Telescope (GBT) H\,{\sc i} Intermediate Galactic Latitude Survey \citep[GHIGLS;][]{Martin2015} and the DRAO H\,{\sc i} Intermediate Galactic Latitude Survey \citep[DHIGLS;][]{Blagrave2017}, but found no significant difference in the results, with the Galactic contribution being no more than 50~per cent (and mostly much smaller).  Among these maps, only the DHIGLS (that is only available for the ELAIS-N1 field) should be an unbiased estimator of Galactic emission, with a resolution of approximately $1$\,arcmin. We find that the level of the correction obtained with the DHIGLS map is similar or smaller on all scales when compared to the other maps tested in the ELAIS-N1 field. Finally, we also performed tests using the {\it Planck\/} CIB map from \citet{Lenz2019} (see Fig.~\ref{fig_SPIRE_Lenz}), who separated the CIB component from the {\it Planck\/} maps by using regressions between the {\it Planck\/} maps and H\,{\sc i} surveys (EBHIS for the Northern sky). Because of the lower 5\,arcmin resolution of their map, which is limited by the {\it Planck\/} beam, the reliable range of scales for cross-power spectra analysis is restricted and the uncertainties when compared to our analysis are unclear. Residual Galactic signal on smaller scales, for example, could still be present in the map, contaminating the cross-power spectra. Aside from such uncertainties, we confirm there are correlations between our CFIS/SPIRE data and the map from \citet{Lenz2019}. The clear visual agreement between the SPIRE (even before cirrus subtraction) and CIB map from \citet{Lenz2019}, in contrast to the much weaker correlation with the EBHIS H\,{\sc i} map, as seen in Fig.~\ref{fig_SPIRE_Lenz}, reassures us that the fluctuating background is dominated by the CIB rather than Galactic emission. Figure~\ref{fig_SPIRE_Lenz} also shows the visual correlation between the CFIS LSB and EBHIS map, although such correlation is significantly disturbed visually by the bright individual sources. 

It is worth noting that \citet{Delchambre2022} recently reported that the extinction from {\it Gaia \/} is offset with respect to that from {\it Planck \/} (their figure~26). The discrepancy, however, is not expected to impact our analysis, because we find from the fits that the extinction due to the Galaxy is dominated by the Galactic emission in Eq.~\ref{eq_cirrus_fit}, with the Galactic features appearing bright as seen in Fig.~\ref{fig_Galactic_map_FLS}. Furthermore, the offset reported in \citet{Delchambre2022} is more or less constant over the relevant regime, which can hence be properly accommodated by the linear fitting of the emission term. 

Because the H\,{\sc i} emission is predominantly from the warm neutral medium (WNM) of the ISM \citep[e.g.][]{Hennebelle2012}, there could be dust emission present that is not traced well by the H\,{\sc i} maps, such as that associated with molecular gas (sometimes called `dark gas') or the warm ionized medium (WIM). While quantifying this `missing' emission of the Galaxy in the H\,{\sc i} maps observationally is not trivial \citep[see e.g.][]{Lagache2000}, those components can be neglected for regions with a column density as low as the fields chosen in this study.

We also investigated the use of second-order polynomials, instead of linear relations, for $f(D)$, which might be able to account for (part of) the dust emission associated with molecular gas in regions of high column density such as the filament in the FLS field; however, we found no significant changes in our results. This is demonstrated in Figs.~\ref{fig_regression_CFIS} and \ref{fig_regression_SPIRE}, which show that correlations between the EBHIS and CFIS/SPIRE maps are described reasonably well by linear fits. Moreover, tracers of the Galactic gas components in different phases are shown to present much steeper slopes (ranging from $-2.5$ to $-3.5$) than the slopes in our measurements \citep{Hennebelle2012}, meaning that the detected signal from our analysis (after the subtraction attempt in particular) cannot be explained by Galactic cirrus alone. 

Additionally, the lack of strong variation in the cross-power spectra among the individual fields (Fig.~\ref{fig_Xpower_nomask_fields} and \ref{fig_Xpower_mask_fields}), which are at various Galactic latitudes, is a further indication that the signals are predominantly extragalactic, since fluctuations in the power of Galactic cirrus across latitudes is significant, exceeding easily an order of magnitude \citep[e.g.][]{Martin2010}. This is indeed shown from the largely enhanced correlation signals for the FLS field, as seen in Figs.~\ref{fig_Xpower_nomask_fields} and \ref{fig_Xpower_mask_fields}. To further investigate this, we divide the FLS field into three regions: the eastern third of the field (FLS1) where the Galactic contamination is smallest (see Fig.~\ref{fig_FLSmosaic}); the central third (FLS2); and the western third, where the Galactic contamination is largest.  We present the power spectra for the sub-regions separately in Fig.~\ref{fig_Xpower_mask_fields}.  It is clear that FLS1 (the least-contaminated area) shows a signal comparable to that from the other SPIRE fields, while the signal from FLS3 is stronger than the average over the entire FLS field. The slightly stronger signal from FLS1 relative to the other four fields is because even FLS1 still contains more Galactic cirrus compared to the extragalactic fields. The steeper slope of the power spectra of about $-2.4$ (dotted line in Fig.~\ref{fig_Xpower_mask_fields}) found for FLS (or FLS3), compared with $\simeq -1.1$ for the other fields, confirms that the main contributor to the signal in the FLS field has a different origin than for the other fields (where the slopes are consistent). Note, however, that the Galactic component is never entirely negligible, as can be seen in Fig.~\ref{fig_Gal_sub} and estimated earlier, in the worst case accounting for about 50~per cent (but more typically about 20~per cent) of the total correlation signal from the background.  We also find that the typical scatter is about 40~per cent between the results coming from different Galactic tracer maps (more specifically, the contamination ranges from 12 to 28~per cent of the CIB signal), which indicates an uncertainty in the assessment of systematic effects related to each map.

\begin{figure*}
\includegraphics[width=0.75\linewidth]{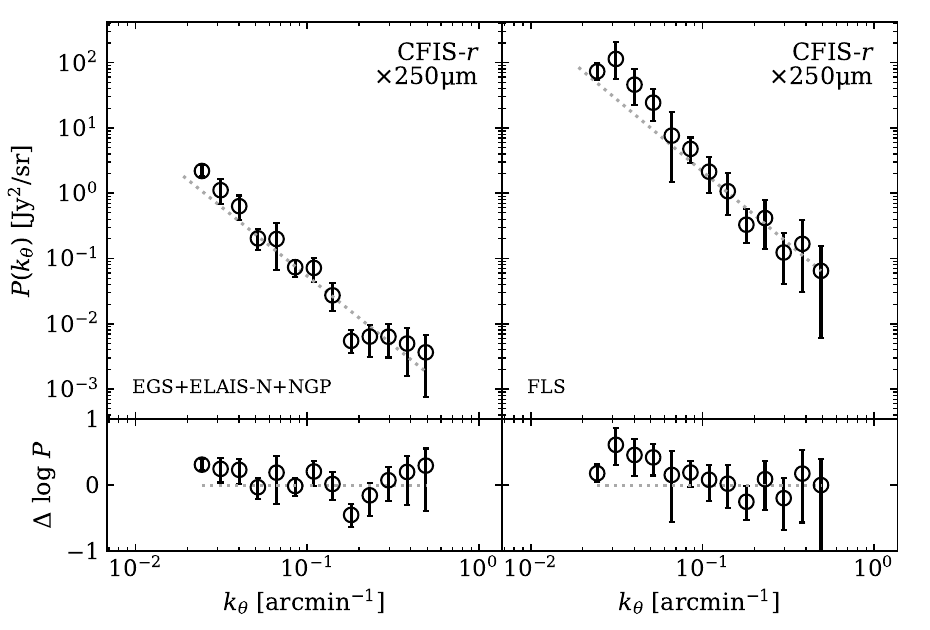}
\caption{Cross-power spectra between the Galactic $r$-band maps and our estimated `cirrus-free' SPIRE maps at $250\,\micron$, the former (latter) being estimated by performing a polynomial fit between the pixels in the CFIS (SPIRE) maps and the EBHIS H\,{\sc i} column-density map \citep{Winkel2016}. Only the total net power spectra are shown, without separation into the positive and negative correlations. The left-panel shows the results after combining the EGS, ELAIS-N1, ELAIS-N2 and HATLAS-NGP fields, while the right panel is for the FLS field. The excess in the cross-power spectra with respect to the noise level (dotted line, obtained by randomizing the phase angles of the Fourier transforms) for the FLS field indicates that our correction for the Galaxy is not sufficient and the maps after the subtraction attempt are actually not free from Galactic cirrus. For the other fields, on the other hand, the net spectra are consistent with the noise level estimated in the case of no correlation, implying that there is no Galactic correlation after the subtraction.
}
\label{fig_Xpower_nomask_cirrus_only}
\end{figure*}

We find that the FLS field contains a substantial amount of Galactic emission even after our subtraction attempts. This can be observed in Fig.~\ref{fig_Xpower_nomask_cirrus_only}, where we show the results from the following test. We first estimate the `cirrus-free' SPIRE maps, which result from the subtraction of the same polynomial fits as above from the original SPIRE maps. We then cross-correlate those with the Galactic $r$-band maps, obtained from the polynomial fitting of the original CFIS maps to the EBHIS map.  If our method used for correcting for the Galaxy is not sufficient and leaves residual emission even after the correction, we would see a correlation between the two maps, i.e.\ the net cross-power in excess of the noise level estimated from randomizing the phase angles of the Fourier transforms. As can be seen in Fig.~\ref{fig_Xpower_nomask_cirrus_only}, there is indeed a strong positive correlation (the result of residual Galactic correlations) in the FLS images. In contrast, there is barely any correlation in the other fields, meaning that the residual Galactic emission after the subtraction is insignificant. The reduced $\chi^2$ of the data relative to the noise level is calculated to be 2.61 for the FLS field in the regime where the resolution of EBHIS is valid, while it is 0.89 for the other fields combined. Apparently, the failure of the correction for the FLS field is due to the complex substructures of the Galactic filament present in the southwest corner of the field. We checked that we can still discern residual structure from the filament after subtraction using the maps of higher resolution, such as the {\it WISE\/} map. For that reason, and because the improvement in the statistics by adding the FLS is not appreciable (due to its relatively small area), we exclude the FLS for our combined estimate of the extragalactic cross-power spectra. Note that the amplitude of the cross-power spectra for the other fields is much smaller, by more than an order of magnitude, compared to those in Fig.~\ref{fig_Xpower_nomask}. This implies that the overall amplitude of the Galactic emission (after attempts to correct for it) is negligible relative to our signal in these other fields. On the other hand, the much higher overall amplitude found for the FLS field seen in Fig.~\ref{fig_Xpower_nomask_cirrus_only} relative to the other fields, reassures that the field is highly contaminated by the Galactic cirrus even after the subtraction method. 

Given the strong detection seen in the measurements, one might expect to see the correlations directly from the images by eye. This is indeed the case for the full images (with no sources masked), as can be seen in Figs.~\ref{fig_visualcorr_EGS}--\ref{fig_visualcorr_N2} and discussed in Appendix~\ref{sec_appC}.  On the other hand, when the detected galaxies are masked out, the cross-correlation is too weak to see by eye in individual patches.

\section[test]{Testing with simulated maps}
\label{sec_test}

Now we test our method for estimating the cross-power spectra (see Sect.~\ref{sec_data}) using simulated maps, in order to assess the biases potentially introduced by masking, filtering, instrumental beam effects, noise, Galactic cirrus and extragalactic dust obscuration.

\subsection{SIDES light cone}
\label{ssec_SIDES}
We perform tests based on the light cone from the Simulated Infrared Dusty Extragalactic Sky \citep[SIDES;][hereafter B17]{Bethermin2017}, which is publicly available.\footnote{\url{http://cesam.lam.fr/sides/}} This is constructed from a dark matter-only simulation, which in turn is based on the Bolshoi-{\it Planck\/} simulation \citep{RodriguezPuebla2016}. The assumed cosmology is consistent with \citet{PlanckXIII2016}. The light cone is 1.4\,deg$\times$\,1.4\,deg, with a total comoving volume of 
approximately 0.17\,${\rm Gpc}^3$. 
The SIDES simulation contains populations of star-forming galaxies that are consistent with number counts at a wide range of wavelengths and also have realistic clustering properties, which are an important considerations for our analysis.
SIDES gives the far-IR/submm properties of galaxies predicted from empirical modelling, but it does not include any optical properties. In order to assign optical properties to the mock galaxies, we use a simple empirical relation motivated from observational data, as described in Sect.~\ref{ssec_KiDSr}. For the submm properties, we use the predicted values from SIDES as they are, since they were shown to match some of the key submm properties from observations well (see B17). 

The SIDES simulation provides a galaxy catalogue extending down to very faint objects. To assign the submm properties, the authors first allocated stellar masses via an abundance matching technique \citep[see e.g.][]{Conroy2006, Guo2010, Behroozi2013, Moster2013, Lim2017}. Then at a given stellar mass, they simplified galaxies into three categories -- quenched, main-sequence (MS) star-forming, and starburst galaxies -- each of which have their own distributions and evolution, as determined from empirical relations derived from observations. By assigning the galaxy types and levels of scatter purely based on stellar mass, the SIDES simulation neglects any secondary dependencies, such as the environment. However, the main goals of our tests using these simulations are to check the robustness of our method for extracting the power spectra, as well as to check potential systematic effects due to Galactic cirrus, rather than accurately modelling the CIB. As will be seen later in Sect.~\ref{ssec_modelfit}, this lack of more elaborate dependencies is not an issue in practice, since the one-halo term is poorly constrained by the current data.

\subsection{Assigning \textit{r}-band magnitudes}
\label{ssec_KiDSr}

\begin{figure}
\includegraphics[width=0.95\linewidth]{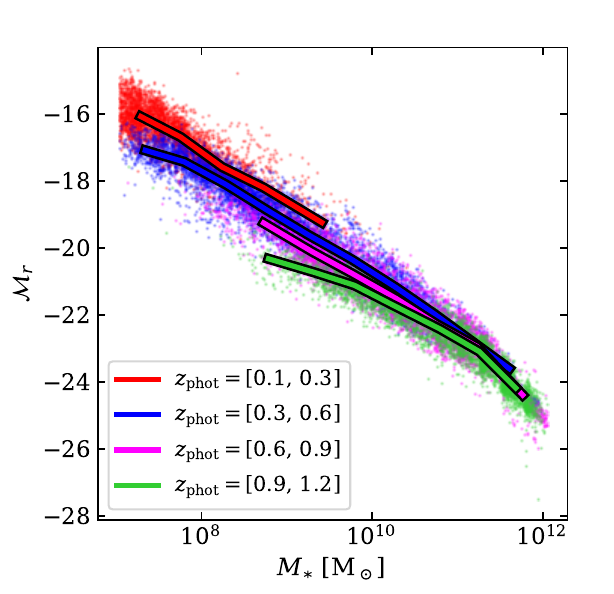}
\caption{Stellar mass and $r$-band magnitude at different photometric redshifts from the fourth data release of the Kilo-Degree Survey (KiDS; \citealt{Kuijken2019}) galaxy catalogue. The solid lines indicate the average, which we use to assign the $r$-band magnitudes to the mock galaxies in the SIDES light cone via interpolation. We assume no evolution at $z>1.2$ for the mock galaxies, using the same average relation at $z_{\rm phot}=1.2$ from the KiDS samples for the galaxies at $z>1.2$.}
\label{fig_KiDS}
\end{figure}

While the SIDES light cone provides redshifts, positions, halo masses, stellar masses, SFRs and mid-to-far IR fluxes for all the galaxies, it contains no information at shorter wavelengths that is needed to test our cross-correlation method. To assign an $r$-band magnitude, we use the mean relation between stellar mass and $r$-band magnitude at different photometric redshifts, $z_{\rm phot}$, derived from the fourth data release (DR4) of the Kilo-Degree Survey (KiDS; \citealt{Kuijken2019}). The stellar masses of the KiDS catalogue used here are constructed in the same manner as described in \citet{Wright2019}, and cover the full KiDS DR4 footprint. The stellar masses are matched to the KiDS shear samples. As in \citet{Wright2019}, both the stellar masses and $r$-band magnitudes have aperture corrections applied to account for the limited aperture used by the observation to estimate the totals. Finally the stellar masses are converted to match the {\it Planck} cosmology adopted throughout this paper, although the impact is almost negligible. The $r$-band magnitudes used here, as well as used later in Sect.~\ref{ssec_submm} for the parametrization of the modelling, are $K$-corrected ones. We neglect the difference in bandwidth and response functions between the CFIS and KiDS $r$ bands, since our tests here are aimed at checking the robustness of our method against systematic effects, rather than to construct a fully accurate model of the various populations. The KiDS catalogue contains a total of 21 million objects from its 1006 tiles, each of size $1\,{\rm deg}\times1\,{\rm deg}$. We bin the galaxies from the catalogue according to their photometric redshifts and stellar masses in such a way that every bin contains at least 100 galaxies. This eliminates some of the massive bins at the low redshift, and some of the low mass bins at the high redshifts, while the bin widths are mostly 0.5\,dex. Most of the KiDS galaxies lie between $z_{\rm phot}=0.1$ and $1.2$, with stellar masses ranging from $10^7\,{\rm M_\odot}$ to $10^{12}\,{\rm M_\odot}$. Using the average relation between the $r$-band magnitude and stellar mass for the bins (Fig.~\ref{fig_KiDS}), we linearly interpolate to assign an $r$-band flux density to each of the SIDES galaxies.  For galaxies outside the interpolation range probed by the KiDS catalogue, we assume no evolution, namely we assume that galaxies at $z>1.2$ in the simulation follow the same relation as at $z_{\rm phot}=1.2$ from the KiDS samples. We confirmed that different prescriptions for treating galaxies beyond $z=1.2$ do not change our conclusions regarding the robustness of our method and sensitivity to systematic effects.

\subsection{Map preparation and power spectra}
\label{ssec_map}

\begin{figure}
\includegraphics[width=0.93\linewidth]{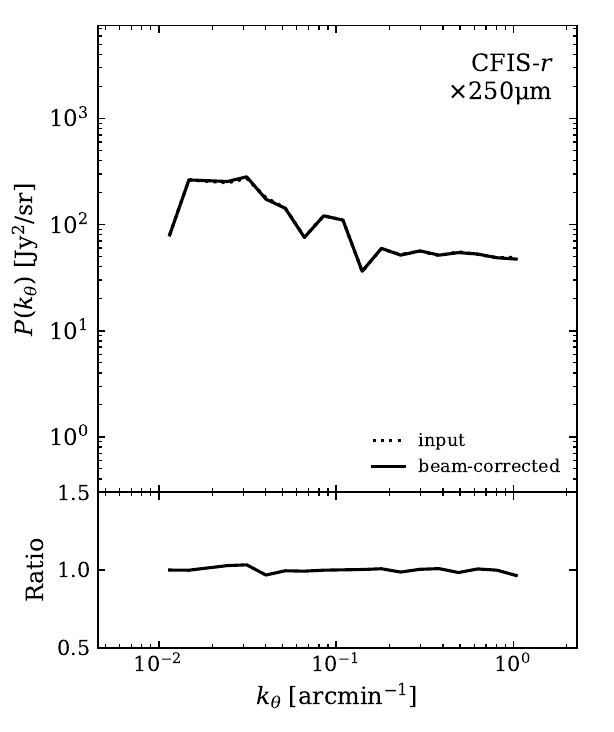}
\caption{Cross-power spectra between the simulated CFIS $r$-band and SPIRE maps. The dotted lines show the input power spectra obtained with no beaming effect and a much smaller pixel size than the observational data. The solid lines indicate the power spectra obtained with the same pixel size and beam as the data and then corrected for the beam following the method presented in Sect.~\ref{ssec_Xcorr}. While we only present results from the SPIRE map at 250\,$\micron$, we find similar results at 350 and 500\,$\micron$. The lower panel shows the ratios of the beam-corrected spectra to the input spectra.}
\label{fig_sim1}
\end{figure}

Using the SIDES light cone catalogue of galaxies, we construct two-dimensional maps for the CFIS $r$ band and the three SPIRE bands. We insert the mock galaxies into the same grid as the observational data (i.e.\ a grid with the same pixel size as the SPIRE maps), then smooth them by the SPIRE beam to assign values to the pixels, 
\begin{equation}\label{eq_smth}
S_{i,{\rm smth}} = \sum_{j=1}^{N_{\rm gal}}{\int{B(\boldsymbol{\theta}) S_j \delta(\boldsymbol{n_i}-\boldsymbol{\theta}-\boldsymbol{n_j}) \, d^2\theta}}\,, 
\end{equation}
where $S_{i, {\rm smth}}$ is the flux density assigned to pixel `$i$' in question from the smoothing, $\boldsymbol{\theta}$ is the projected position relative to the centre of the pixel `$i$', $B(\boldsymbol{\theta})$ is the SPIRE beam, which we approximate as a Gaussian with the FWHM at a given frequency, and $S_j$ is the flux of galaxy `$j$', which we treat like a point source.  We find that the correction for the instrumental beam, as described in Sect.~\ref{ssec_Xcorr}, recovers the true underlying cross-power spectra (calculated from the maps with a much smaller pixel size and without smoothing) between the simulated CFIS $r$-band and SPIRE maps, to within a few per cent for $k=[0.02, 0.6\times 250\,\micron/\lambda]\, {\rm arcmin}^{-1}$, the range in $k$-space probed in our analysis.  We therefore use this map as the basis for the following tests. The power spectrum measured from the map at 250\,$\micron$ is shown in Fig.~\ref{fig_sim1}. 

\subsection{Instrumental noise}
\label{ssec_noise}

In principle, if the cross-correlation analysis was performed on an infinite number of realizations, the impact of noise would be zero because the two maps are taken completely independently and with different instruments. In practice, however, there still could be residual deviations from zero that are purely statistical due to the limited number of modes contained in each map. Depending on the scientific signal of interest, the residual cross-correlation between the noise sources could potentially be substantial enough to affect our estimates of the total power spectra. 

\begin{figure}
\includegraphics[width=0.93\linewidth]{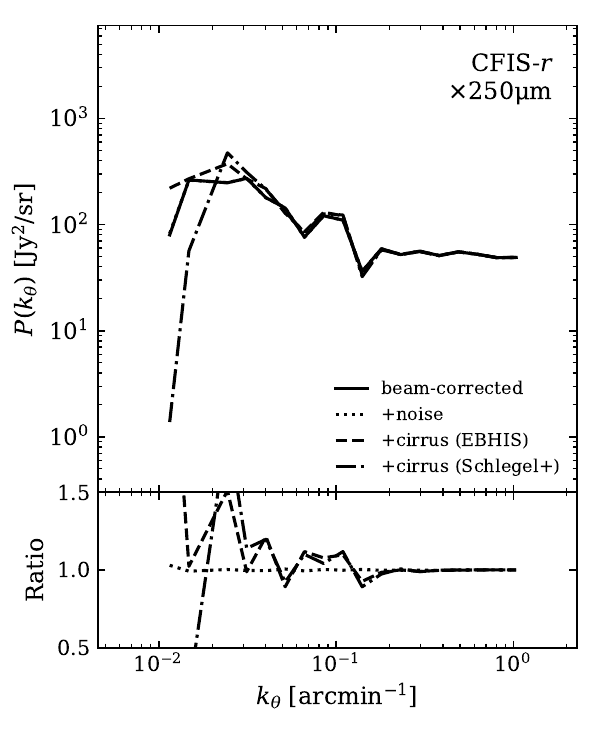}
\caption{The same as Fig.~\ref{fig_sim1}, but for the maps without noise (solid) and with noise (dotted), as well as with cirrus added using the the Effelsberg-Bonn H\,{\sc i} Survey \citep[][dashed]{Winkel2016} and using the reddening based on the maps by \citet{Schlegel1998} (dot-dashed). We use the same Galactic components estimated for CFIS and SPIRE as in Sect.~\ref{ssec_cirrus_res}, which were obtained through the linear fit to CFIS and SPIRE. Specifically, the Galactic component estimated for the EGS field was used to add the cirrus to the simulated maps; however, we find the same conclusions when using the cirrus map estimated for any other field (other than the FLS). As can be seen, the impact of the Galaxy, on the net spectra in particular, is only moderate, increasing the amplitude of the spectra by at most a factor of 2. This is consistent with our finding for the observational data in Sect.~\ref{ssec_res}. The lower panel shows the ratios with respect to the beam-corrected spectra without the noise and the Galactic emission.}
\label{fig_sim2}
\end{figure}

To check the impact of the noise on the power spectra, we use the error maps included in the HerMES data release of the SPIRE maps (and calculated as part of the mapmaking process). For the CFIS $r$-band data, such an error map does not exist. We thus calculate the standard deviation of the CFIS mosaic maps after masking all identified sources including the galaxies down to a 5$\,\sigma$ depth of 24.85 mag, where $\sigma$ is predominantly from night-sky emission. We then generate random Gaussian maps in which the pixel values are drawn randomly and independently (i.e.\ no spatial correlations) from a Gaussian distribution where the dispersion is the measured standard deviation. This assumes that the contribution from unidentified galaxies in the CFIS $r$-band is insignificant. While we do not have a direct way to check the degree to which this assumption is fair, the CFIS noise map generated in this way can be considered as an upper limit for the impact of the noise. We add the resulting noise maps to the simulated maps, and repeat the analysis to estimate the power spectra. As shown in Fig.~\ref{fig_sim2}, the impact of the residual correlation of noise on the power spectra is negligible compared to the signal. 

\subsection{Galactic contamination}
\label{ssec_cirrus}

As discussed earlier in Sect.~\ref{ssec_cirrus_res}, the measurement of the extragalactic cross-power spectrum can be contaminated by the Galaxy, either adding a positive correlation (via its dust emission and stellar light), or a negative correlation (via the attenuation of background galaxies by its cirrus). Because there is no correlation expected between the Galactic and extragalactic sources, the effects of the Galaxy in any given field can be simply added to the simulated maps, to test for the amplitude of the effects of contamination. Because our correction for the Galaxy is not very successful for the FLS field (Sect.~\ref{ssec_res}), we choose the Galactic maps of the other fields to add to the simulated maps to evaluate the impact of the Galaxy. To provide an upper limit to the impact, we present the results based on the Galactic map of EGS field, which we find has the second-strongest Galactic contamination after the FLS among all our fields (and therefore in that sense is conservative); however, we confirmed that using the Galactic maps of the other fields does not change our conclusions. To estimate the additional emission from the Galactic cirrus in the SPIRE map, we use the Galactic emission map for SPIRE estimated for the EGS field in Sect.~\ref{ssec_cirrus_res}, which was obtained from the linear relation between the SPIRE map and an external map. To account for the impact on the CFIS map, we apply the extinction to the simulated CFIS map as in the first term on the right-hand side of Eq.~\ref{eq_cirrus_fit}, and use the polynomial fit with the same parameters obtained in Sect.~\ref{ssec_cirrus_res} to add the Galactic emission. Figure~\ref{fig_sim2} shows the impact of the Galaxy on the cross-power spectrum of the simulated maps, using the SFD and EBHIS maps for the EGS field, while we confirm the same conclusion using the GNILC and {\it WISE\/} maps. Clearly, the Galaxy has only a small to moderate impact on the measurements of the cross-correlation.

\begin{figure}
\includegraphics[width=0.93\linewidth]{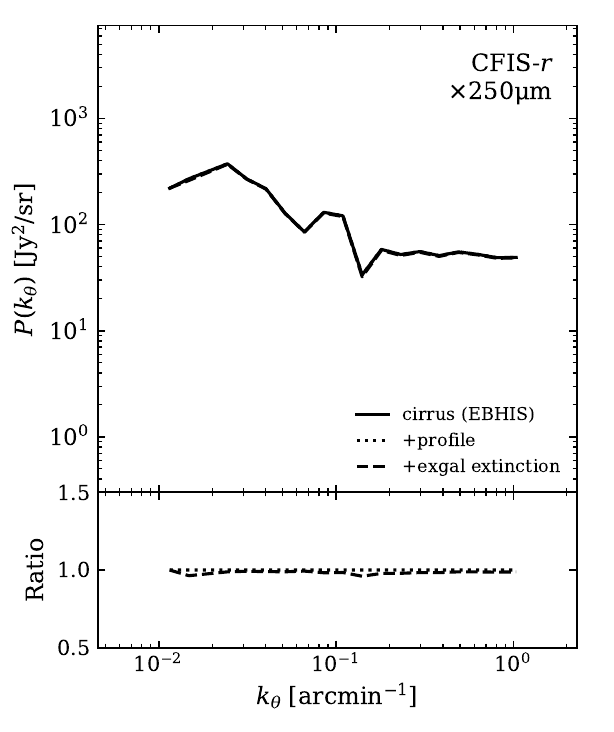}
\caption{The same as Fig.~\ref{fig_sim1}, but for the maps where sources are treated as point-like (solid), exponential discs ($\Sigma(r)=\Sigma_0\,\exp (-r/r_{\rm s})$) for both dust and starlight (dotted), and where dust is attenuated by foreground galaxies (dashed). To model the extragalactic extinction, we follow \citet{Popping2015} to estimate the total gas mass for the mock galaxies, and assume an exponential profile and extinction law to calculate the optical depth and extinction. Based on these results, we neglect the impact of profiles and extragalactic extinction of galaxies on our observational measurement, as well as on our modelling. While we present the results only from the SPIRE map at 250\,$\micron$, we find similar results for the 350 and 500\,$\micron$ maps. The lower panel shows the ratios with respect to the spectra obtained without taking into account the profile and the extragalactic extinction. }
\label{fig_sim2_2}
\end{figure}

\subsection{Impact of resolved sources}
\label{ssec_profile}

We have thus far assumed that all sources in the maps are point-like, i.e.\ unresolved by the survey instruments. While this is expected to be a fair assumption for most sources included in our analysis (given the relatively large beam of {\it Herschel\/}), here we explore how the measurements of cross-correlation are affected if we adopt resolved profiles for the galaxies. We model the shapes of galaxies as exponential discs, with surface density $\Sigma(r)=\Sigma_0\,\exp (-r/r_{\rm s})$, for both stellar emission and submm emission. Here $r$ is the projected distance from the galaxy centre and $r_{\rm s}$ is the disc scale length. We use the stellar scale length $r_\ast$ measured by \citet{vanderWel2014}. The median value of $r_\ast$ for the SIDES galaxies is 1.3\,kpc. We adopt a dust-to-stellar scale length ratio of 2.6 \citep{Kravtsov2013,Popping2015}, and take this to be the scale length of submm emission. Using the exponential profile with these scale lengths, we distribute the total submm and CFIS $r$-band flux density around the positions of the galaxies, and recalculate the cross-power spectra; the results are shown in Fig.~\ref{fig_sim2_2}. It can be seen that the impact of resolved sources is fairly negligible.  This is due to the fact that the scale lengths for most of the sources are much smaller than the resolution of the SPIRE data. For the same reason, we find that the choice of the ratio between the stellar and dust scale lengths has no significant impact on the results, unless we increase the sizes by at least an order of magnitude. Also, again for the same reason, the choice of $r_\ast$ has no impact on the conclusion, despite likely differences of a certain degree between $r_\ast$ estimated from the CFIS $r$-band and \citet{vanderWel2014}, unless they differ by more than an order of magnitude. Based on these tests, we neglect the impact of galaxy profiles on our observational measurements, as well as the modelling later in Sect.~\ref{sec_model}.

\subsection{Extragalactic dust obscuration}
\label{ssec_excirrus}

Obscuration of distant stellar light can also occur when the light passes through foreground galaxies. To estimate the potential bias due to this effect in the measurement and interpretation of the total power spectra, we model the extinction in a similar way as was done for Galactic cirrus (Sect.~\ref{ssec_cirrus}), in combination with the prescriptions for dust profiles as described in Sect.~\ref{ssec_profile}. One missing component is the total gas mass, $M_{\rm gas}$, for each of the mock galaxies. To assign a value for $M_{\rm gas}$, we follow the method presented in \citet{Popping2015}. \citet{Popping2015} combined empirical relations between gas surface density and star formation, and between the pressure and the molecular fraction of cold gas, in order to infer the gas mass for a galaxy of a given SFR. Similarly, we estimate the gas mass by iteratively seeking a solution that satisfies the empirical star-formation relation used in \citet{Popping2015} and the SFR value from SIDES. Then, combining the gas-mass estimate with the exponential profile of the gas scale-length (taken to be the same as the dust scale-length), we distribute gas mass over the simulated light cone. Finally, we calculate the foreground gas column density along the line of sight for each of the mock galaxies. As was done for the estimation of Galactic extinction, we assume an empirical relation of the form $N_{\rm H}\,/ \, A_V = 1.8\times10^{21}\, {\rm [mag^{-1}\, cm^{-2} ]} $ and the conversions from \cite{Schlafly2011} to infer the extinction in the $r$ band. Note that, unlike for Galactic cirrus, the dust emission from each galaxy is automatically accounted for via the simulation, and that self-extinction by the dust of a target galaxy has also been taken into account through the empirical relation used to assign the $r$-band magnitudes. Figure~\ref{fig_sim2_2} shows that dust obscuration by foreground galaxies does not noticeably change the measurements of the power spectra, and thus can be safely neglected in the interpretation of the data or modelling.

\subsection{Investigating the positive and negative cross-power spectra}
\label{ssec_posneg}

\begin{figure*}
\includegraphics[width=0.99\linewidth]{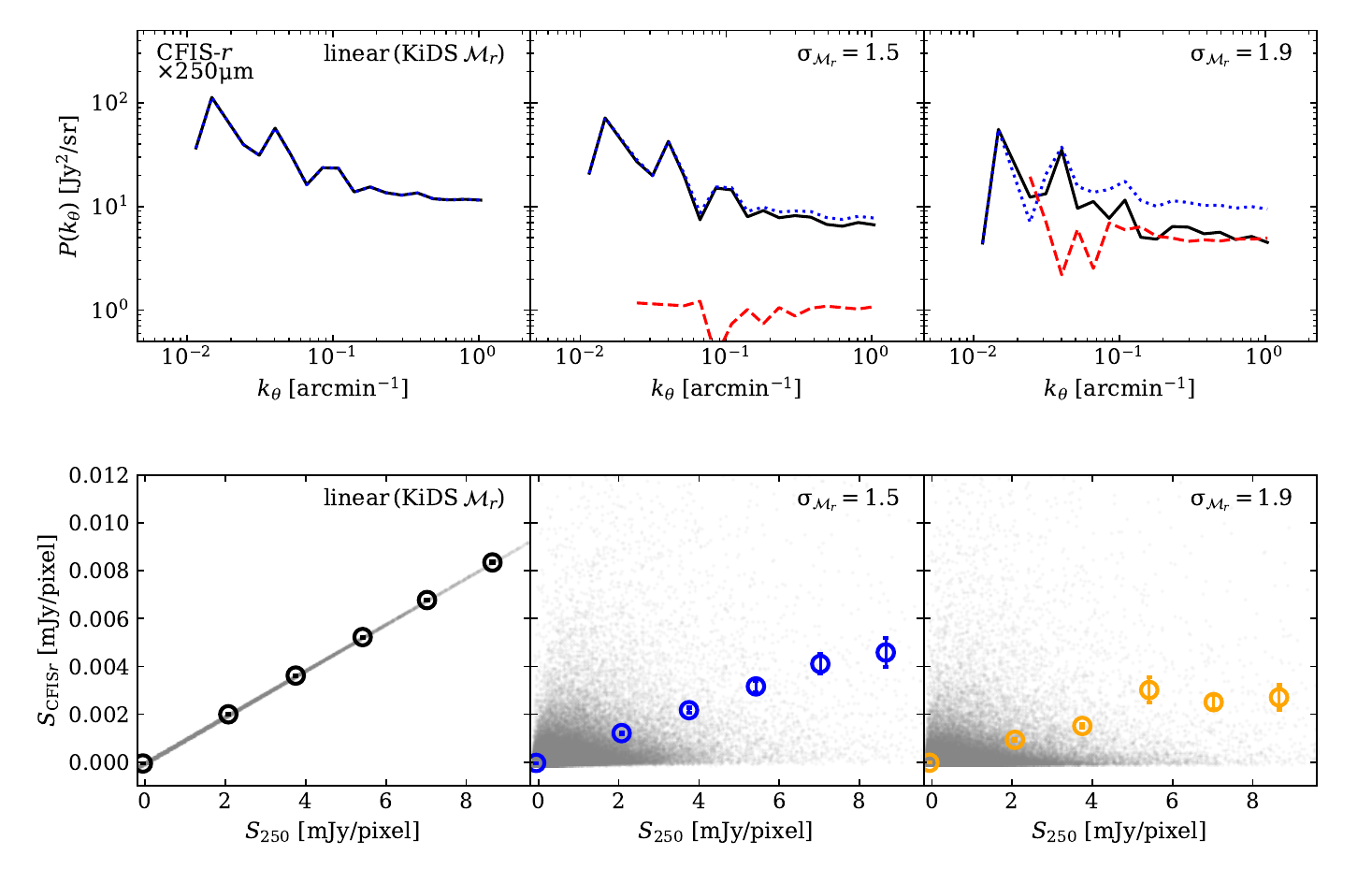}
\caption{\textit{Upper panels}: Net (black), positive (blue) and negative (red) cross-power spectra, as defined in Eqs.~\ref{eq_Xpower} and \ref{eq_Xpower_pos_neg}, measured from test cases using simulated maps. From left to right: the test case where there is a perfect linear relation between $r$-band and submm flux density with no scatter; and the test case where the scatter between the $r$-band and the submm band is 1.5\,dex (middle) and 1.9\,dex (right), respectively. \textit{Lower panels}: Scatter plot (dots) and binned averages (circles with error bars) of pixel values for each of the test cases from the upper panels. The error bars represent the errors of the means, calculated from $10{,}000$ bootstrap samples.}
\label{fig_Xpower_sim3}
\end{figure*}

Here we explore the interpretation of the separation of power spectra into positive and negative components by investigating what they trace and how they are affected by changes in the underlying cross-correlation properties. Due to statistical uncertainty arising from limited sky coverage and thus limited realisations, the net total power spectra will always deviate from zero even in the case of no physical correlation. This means that it is not sufficient to only measure the net power spectrum in order to ensure that what it captures is a signal of physical origin rather than purely statistical noise. Measuring the positive and negative correlations separately helps in this regard. If the net measurement is dominated by statistical noise, then we would expect that the positive and negative power spectra would be consistent with each other (as indeed shown for the null tests in Appendix~\ref{sec_appB}). On the other hand, if the net power is dominated by a physical correlation (or anti-correlation), then the positive and negative cross-power spectra will differ appreciably. 

To investigate more quantitatively how the positive and negative power spectra behave, we carry out tests of a few difference models. In the first test case, we assume that the $r$-band flux density has a perfect linear relation with submm flux. We fit the $r$-band flux density from the KiDS catalogue and the submm flux density from SIDES with a linear relation to find the best-fitting proportionality constant between the two fluxes. We then use this constant to assign the $r$-band flux density to the galaxies in our simulated map. We measure the net, positive and negative cross-power spectra of the resulting map pair. As shown in the upper panel of Fig.~\ref{fig_Xpower_sim3}, there is no negative correlation here and the net power spectrum equals the positive power spectrum because the two maps are identical except for a normalization factor. Also shown in the lower panel of Fig.~\ref{fig_Xpower_sim3} is the scatter of pixel values between the two maps, showing a perfect linear relation, with the slope being the proportionality constant obtained from the fitting.

In the second test case, we randomly assign a scatter of 1.5 (in magnitude, 0.6\,dex in flux) to the $r$-band fluxes of galaxies, while keeping the same mean relation between the two fluxes as in the previous case. As can be seen in the middle panel of Fig.~\ref{fig_Xpower_sim3}, introducing such scatter increases the power from the out-of-phase Fourier transforms (red curves), resulting in a decrease in both the net and in-phase (positive) cross-power spectra. It can be understood that, as the scatter increases, there will be essentially no correlation between the two fluxes eventually, thus the two maps will be fully uncorrelated, and the net cross-power spectra will only measure statistical noise. 

\begin{figure}
\includegraphics[width=0.95\linewidth]{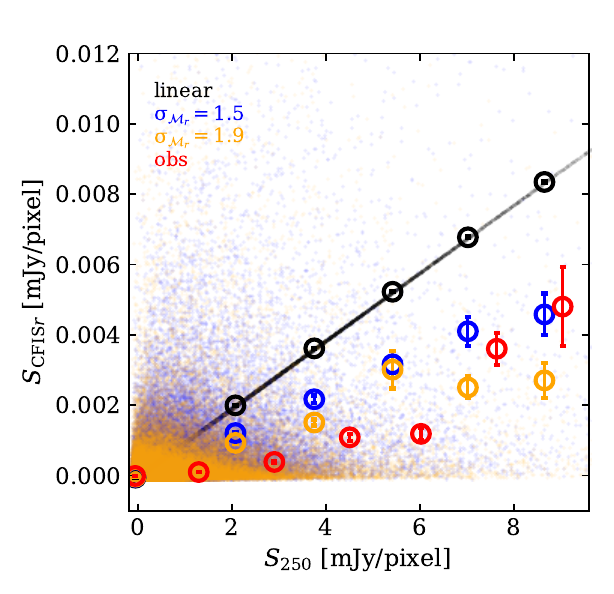}
\caption{Scatter plot (dots) and binned averages (circles with error bars) of pixel values from the observational data (red), and from test cases using simulated maps, including the case of a perfect linear relation assumed between $r$-band and submm flux density with no scatter (black), and the case where the scatter between $r$-band and submm flux density is 1.5\,dex (blue) and 1.9\,dex (orange). The error bars represent the errors of the means, calculated from $10{,}000$ bootstrap samples. With no flux density cut applied in the plot, the increasing mean in $r$-band flux density with increasing submm flux density indicates that there is a positive correlation between the map pairs.}
\label{fig_scatter}
\end{figure}

Finally, in the third case, similar to the second, we add a random scatter of 1.9 (0.76\,dex in flux) to the same mean relation to assign the $r$-band fluxes to the galaxies. The scatter of 1.9 is approximately the scatter found between the fluxes from the KiDS and the SIDES catalogs, thus making this test case more realistic than the second case (as will be seen shortly below, the scatter plot of pixel values shows great similarity between the observation and the third test case, see Fig.~\ref{fig_scatter}). Compared to the second case where the scatter is 1.5\,dex, the increased scatter further elevates the out-of-phase cross-power between the Fourier transforms while moderately lowering the net spectra. 

In Fig.~\ref{fig_scatter}, we show a scatter plot of pixel values from the observational data, together with the results from the test cases described above. The data used in this plot are only from the EGS field, but we find no significant change in the results when other fields are used. It can be seen that the test case of 1.9\,dex scatter around the best-fitting linear relation is in good agreement with the observational data. This is not surprising, since both the linear relation and the scatter used to assign the $r$-band and submm flux density to the mock galaxies were motivated by observations from KiDS. Note that although no cut was applied when making the plot, the $r$-band flux density distribution from some test cases in the scatter plot is so skewed that it looks almost flat around zero. If there is no real correlation and the cross-power measurement is dominated by noise of any type, the scatter distribution will appear symmetric around zero and the mean $r$-band flux density at given submm flux density will be flat around zero. The increasing mean in $r$-band flux density with increasing submm flux density, as seen for all cases considered here and including in the observation, indicates that there is a positive correlation between the two maps, consistent with our interpretation of the positive and negative cross-power spectra. In other words, the main effect of Galactic dust in the optical images is to add (rather than absorb) light that is correlated with the submm emission.

\section[model]{Halo-model fitting}
\label{sec_model}

\subsection{Cross-power spectrum formalism}
\label{ssec_formalism}

We now consider a halo-based model that we can fit to the data in order to interpret the cross-correlation results. The halo model we adopt here is similar to the previous work by \citet{Bethermin2013}, implemented in the context of a more general halo-model formalism, described by e.g.\ \citet{Cooray2002}. The cross-power spectrum between two frequencies, $\nu$ and $\nu'$, can be expressed as a sum of the one-halo (1h), two-halo (2h) and Poisson terms (also known as `shot noise'), 
\begin{equation} \label{eq_Cl_gen}
\boldsymbol{P_{\nu\nu'}(k) = P_{\nu\nu'}^{\rm 1h} + P_{\nu\nu'}^{\rm 2h} + P_{\nu\nu'}^{\rm shot}.} 
\end{equation}
Taking the small-sky limit \citep{Limber1953}, the one-halo term can be expressed as, 
\begin{equation} \label{eq_Cl_1h}
\boldsymbol{P_{\nu\nu'}^{\rm 1h}(k)} =
\int{
dz\frac{dV}{dz}
}
\int{
dM_{\rm h}\frac{dn_{\rm h}}{dM_{\rm h}}\, \overline{u}_k^\nu(M_{\rm h},z)\,  \overline{u}_k^{\nu'}(M_{\rm h},z),
}
\end{equation}
where $dV$ is the cosmological volume element, $M_{\rm h}$  is the halo mass, $dn_{\rm h}/dM_{\rm h}$ is the halo mass function at redshift $z$ and $\overline{u}_k^{\nu}$ is the Fourier transform of the profile of the observable (flux density profile, in this case) within a halo at frequency $\nu$. The two-halo term can be expanded as \citep[e.g.][]{Cooray2002, Addison2012} 
\begin{equation} \label{eq_Cl_2h}
\boldsymbol{P_{\nu\nu'}^{\rm 2h}(k)} = 
\int{
dz\frac{dV}{dz}\, \overline{b}_k^\nu(z)\, \overline{b}_k^{\nu'}(z) P_{\rm m}(2\pi k/\chi, z)
} ,
\end{equation}
where $P_{\rm m}$ is the linear matter power spectrum, with $\chi$ being the comoving distance.  The coefficient $\overline{b}_k^\nu$ is an effective linear bias of an observable at frequency $\nu$, defined by 
\begin{equation} \label{eq_bl}
\overline{b}_k^\nu = 
\int{
dM_{\rm h}\frac{dn_{\rm h}}{dM_{\rm h}} \overline{u}_k^\nu(M_{\rm h}, z) b_{\rm h}(M_{\rm h}, z)
},
\end{equation}
where $b_{\rm h}(M_{\rm h}, z)$ is the linear bias of haloes with mass $M_{\rm h}$ at redshift $z$. Finally, the shot-noise term can be written as \citep[e.g.][]{BondCarrHogan1991,ScottWhite99,Knox2001}
\begin{equation} \label{eq_Cl_shot}
\boldsymbol{P_{\nu\nu'}^{\rm shot}} = \int\int{S_\nu S_{\nu'} \frac{d^2N}{dS_\nu dS_{\nu'}}dS_\nu dS_{\nu'}},
\end{equation}
where $S_\nu$ is the flux density, and $d^2N/dS_\nu dS_{\nu'}$ is the differential number count of sources in given flux density bins of two observables. 

By considering the contributions to $\overline{u}_k^{\nu}$ separately from central and satellite galaxies, the one-halo term can be further expanded as 
\begin{align} \label{eq_Cl_1h_sep}
\boldsymbol{P_{\nu\nu'}^{\rm 1h}(k)} &= \int{dz\frac{d\chi}{dz}\chi^2}
\int {dM_{\rm h}\frac{dn_{\rm h}}{dM_{\rm h}}} \nonumber \\ 
&\ \times \bigg\{\overline{S}_{\rm \nu, cen} \overline{S}_{\rm \nu', sat} u_{{\rm gal},k}(M_{\rm h},z) \bigg. \nonumber \\ 
&\quad + \bigg. \overline{S}_{\rm \nu, sat} \overline{S}_{\rm \nu', cen} u_{{\rm gal},k}(M_{\rm h},z) \bigg. \nonumber \\
&\quad + \bigg. \overline{S}_{\rm \nu, sat} \overline{S}_{\rm \nu', sat} u^2_{{\rm gal},k}(M_{\rm h},z) \bigg\},
\end{align}
where $u_{{\rm gal},k}$ is the Fourier transform of the profile of the distribution of galaxies within a halo, while $\overline{S}_{\rm \nu, cen}$ and $\overline{S}_{\rm \nu, sat}$ are the average flux densities of galaxies at frequency $\nu$ from centrals and satellites, respectively, integrated within a halo of mass $M_{\rm h}$. This equation assumes that `central' galaxies are at the centre of haloes. Cross-terms, namely dust emission in the $r$ band or stellar emission in the submm bands, do not appear in the equation, since we are assuming that they can be neglected; this is a reasonable approximation given the nature of these emission processes and how far apart the frequency bands are. Similarly, by considering the contributions to the effective linear bias, $\overline{b}_k^\nu$, from central and satellite galaxies separately, we have
\begin{equation} \label{eq_bl_sep}
\overline{b}_k^\nu = 
\int{
dM_{\rm h}\frac{dn_{\rm h}}{dM_{\rm h}} b_{\rm h}(M_{\rm h}, z) \bigg\{\overline{S}_{\rm \nu, cen} + \overline{S}_{\rm \nu, sat} u_{{\rm gal},k}(M_{\rm h},z)\bigg\},
}
\end{equation}
where the two-halo term is given by 
\begin{align} \label{eq_Cl_2h_sep}
\boldsymbol{P_{\nu\nu'}^{\rm 2h}(k)} &= \int{dz\frac{d\chi}{dz}\chi^2}
\int {dM_{\rm h}\frac{dn_{\rm h}}{dM_{\rm h}}}
\int {dM'_{\rm h}\frac{dn_{\rm h}}{dM'_{\rm h}}} \nonumber \\ 
&\ \times \bigg\{\overline{S}_{\rm \nu, cen} + \overline{S}_{\rm \nu, sat} u_{{\rm gal},k}(M_{\rm h},z) \bigg\} \nonumber \\ 
&\ \times \bigg\{\overline{S}_{\rm \nu', cen} +  \overline{S}_{\rm \nu', sat} u_{{\rm gal},k}(M'_{\rm h},z) \bigg\} \nonumber \\
&\ \times b_{\rm h}(M_{\rm h}, z)b_{\rm h}(M'_{\rm h}, z) P_{\rm m}(2\pi k/\chi, z). 
\end{align}
Finally, the Poisson term can be expressed as 
\begin{align} \label{eq_Cl_shot_sep}
\boldsymbol{P_{\nu\nu'}^{\rm shot}} &= \int{dz\frac{d\chi}{dz}\chi^2}
\int {dM_{\rm h}\frac{dn_{\rm h}}{dM_{\rm h}}} \nonumber \\ 
& \times \bigg\{\overline{S_{\rm \nu, cen} S_{\rm \nu', cen} \strut}  + \overline{S_{\rm \nu, sat} S_{\rm \nu', sat} \strut} \bigg\}. 
\end{align}

To compute the cross-power spectra with this formalism, we adopt the halo mass function of \citet{Tinker2008}, the fitting function for the linear halo bias from \citet{Tinker2010}, and the linear matter power spectrum calculated using {\sc camb}.\footnote{\url{https://camb.info/}} Finally, we assume that the distribution of galaxies within haloes follows an NFW profile \citep{Navarro1997}, giving us $u_{{\rm gal},k}(M_{\rm h}, z)$. With this formalism and the assumptions and fitting functions described above, the only components left to be addressed to calculate the power spectra are $S_{\rm \nu, cen}$ and $S_{\rm \nu, sat}$,  the flux densities in the SPIRE submm and CFIS $r$ band from haloes of a given mass.

\subsection{Star formation and stars in haloes}
\label{ssec_submm}

Here we derive an expression for the average flux density in haloes with a set of model parameters. We adopt a similar methodology to the approach taken by \citet{Bethermin2012, Bethermin2013, Bethermin2017}, where a galaxy of given mass belongs to one of three populations: quenched galaxies (assumed to have zero star-formation rate); main-sequence (MS) star-forming galaxies; and starbursting (SB) galaxies.  The latter two classes are collectively referred to as `star-forming galaxies' throughout.

\subsubsection{Linking stellar mass to halo mass}

We begin from the halo mass function, for which we adopt the parametrization of \citet{Tinker2008}, as mentioned above. To connect stellar mass to each halo of given total mass, we use the stellar mass-to-halo mass (SMHM) ratio of \citet{Behroozi2013}. Using observational constraints, such as the stellar mass function (SMF), SFR and cosmic SFR density, \citet{Behroozi2013} constrained a parametrized ratio between stellar mass and halo mass over a redshift range of $z=0$ to $8$. The SMHM relation has a total of five free parameters (a characteristic mass, a normalization, the low-mass-end and massive-end slopes, and a parameter controlling the transition between the two asymptotic slopes), and the evolution of each parameter is further modelled with an additional parametrization as a function of redshift. We fix the parameter values of the SMHM ratio to those in \citet{Behroozi2013}.

\subsubsection{Quenched fraction}
Because the stellar mass function obtained above includes quenched galaxies, which we assume to have zero SFR and thus zero contribution to the submm flux, we need to account for the quenched fraction at any given stellar mass and redshift, $f_{\rm Q}(M_\ast, z)$, when computing the average submm flux density from haloes. We adopt the quenched fraction estimated by \citet{Bethermin2017}, which is an analytic formulation (containing a complementary error function) to the non-quenched fraction of galaxies from the observation of \citet{Davidzon2017}. We use this quenched fraction with the parameters fixed to the best-fitting values from the original paper. The assumption that the quenched population has practically no contribution to the observed submm flux density is supported by observational evidence that these galaxies in general have infrared luminosities that are lower by more than a factor of 10 compared to MS galaxies \citep[see][]{Viero2013,Man2016}.

\subsubsection{Star-formation rate}
\label{sssec_SFR}

Following \citet{Bethermin2017}, we adopt the SFR approach proposed by \citet{Schreiber2015}, who fit a parametrized function to observational measurements for MS galaxies, 
\begin{align} \label{eq_SFR}
\log\frac{{{\cal S}_{\rm MS}}(M_\ast, z)}{\rm M_\odot\,yr^{-1}} &= m -m_0- a_1\left[ \max (0, m-m_1-a_2\eta) \right]^2 \nonumber \\
& + a_0\eta- 0.1\times\frac{0.5-\min (0.5, z) }{0.5-0.22},
\end{align}
where $\eta=\log(1+z)$, $m=\log(M_\ast/10^9{\rm M_\odot})$, and the last term is a correction suggested by \citet{Bethermin2017}, for an offset of $\simeq 0.1$\,dex found relative to another set of observations at lower redshift by \citet{Sargent2014}. The parameters $m_0$, $m_1$, $a_0$, $a_1$ and $a_2$ are free parameters that we keep in our model to be constrained by fitting to the cross-correlation measurements. For starburst galaxies, we assume that the SFRs are higher than MS galaxy SFRs by a factor of $\alpha_{\rm SB}$, which is treated as a free parameter to be constrained by the data. We assume the same factor of $\alpha_{\rm SB}$, regardless of mass and redshift. 

We also model the scatter around the mean SFR at a given stellar mass and redshift. Motivated by the observational findings of \citet{Schreiber2015}, we fix both levels of scatter for MS and SB galaxies to be 0.31\,dex.\footnote{The same amount of scatter is assumed for MS and SB galaxies by design. If this condition is not imposed, the resulting scatter from fitting to the observational data can significantly differ.  Fixing the scatter to 0.31\,dex is somewhat arbitrary, but is similar to what is observed; see \citet{Schreiber2015}.} To account for the difference between the means in log-normal space and linear space, we correct for the offset. For a given log-normal distribution (with a base of 10) with a mean of $\mu$ and a scatter of $\sigma$, the mean in linear space is $10^\mu\exp\{(\sigma\ln10)^2/2\}$, i.e.\ larger than the `targeted' mean by a factor of $\exp\{(\sigma\ln10)^2/2\}$. We thus subtract $(\sigma^2\ln10)/2$ from the $\log{\cal S}$ values when considering SFR in the calculations. In this way, the SFR computed later with the best-fitting parameters, through Eq.~\ref{eq_SFR} for instance, is the actual mean SFR for MS galaxies. 

As in \citet{Bethermin2017}, and motivated by observations, we model the ratio of the MS to SB populations, $f_{\rm SB}$, as a function that linearly increases with redshift until $z=1$, after which it is assumed not to change, and it is also considered to be independent of mass \citep[e.g.][]{Elbaz2011, Sargent2012}, i.e.\
\begin{equation} \label{eq_f_SB}
f_{\rm SB}(z) = 0.015 \times \left[1 + {\rm min}  (z,1)\right]. 
\end{equation}

\subsubsection{Submillimetre flux density}

Combining the model parametrizations presented above, and using the SFR-to-$L_{\rm IR}$ conversion factor from \citet{Kennicutt1998} for a Chabrier IMF, $K={\rm SFR}/L_{\rm IR}=1\times 10^{-10}\,{\rm M_\odot yr^{-1} L_\odot^{-1}}$ (where $L_{\rm IR}$ is the luminosity integrated over 8 to 1000\,$\micron$), the average submm flux density from central galaxies in haloes can be written as
\begin{align} \label{eq_Fcen}
\overline{S}_{\rm \nu, cen}(M_{\rm h}, z) &= \frac{{{\cal S}_{\rm MS}}(M_\ast, z)}{K}\nonumber \\
&\quad \times (1 - f_{\rm Q}(M_\ast, z)) \times \overline{f}_{\rm IR\mh to \mh\nu}.
\end{align}
Here $M_\ast$ is the stellar mass for haloes of mass $M_{\rm h}$ from the SMHM relation, and $\overline{f}_{\rm IR\mh to \mh\nu}$ is the average conversion factor from the total IR luminosity to submm flux density at frequency $\nu$. We follow \citet{Bethermin2013} (their equations 14 and 15, in particular), who compute $\overline{f}_{\rm IR\mh to \mh\nu}$ using the results of \citet{Magdis2012}, but using the updated constraints on the mean radiation field from \citet{Bethermin2017}. Assuming a different IMF only affects the modelling by changing the SFR-to-$L_{\rm IR}$ conversion factor. For example, the resulting SFR from the model fitting is normalized by a factor of $1.72$ when a \citet{Salpeter1955} IMF is assumed, from $K=1.72\times 10^{-10}\,{\rm M_\odot yr^{-1} L_\odot^{-1}}$ for such an IMF. 

Similarly, the average integrated submm flux density from satellite galaxies in haloes of mass $M_{\rm h}$ is given by
\begin{align} \label{eq_Fsat}
\overline{S}_{\rm \nu, sat}(M_{\rm h}, z) &= \int dm_{\rm sub}\frac{dN_{\rm sub}}{dm_{\rm sub}}(m_{\rm sub}|M_{\rm h})  \nonumber \\
&\times \frac{{{\cal S}_{\rm MS}}(M_{\rm \ast, sub}, z)}{K} \nonumber \\ 
&\times \left[1 - f_{\rm Q}(M_{\rm \ast, sub}, z)\right] \overline{f}_{\rm IR\mh to \mh\nu}, 
\end{align}
where ${dN_{\rm sub}}/dm_{\rm sub}$ is the subhalo mass function and $M_{\rm \ast, sub}$ is the stellar mass corresponding to a subhalo of mass $m_{\rm sub}$ from the SMHM relation. Subhaloes are haloes inside a bigger (host) halo that were once independent before being captured and absorbed by the host. Subhaloes are also where satellite galaxies are generally assumed to reside. We use the subhalo mass function of \citet{TinkerWetzel2010}.

\subsubsection{CFIS $r$-band flux density}

We model the CFIS $r$-band brightness for galaxies in haloes via another parametrization. Specifically, motivated by the relation between stellar mass and absolute magnitude in the $r$-band (${\cal M}_r$) for the KiDS samples (\citealt{Kuijken2019}; see Sect.~\ref{ssec_KiDSr}), we model ${\cal M}_r$ by a broken linear relation described by four free parameters (which thus is a broken power law in linear flux versus mass space): 
\begin{align}\label{eq_Mr}
{\cal \overline{M}}_r = 
\begin{cases} 
\alpha_{\rm lo} \log(M_\ast / M_{\rm \ast, piv}) + {\cal M}_{r, {\rm piv}}, 
& \mbox{if } M_\ast \le M_{\rm \ast, piv}; \\ 
\alpha_{\rm hi} \log(M_\ast / M_{\rm \ast, piv}) + {\cal M}_{r, {\rm piv}}, 
& \mbox{if } M_\ast > M_{\rm \ast, piv}.
\end{cases}
\end{align}
Here $\alpha_{\rm lo}$ and $\alpha_{\rm hi}$ are the low-mass-end and massive-end slopes, respectively, and $M_{\rm \ast, piv}$ is the pivot mass where the relation changes its slope, while ${\cal M}_{r, {\rm piv}}$ is the $r$-band magnitude at the pivot. Because the observational relation does not show any strong hint for evolution (Fig.~\ref{fig_KiDS}), we assume that there is no redshift dependence, and that the luminosity is solely determined by stellar mass. We find that more sophisticated functional forms, such as an exponentially-falling relation at the massive end or a double power-law with a smooth transition at the pivot, do not improve the fit greatly, and thus we choose this simple relation for the model fitting. We also include the scatter around the mean relation in the model by introducing another free parameter, $\sigma_{{\cal M}_r}$. Similar to the submm flux density, the average CFIS $r$-band flux density from central and satellite galaxies in a halo are given by 
\begin{align}\label{eq_Fr}
\overline{S}_{r, {\rm cen}}(M_{\rm h}, z) &= \overline{S}_r(M_\ast, z),  \nonumber \\
\overline{S}_{r, {\rm sat}}(M_{\rm h}, z) &= \int dm_{\rm sub}\frac{dN_{\rm sub}}{dm_{\rm sub}}(m_{\rm sub}|M_{\rm h})  \nonumber \\
&\quad\times \overline{S}_r(M_{\rm \ast, sub}, z),
\end{align}
where $\overline{S}_r$ is the average $r$-band flux density for an object of given mass at redshift $z$ from Eq.~\ref{eq_Mr}.

\subsubsection{Poisson noise}

In principle, the computation of the Poisson term can be more complicated than that of the clustered components. This is because it requires calculating pairs of submm and optical flux densities over all individual galaxies from the flux density distributions (see Eq.~\ref{eq_Cl_shot_sep}), which may potentially involve correlations between the flux densities in the given two bands at given mass and redshift. However, in the case of no correlation between the two observables at given mass and redshift, it is greatly simplified, and can be written as a multiplication of the two averages obtained above: 
\begin{align}\label{eq_Fshot_cen}
\overline{S_{\rm \nu, cen} S_{\rm \nu', cen} \strut} &= \overline{S}_{\rm \nu, cen} \overline{S}_{r, {\rm cen}} \nonumber \\
&= \frac{1}{\sqrt{2\pi}\sigma_{{\rm M}_r}} 
\int{
d{\rm M}_r S_r \exp\left\{-\frac{({\rm M}_r-{\rm \overline{M}}_r)^2}{2\sigma^2_{{\rm M}_r}}\right\}
} \nonumber \\
&\times \int
d({\cal S}) \left\{p_{\rm MS}\int{d\langle U \rangle f_\nu^{\rm MS} \psi_{\rm MS}(\langle U \rangle)} \right. \nonumber \\
&\quad \left. + p_{\rm SB}\int{d\langle U \rangle f_\nu^{\rm SB} \psi_{\rm SB}(\langle U \rangle)}\right\} \nonumber \\ 
& \times \frac{\cal S}{K} \times (1-f_{\rm Q}), 
\end{align}
where $p_{\rm MS\,(SB)} (M_\ast, z)$ is the probability distribution function (PDF) of SFRs for MS (SB) galaxies according to Sect.~\ref{sssec_SFR}, $\langle U \rangle$ is the mean radiation field of a source \citep[see e.g.][]{Bethermin2012}, $f_\nu^{\rm MS\,(SB)} (\langle U \rangle, z)$ is the conversion factor from $L_{\rm IR}$ to submm flux density at frequency $\nu$, and $\psi_{\rm MS\,(SB)} (\langle U \rangle, z)$ is the PDF of $\langle U \rangle$ from \citet{Bethermin2017}, which is based on an empirical model. 

If there is a correlation between the submm and $r$-band flux densities at a fixed mass and redshift, the effective value of the left-hand side (LHS) of the equation will deviate from the value obtained from the right-hand side (RHS). To explore this possibility, we introduce a free parameter in our model to be constrained, $\kappa_{\rm corr}$, defined as, $({\rm LHS}) = ({\rm RHS}) \times \kappa_{\rm corr}$. The coefficient $\kappa_{\rm corr}$ is greater (smaller) than unity if there is a positive (negative) correlation between the fluxes at given mass and redshift, while $\kappa_{\rm corr}$ is equal to unity in the case of no correlation. Note that $\kappa_{\rm corr}$ only investigates further correlations at a fixed mass and redshift in excess of the average correlations at the mass and redshift, which are already reflected in the RHS of the equation even in the case of $\kappa_{\rm corr}=1$. 

Assuming that the same value of $\kappa_{\rm corr}$ describes the potential correlation between satellite galaxies, then
\begin{align}\label{eq_Fshot_sat}
\overline{S_{\rm \nu, sat} S_{\rm \nu', sat} \strut} &= 
\kappa_{\rm corr}\int dm_{\rm sub}\frac{dN_{\rm sub}}{dm_{\rm sub}}(m_{\rm sub}|M_{\rm h}) \nonumber \\
& \times \overline{S}_r(M_{\rm \ast, sub}, z) \frac{{{\cal S}_{\rm MS}}(M_{\rm \ast, sub}, z)}{K} \nonumber \\ 
& \times \left[1 - f_{\rm Q}(M_{\rm \ast, sub}, z)\right] \overline{f}_{\rm IR\mh to \mh\nu}.  
\end{align}
By substituting Eqs.~\ref{eq_Fshot_cen} and \ref{eq_Fshot_sat} into Eq.~\ref{eq_Cl_shot_sep}, we obtain the model prediction for the shot noise. 

\subsection{Model fitting and results}
\label{ssec_modelfit}

\begin{table*}
 \renewcommand{\arraystretch}{1.2} 
 \centering
 \begin{minipage}{175mm}
  \caption{List of the 12 parameters in our model, fit to the cross-correlation measurements. The prior ranges, best-fitting values (defined as the model with the smallest $\chi^2$) and descriptions are summarized here.}
  \begin{tabular}{clccc}
\hline
Parameter & Description & Equation & Best-fit & Prior \\
\hline
\hline
$m_0$ & Normalization of the star-forming main-sequence (MS) SFR & Eq.~\ref{eq_SFR} & 0.47 & [$-5$, 5] \\  
$a_0$ & Redshift dependence in normalization of the MS SFR &  Eq.~\ref{eq_SFR} & 1.61 & [0, 10] \\  
$m_1$ & Pivot stellar mass in the MS SFR & Eq.~\ref{eq_SFR} & 0.36 & [$-10$, 10] \\  
$a_1$ & Normalization of the MS SFR beyond $m_1$  & Eq.~\ref{eq_SFR} & 0.29 & [0, 5] \\  
$a_2$ & Redshift dependence of $m_1$ & Eq.~\ref{eq_SFR} & 1.79 & [$-10$, 10] \\  
$\alpha_{\rm SB}$ & Normalization of the SFR for starbursting (SB) galaxies relative to the MS SFR & & 4.49 & [0, 20] \\  
$\log M_{\rm \ast, piv}$ & Pivot stellar mass in the $r$-band magnitude(${\cal M}_r$) & Eq.~\ref{eq_Mr} & 11.57 & [10, 15] \\  
$\alpha_{\rm lo}$ & Slope of ${\cal M}_r$ at $M_\ast \le M_{\rm \ast, piv}$ & Eq.~\ref{eq_Mr} & $-1.98$ & [$-5$, 0] \\  
$\alpha_{\rm hi}$ & Slope of ${\cal M}_r$ at $M_\ast > M_{\rm \ast, piv}$ & Eq.~\ref{eq_Mr} & $-3.08$ & [$-5, 0$] \\  
${\cal M}_{r, {\rm piv}}$ & Normalization of ${\cal M}_r$, measured at $M_{\rm \ast, piv}$ & Eq.~\ref{eq_Mr} & $-23.23$ & [$-25$, $-20$] \\  
$\sigma_{{\cal M}_r}$ & Scatter in ${\cal M}_r$ around the mean relation  & & 0.33 & [0, 1] \\  
$\kappa_{\rm corr}$  & Coefficient measuring correlation between the submm and $r$-band flux densities & Eq.~\ref{eq_Fshot_sat} & 0.99 & [0, 5] \\  
\hline
\\
\vspace{-8mm}
\end{tabular}
\vspace{-5mm}
\label{tab_params}
\end{minipage}
\vspace{0mm}
\end{table*}

\begin{figure*}
\includegraphics[width=0.99\linewidth]{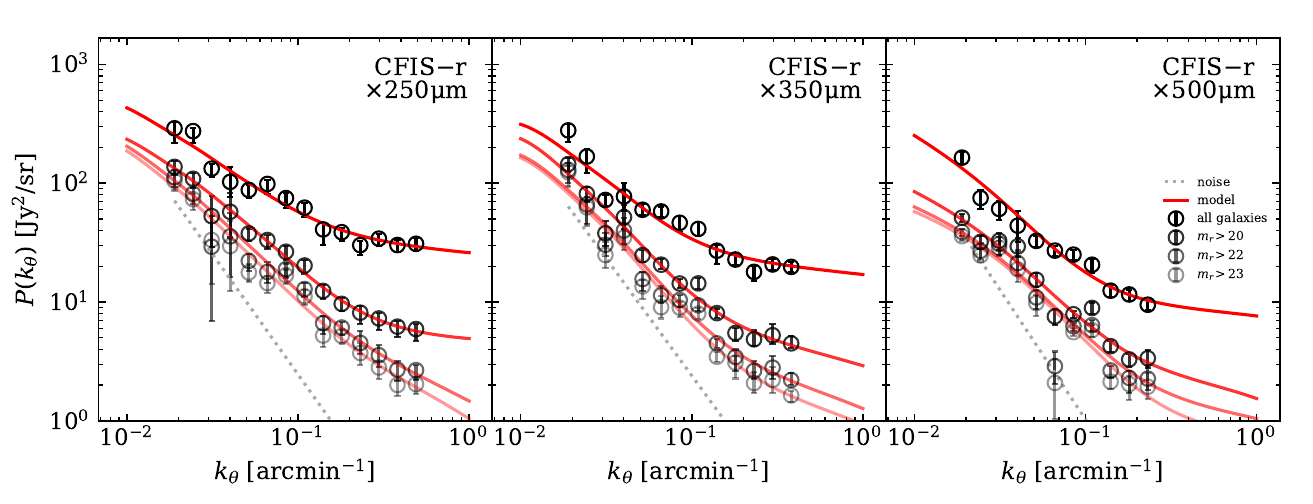}
\caption{Best-fitting halo model (lines) to the total cross-power spectra measurements (symbols) for the cases that no detected galaxy is masked, and galaxies brighter than $m_r=20$, 22 and 23 are masked, as labelled in the legend. The best-fitting parameter values, together with their priors, are provided in Table~\ref{tab_params}, while the 68 and 95~per cent confidence intervals of the posterior distributions are presented in Fig.~\ref{fig_posterior}. The dotted lines denote the noise levels for the background maps from Fig.~\ref{fig_Xpower_mask}.}
\label{fig_model_fit_magcut}
\end{figure*}

\begin{figure*}
\includegraphics[width=0.99\linewidth]{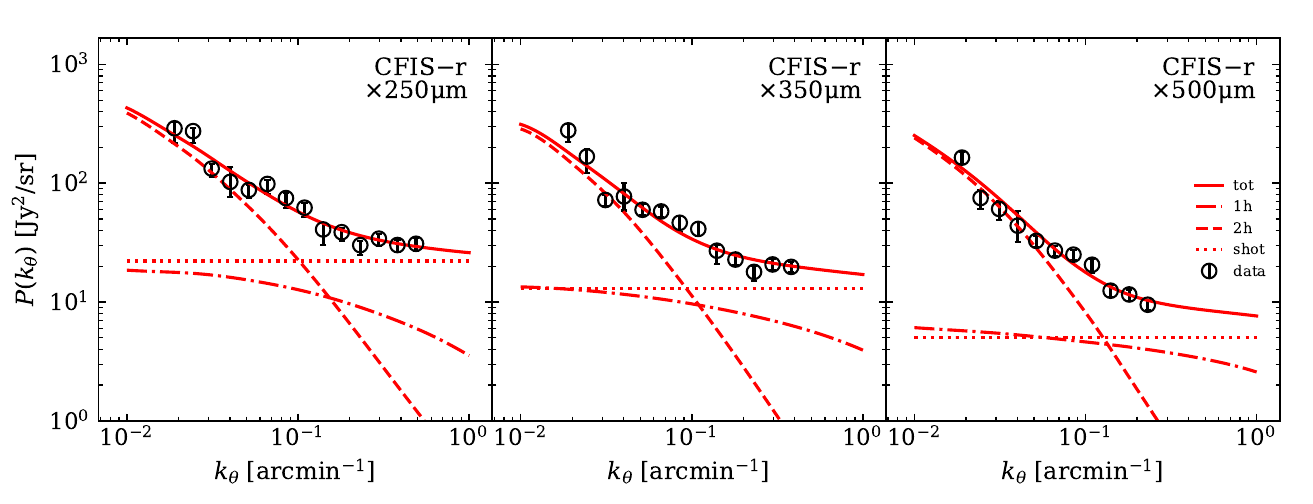}
\caption{Best-fitting halo model (lines) to the total cross-power spectra measurement (symbols) for the case that no detected galaxy is masked. Each component of the model is shown separately: the one-halo term by the dot-dashed lines; the two-halo term by the dashed lines; the shot noise by the dotted lines; and the combined spectra by the solid lines. }
\label{fig_model_fit}
\end{figure*}

\begin{figure*}
\includegraphics[width=0.99\linewidth]{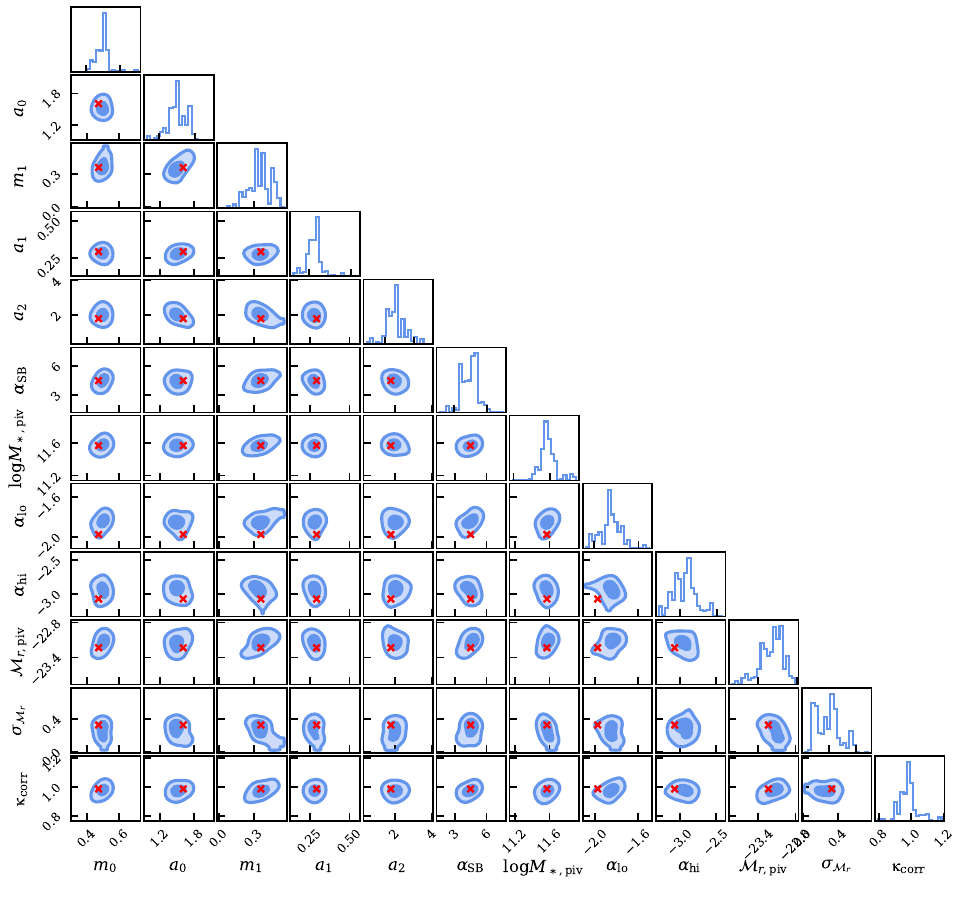}
\caption{Distributions of the free parameters in the model from fitting to the cross-correlation measurements using an MCMC method. The contours show 68 and 95~per cent ranges of the posterior distributions, while the red crosses mark the best-fitting model (defined as the parameters with the smallest $\chi^2$). See also Table~\ref{tab_params} for the priors, best-fitting values, medians of the posterior distributions and physical meanings of the parameters. 
}
\label{fig_posterior}
\end{figure*}

\begin{figure}
\includegraphics[width=0.9\linewidth]{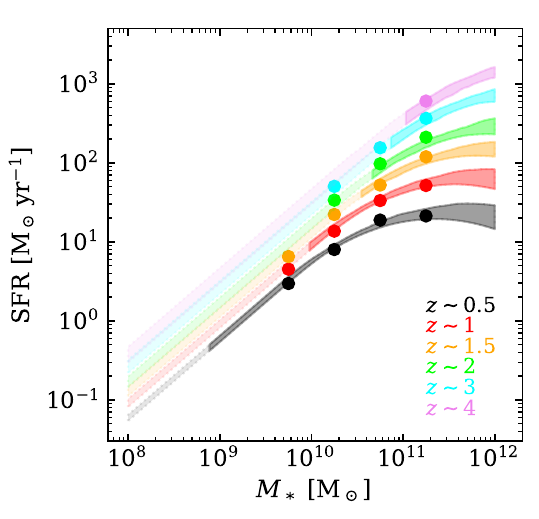}
\caption{SFR as a function of mass at various redshifts, predicted by the 68 per cent range of the posterior distribution for the model parameters. The dark-shaded regions show the constraints in the parameter space above the observational detection limit of individual sources, while the light-shaded regions indicate those indirectly constrained only through the parametrization of the modelling. The prediction is compared with the SFRs measured by \citet{Schreiber2015}, shown by the circles.}
\label{fig_SFR}
\end{figure}

Substituting Eqs.~\ref{eq_Fcen}--\ref{eq_Fshot_sat} into Eqs.~\ref{eq_Cl_1h_sep}, \ref{eq_Cl_2h_sep} and \ref{eq_Cl_shot_sep}, we fit the cross-power spectra of the model to the observational data. In addition to the fiducial measurements in Fig.~\ref{fig_Xpower_nomask}, we also compute the cross-power spectra when detected galaxies of $m_r\,{<\,}20$, 22, and 23 are masked in the CFIS $r$-band mosaics, and use them as joint constraints for the model. The covariance between the measurements are calculated by jackknife resampling from division of the fields into 200 sub-regions, as in Sect.~\ref{ssec_Xcorr}, and used this to estimate likelihoods in the fitting process. As described above, our modelling includes a total of 12 free parameters, the descriptions of which are summarized in Table~\ref{tab_params}. We perform a Markov chain Monte Carlo (MCMC) to find the best-fitting values for the parameters. As we mentioned earlier in Sect.~\ref{sec_data}, we multiply the Hartlap factor by the covariance matrix when calculating the likelihood, to account for a bias in the inverse of the matrix. For all parameters, we employ uniform priors (in linear space), and report in Table~\ref{tab_params} the model with the smallest $\chi^2$ as the best-fitting value. Figure~\ref{fig_model_fit_magcut} presents the best-fitting model in comparison to the full set of observational data. We only compare the total cross-power spectra without dividing them into the components with positive and negative correlations, since the halo model formalism used in our analysis only predicts the amplitudes of the Fourier transform, and cannot make predictions about the phase information. As can be seen, the model provides a good description of the data for each of the magnitude cuts. 

Figure~\ref{fig_model_fit} shows the best-fitting model with each of the terms (one-halo, two-halo and shot noise) separately, in comparison to the fiducial measurements. We see that the power spectra are dominated by the two-halo term on large scales, and by the Poisson term on small scales, and the contribution from the one-halo term appears to always be subdominant, regardless of scale and frequency. Thus although the one-halo term is poorly-constrained and neglects environmental effects such as quenching of satellite galaxies in massive haloes, we do not expect this term to strongly affect our results. On scales where the one-halo term contributes most significantly ($k \simeq 0.1\,{\rm arcmin}^{-1}$) we find relatively poor agreement, giving a hint that the current prescription for the one-halo term in the model may not be enough and a more sophisticated model accounting for environmental effects might provide a better fit to the data. 

Figure~\ref{fig_posterior} presents the posterior distribution, specifically 68 and 95~per cent ranges, of the model parameters. The posterior distribution of $\alpha_{\rm SB}$ implies that the mean SFR of SB galaxies is roughly 5 times higher than that of MS galaxies, broadly consistent with previous studies, including \cite{Schreiber2015}. The parameters describing the SFR of MS galaxies also agree with those obtained by \cite{Schreiber2015} well within 1$\,\sigma$. We find no significant changes in the best-fitting model, being in agreement within 1$\,\sigma$, when only the free parameters for describing the SFR (namely, $m_0$, $a_0$, $m_1$, $a_1$, $a_2$, and $\alpha_{\rm SB}$) are constrained while the other parameters are fixed at the values inferred from the KiDS samples. Also, we find that using only the observational measurements of the cross-power spectra in two bands results in increased scatter of the posterior distribution of typically about 25\,per cent. This indicates that the scatter in the model constraints obtained with the current measurements from three bands may be further reduced by a similar amount when one or two more bands from other observation are added for computing the cross-correlation. Finally, we find that $\kappa_{\rm corr}$ is very close to unity, indicating that there is no strong correlation hinted between submm and $r$-band flux densities of the sources \textit{at a given mass and redshift}, in addition to the average correlation at the mass and redshift. It is unclear whether this reflects the true underlying nature of no such correlation, or future observations can suggest a different conclusion with a better and more accurate constraining power.  

Figure~\ref{fig_SFR} shows the SFR predicted by the 68 per cent range of the posterior distribution for the model parameters at various redshifts, in comparison with the observational measurements of \citet{Schreiber2015}. Here we separate, in the presentation, the parameter space that is directly constrained above the observational limits for individual detection of sources (dark-shaded), from where the constraints are indirect only through the parametrization of the modelling (light-shaded). We used the 5$\,\sigma$ depth of the CFIS $r$-band point source detection, and the detection limit of the SPIRE sources, as the observational limits to this end. As can be seen, although the observational data have not been used as constraints in the fitting process, the model prediction agrees remarkably well with the data over a wide range of redshifts.

\section[summary]{DISCUSSION}
\label{sec_disc}
From the perspective of optical imaging, it is useful to ask about the brightness of the fluctuations that we are detecting. This helps to place this \textit{statistical} study into the context of searches for \textit{individual} low surface-brightness features. It is worth emphasizing that the inference about the COB coming from our cross-correlation measurements is not an average background level but its fluctuation, more specifically the part of it that correlates with the submm emission. We know that the extragalactic background light in the $r$ band has an amplitude of around $7.5\,{\rm nW}\,{\rm m}^{-2}\,{\rm sr}^{-1}$ \citep[e.g.][]{Driver2016}, which corresponds to $27.5\,{\rm mag}\,{\rm arcsec}^{-2}$ (assuming that the background corresponds to $\nu I_\nu$ and using AB magnitudes).  The measured autocorrelation function on arcminute scales in the CIB can be expressed as fluctuations with $\delta I_\nu/I_\nu$ of around the 15~per cent level \citep[see][]{Viero2009}, and so if we assume similar amplitude fluctuations in the COB (although we have not measured such autocorrelations directly), then these have an rms of approximately $29.5\,{\rm mag}\,{\rm arcsec}^{-2}$.  However,
the COB and CIB images that we have investigated are perhaps only around 5~per cent correlated (once bright sources are removed, see Appendix~\ref{sec_appC}).  Hence the cross-correlation signals that we are measuring correspond to surface brightness fluctuations around the level of $32.5\,{\rm mag}\,{\rm arcsec}^{-2}$.  These are of course too faint to detect individually, and we are only able to reach these levels by making a statistical measurement of the fluctuations as a whole, and using the forgiving nature of the cross-correlation technique.

With wide-field imaging now being undertaken across the entire electromagnetic spectrum, there is great power in combining data in different wavebands.
The cross-correlation approach that we have described in this paper is one component in a tool-box of techniques that can be used to combine wide-field images.  The most commonly used approach is to extract objects from images, build catalogues of object properties, and work directly with brightnesses, shapes, etc.  Additionally one can look at the 2-point correlation function of the catalogued sources, which includes studying cross-correlations of sources between wavebands, e.g.\ combining optical and submm catalogues \citep{Blake2006,Hildebrandt2013}.  However, this misses the information about the statistical properties of the galaxies that are {\it not\/} individually detected and yet are still there at a faint level in the images.  So-called `stacking' can be used to extract the properties of subsets of sources \citep[e.g.][]{Viero2015} -- this is essentially a cross-correlation between a map and a catalogue \citep[e.g.][]{Marsden2009}. The information contained in the fluctuating extragalactic background, which is the focus of this paper, is complementary to studies of distinct objects, and reaches down to fainter objects. Provided that images exist with reasonable control of systematic effects, then this information can be extracted `for free', in addition to the more detailed studies carried out on individual galaxies. Moreover, the signal coming from optical-to-far-IR correlations will contain information about the relationship between stars and star formation in galaxy halos, which will be complementary to information extracted from optical or far-IR data alone. 

The data sets that we have used here are modest in size and quality compared with what is expected in the near future.  A prominent example is {\it Euclid\/} \citep{Laureijs2011}, which is scheduled to launch in 2023.  In its imaging part, {\it Euclid\/} will map 15{,}000\,${\rm deg}^2$ to 24th magnitude in several visible and near-IR filters.  Artefacts caused by Earth's atmosphere that plague ground-based observatories (or even low-Earth orbiting facilities such as the {\it Hubble Space Telescope\/}) will be entirely absent in {\it Euclid\/} imaging.  It is therefore expected that {\it Euclid\/} will be excellent for studying low surface-brightness features on the sky, with estimates that it should be possible to reach levels of $29.8\,{\rm mag}\,{\rm arcsec}^{-2}$ (visible;  $28.4\,{\rm mag}\,{\rm arcsec}^{-2}$ for the near-IR bands) for individual features \citep{Scaramella2021}. This is about one magnitude deeper than the CFIS $r$-band data reach. This also makes {\it Euclid\/} ideal for studying statistical correlations in the visible and near-IR sky \citep[see also][]{Kashlinsky2018}, especially using cross-correlation techniques. As well as {\it Euclid\/} in space, we are looking forward to improvements in optical imaging from the ground via the Vera Rubin Observatory \citep[formerly LSST][]{LSST}, which is expected to begin survey operations in the next couple of years. This observatory will house a large-aperture telescope with a field-of-view of 9.6\,${\rm deg}^2$. Its main survey will cover approximately 18{,}000\,${\rm deg}^2$ mainly of the southern sky in six filters, down to 24.5 magnitude in $r$-band. Its wide field of view, combined with the wide survey area and high sensitivity, is expected to bring a significantly improved view of the southern sky, complementing surveys such as CFIS that are focused on the northern sky. Also, current surveys such as the Dark Energy Survey \citep{DES} and UNIONS, each eventually covering about 5{,}000\,${\rm deg}^2$ down to a depth matching that of {\it Euclid\/}, already provide useful data to explore galaxy evolution from statistical correlations, as demonstrated by our results. As stated in Sect.~\ref{sec_intro}, while there are many known issues for ground-based measurements, including variable optics effects, as well as sky lines, etc. \citep[see e.g.,][]{Leinert1998, Odenwald2003}, how efficiently those issues may be overcome by cross-correlation analysis between different wavelengths (like the analysis presented here) have not been investigated thoroughly yet. Studies along the lines described in this paper will undoubtedly become more powerful in the next few years as more data are gathered. For the far-CIB side, on the other hand, CMB-type experiments have already mapped the millimetre sky over thousands of square degrees with beamsizes of around $1\,{\rm arcmin}$ \citep[e.g.][]{ACT2021,SPT2022} and the entire sky with {\it Planck}'s 5-arcmin beam. In a few year's time, the Prime-Cam instrument on CCAT-prime/FYST \citep{CCATp} will carry out a survey of 20{,}000\,${\rm deg}^2$ of sky from Chile, over several millimetre-to-submillimetre bands, with a beamsize of 15\,arcsec at $350\,\mu$m. It should thus be possible to extend the modelling to more sophisticated parametrizations that tell us more about the evolution of star-forming galaxies. Also, some of the parameter degeneracies shown in Fig.~\ref{fig_posterior} will presumably be broken when much better data are available, extending over more wavebands and angular scales.

\section[summary]{CONCLUSIONS}
\label{sec_sum}

In this study we have investigated correlations between the cosmic infrared and optical backgrounds (CIB and COB, respectively) by performing an analysis of cross-power spectra between {\it Herschel}-SPIRE and CFIS $r$-band images. Specifically, for the submm data, we have used the SPIRE images at 250, 350, and 500\,$\micron$ from the HerMES survey, and from the {\it Herschel\/} Extragalactic Legacy Project (HELP). For the optical data, we have constructed CFIS $r$-band mosaic maps based on the individual Low Surface Brightness (LSB) tiles from the survey, and have used them for our analysis. The CFIS mosaic maps were constructed in two versions: `galaxy maps', in which only stars and artefacts are masked, while galaxies are not masked; and diffuse `background' maps, where every galaxy detected in the CFIS image is also masked. Our analysis has focused on a total of five SPIRE fields in the northern sky that significantly overlap with the current coverage of the CFIS data, namely the EGS, FLS, ELAIS-N1, ELAIS-N2 and HATLAS-NGP fields. The total combined area of sky used for the analysis is 91\,${\rm deg}^2$. 

For each of the five fields, we have calculated one-dimensional cross-power spectra, azimuthally averaged from two-dimensional Fourier transforms for each annulus in $k$-space. The true underlying power spectra have been recovered by accounting for the instrumental beam, the transfer function of the mapmaker, and the masking. The results from each of the fields have been combined (weighted with the errors) to obtain the total average spectra and the errors. We also separated the in-phase and out-of-phase cross-power spectra to test that the detection is a physical signal, not noise. Because the measured cross-power spectra are the absolute values and come from a finite number of pixels and realisations, they are positive even when there is no real signal and the cross-correlation is purely from statistical noise. However, because the in-phase component is stronger than the out-of-phase part, we have demonstrated the existence of a very strong CIB-COB emission correlation. As a test, we also have estimated the noise level in the case of no correlation and compared that with the measurements to show that the signal we detect has a significant excess over the noise, overall at $\gtrsim 18\,\sigma$ ($\gtrsim 14\,\sigma$ for the `background' map) in each of the SPIRE bands. 

A crucial issue to tackle in this study is the potential contamination from our own Galaxy. We have found that the signal from one of the fields, namely FLS, is dominated by a filament of Galactic cirrus. We have excluded the FLS field for this reason for the calculation of average power spectra, but at the same time it is useful for showing what dominant cirrus contamination looks like. For the other fields, the foreground contamination is found to be typically 10 to 20\,per cent and always smaller than 50\,per cent. This comes from a linear regression with external maps from other surveys, such as the SFD, EBHIS, {\it Planck}-GNILC and {\it WISE\/} 12-$\micron$ maps. Among these, the EBHIS map (like any other H\,{\sc i}-based map) is confined to the Galaxy, with no extragalactic contribution included, enabling a reliable subtraction of the Galactic cirrus without any loss of signal from the background \citep{Chiang2019}. The results from the {\it WISE\/} and DHIGLS maps, which have higher angular resolution (comparable to SPIRE), have shown that the impact on small scales due to the relatively poor resolution of some of the other maps (including EBHIS) is insignificant. This is because of a much steeper slope of the power spectra of Galactic cirrus ($P(k)\propto 1/k^\beta$, with $\beta\simeq2.5$--3.5) than that of the CIB. The shallower slope found from our results ($\beta\simeq1.0$--1.5) also confirms that the detected signal from our analysis is predominantly extragalactic, particularly after attempting to subtract Galactic contamination. We have found no significant difference in the results from using the different external survey maps, despite potential systematic effects due to the independent tracers and the processing and creation of each of the maps. The (pre)dominance of extragalactic signal in the detection is further supported by the lack of strong variations in the results among the fields (except for the FLS field). 

We have used the SIDES simulation, in combination with a scaling relation for $r$-band magnitudes from the KiDS data, to demonstrate that our methods and thus results are reliable and not biased regarding corrections for instrumental noise and masking. Using the same set of data, we also have found broad consistency with the inference from observational results that the contribution from the Galactic cirrus is subdominant. Finally, we have shown that the impact on the power spectra from obscuration of optical light by extragalactic dust and from the resolved properties of galaxies are negligible; the main effect of Galactic dust at both of these wavelengths and angular scales is emission, rather than absorption.

We have interpreted our results in a framework of halo-based modelling, with parameters determined by fitting model power spectra to our measurements. Adopting a similar modelling scheme to that of \citet{Bethermin2013, Bethermin2017}, we have presented constraints on the model parameters and the predicted average SFR as a function of mass and redshift, the latter of which is shown to be in good agreement with independent observational measurements. 
In this study, we have shown that a strong correlation exists between optical and submm images, and that this can be used to provide interesting constraints on the star-formation history of objects over a range of masses and redshifts. The image cross-correlation analysis presented in this study enables an exploration of relatively faint objects at higher redshifts compared to catalogue- or source-based analyses. This has been demonstrated in our results with the detection of strong signals from the residual backgrounds, not captured by identified sources.

While we have demonstrated that a very strong correlation signal and therefore useful information can already be extracted from the current data, with improved data from upcoming surveys this cross-correlation will be a powerful tool for exploring galaxy formation and evolution across a wide range of epochs.  In that regard we are particularly excited by the prospects offered by the upcoming {\it Euclid\/} space mission \citep[e.g.][]{Laureijs2011}, where the imaging will be free from terrestrial systematic effects, and the combination of {\it Euclid\/} with CMB-type experiments like CCAT-prime \citep{CCATp} which will extend our analysis to much larger fractions of the sky. Such future studies will find overall signal-to-noise levels $\gg20\,\sigma$ for the CIB-COB cross-correlation in multiple wavebands, thus enabling more sophisticated models to be fit to the data.  In order to fully exploit such data sets it will be important to put effort into improvements in several distinct directions: (1) more careful removal of Galactic cirrus, particularly in order to push to larger angular scales; (2) simulations of the data-taking and reduction process in order to calculate the transfer function of the images in both wavebands; and (3) comprehensive exploration of the parameters of the halo-based model to understand just what physical quantities are best constrained using these cross-correlations, and how that complements what is learned from other kinds of extragalactic investigation.

\section*{ACKNOWLEDGEMENTS}
This research was supported by the Natural Sciences and Engineering Research Council of Canada and by the Canadian Space Agency. Part of this work benefited from the Canadian Advanced Network for Astronomical Research and Compute Canada facilities. This research used the facilities of the Canadian Astronomy Data Centre operated by the National Research Council (NRC) of Canada with the support of the Canadian Space Agency. AHW and AD are supported by an European Research Council Consolidator Grant (No.\ 770935). This publication is part of the project `A rising tide: Galaxy intrinsic alignments as a new probe of cosmology and galaxy evolution' (with project number VI.Vidi.203.011) of the Talent programme Vidi which is (partly) financed by the Dutch Research Council (NWO). This work is also part of the Delta ITP consortium, a program of the Netherlands Organization for Scientific Research (NWO) that is funded by the Dutch Ministry of Education, Culture and Science (OCW). This work is based on data obtained as part of the Canada-France Imaging Survey, a Canada-France-Hawaii Telescope (CFHT) large program of the NRC of Canada and the French Centre National de la Recherche Scientifique (CNRS). Based on observations obtained with MegaPrime/MegaCam, a joint project of CFHT and Commissariat \`a l'\'energie atomique Saclay, at the CFHT which is operated by the NRC of Canada, the Institut National des Science de l'Univers of the CNRS of France and the University of Hawaii. This research has made use of data from the Herschel Multi-tiered Extragalactic Survey (HerMES) project (\url{http://hermes.sussex.ac.uk/}). HerMES is a {\it Herschel\/} Key Programme utilizing guaranteed time from the Spectral and Photometric Imaging Receiver instrument team, European Space Astronomy Centre scientists and a mission scientist. The HerMES data were accessed through the Herschel Database in Marseille (\url{http://hedam.lam.fr}) operated by the Marseille Astrophysics Data Center and hosted by the Laboratoire d'Astrophysique de Marseille. The Herschel Extragalactic Legacy Project is a European Commission Research Executive Agency funded project under the SP1-Cooperation, Collaborative project, small or medium-scale focused research project, FP7-SPACE-2013-1 scheme, Grant Agreement Number 607254. This research made use of {\sc Montage}. It is funded by the National Science Foundation under Grant Number ACI-1440620, and was previously funded by the National Aeronautics and Space Administration's Earth Science Technology Office, Computation Technologies Project, under Cooperative Agreement Number NCC5-626 between NASA and the California Institute of Technology. Based on observations made with the European Southern Observatory Telescopes at the La Silla Paranal Observatory under programme IDs 177.A-3016, 177.A-3017, 177.A-3018 and 179.A-2004, and on data products produced by the Kilo-Degree Survey (KiDS) consortium. The KiDS production team acknowledges support from: Deutsche Forschungsgemeinschaft (DFG), European Research Council, Nederlandse Onderzoekschool Voor Astronomie and Nederlandse Organisatie voor Wetenschappelijk Onderzoek-Medium grants; Target; and the University of Padova  and the University Federico II (Naples). The Effelsberg-Bonn H\,{\sc i} Survey (EBHIS) data are based on observations performed with the 100-m telescope of the Max Planck Institute for Radio Astronomy at Effelsberg; EBHIS was funded by the DFG under the grants KE757/7-1 to 7-3. This publication makes use of data products from the {\it Wide-field Infrared Survey Explorer}, which is a joint project of the University of California, Los Angeles, and the Jet Propulsion Laboratory/California Institute of Technology, funded by the National Aeronautics and Space Administration. We thank the groups involved in the production and release of the {\it Planck\/} generalized needlet internal linear combination map, the reddening map of Schlegel, Finkbeiner \& Davis, the Green Bank Telescope H\,{\sc i} Intermediate Galactic Latitude Survey, the Dominion Radio Astrophysical Observatory H\,{\sc i} Intermediate Galactic Latitude Survey, the extinction maps of Schlafly et al. and Green et al., and the Simulated Infrared Dusty Extragalactic Sky project.

\section*{DATA AVAILABILITY}

The advanced products underlying this article will be shared on reasonable request to the corresponding author.

\bibliographystyle{mnras}
\bibliography{cross-correlation.bib}

\appendix

\section{The cross-power spectra measurements in flux unit}
\label{sec_appA}
\begin{figure*}
\includegraphics[width=0.99\linewidth]{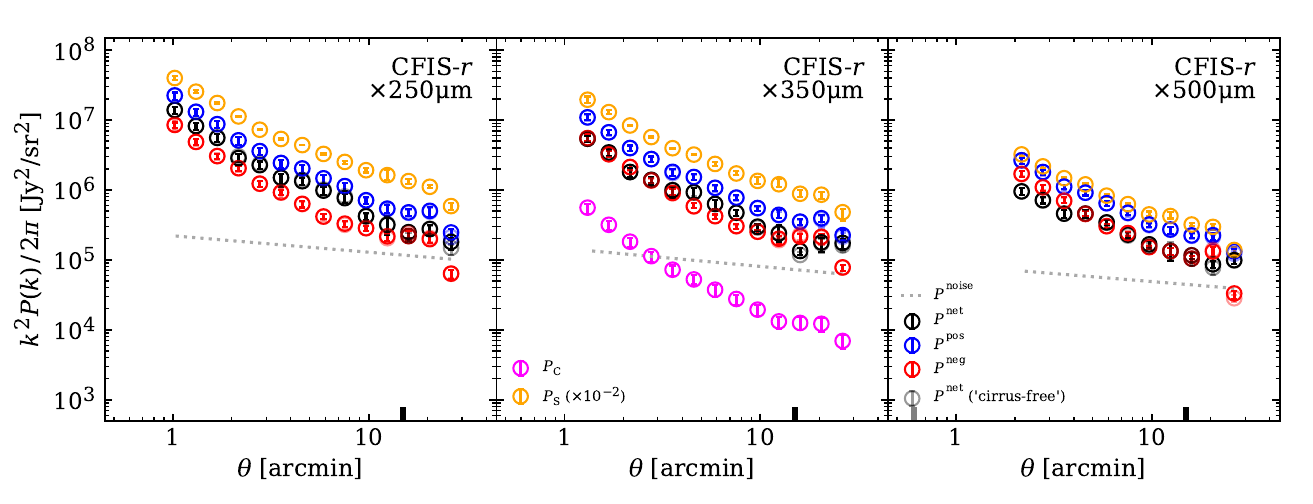}
\caption{Same measurements of the power spectra as in Fig.~\ref{fig_Xpower_nomask} but in flux unit. The black data points show the net cross-power spectra as calculated using Eq.~\ref{eq_Xpower}. The blue and red data points show the positive and negative cross-power spectra (see Eq.~\ref{eq_Xpower_pos_neg} and the main text for more details). The dotted lines show the expected noise level in the case of no correlation, which are obtained by shuffling the modulus and randomizing the phases of the Fourier transforms among pixels. The absolute values of the noise were taken to present only its magnitude, regardless of the sign. The faint symbols are the measurements from the cirrus-free maps, indicating that the impact of Galactic cirrus is negligible. The thick vertical tickmarks indicate the scales of a CFIS tile (black; 0.5\,deg) and the SPIRE beam (grey; 18.1, 25.5, and 36.6\,arcsec at 250, 350 and 500\,$\micron$, respectively). The auto-power spectra of the CFIS (magenta; only shown in the middle panel) and SPIRE (orange) maps estimated using the same approach are also presented, for reference. }
\label{fig_Xpower_nomask_}
\end{figure*}

\begin{figure*}
\includegraphics[width=0.99\linewidth]{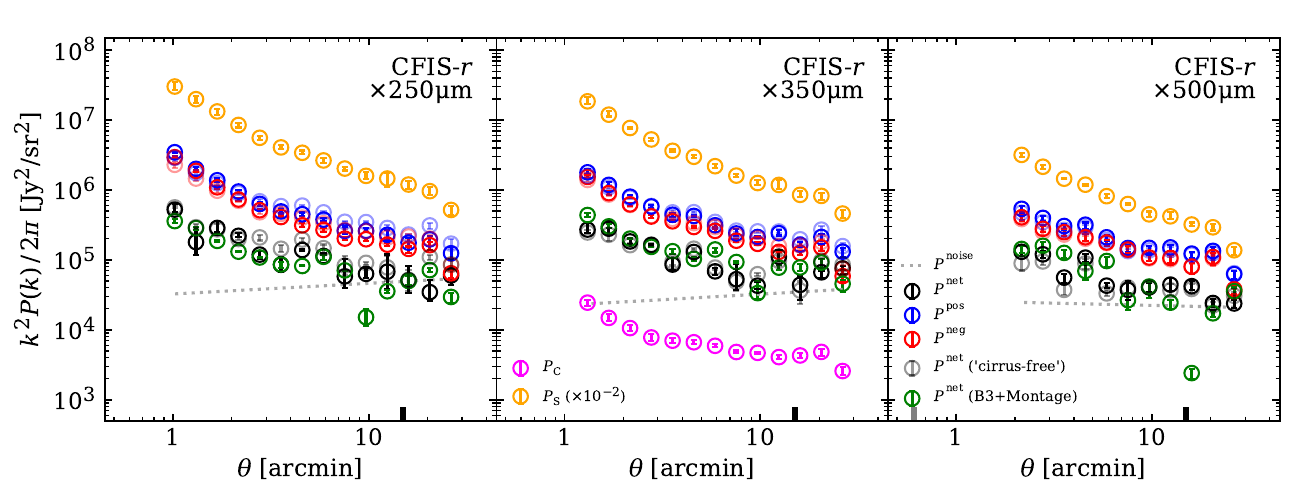}
\caption{Same measurements of the power spectra as in Fig.~\ref{fig_Xpower_nomask_} but after masking all sources detected in the CFIS map, including galaxies. As in Fig.~\ref{fig_Xpower_nomask_}, the faint symbols represent the results from the `cirrus-free' maps. The green symbols show the results based on another version of mosaics (see Sect.~\ref{ssec_mosaic}). The auto-power spectra of the CFIS (magenta; only shown in the middle panel) and SPIRE (orange) maps estimated using the same approach are also presented, for reference. }
\label{fig_Xpower_mask_}
\end{figure*}

Here we present the average cross-power spectra measurements, namely Figs.~\ref{fig_Xpower_nomask} and \ref{fig_Xpower_mask}, but in flux units, as shown in Figs.~\ref{fig_Xpower_nomask_} and \ref{fig_Xpower_mask_}.  This may be useful for those who are more familar with such units. 

\section{Null tests of the cross-power spectra measurements}
\label{sec_appB}
Here we perform null tests of the cross-power spectra measurements between the CFIS $r$-band and SPIRE maps, to ensure that our results and interpretations are not biased by our method. For such null tests, we take the measurements from the following four cases: translating one map relative to the other map by one-fourth the size of a field along the major axis (the axis along which the non-blank data are most contiguous) of the field each time (thus repeating three times in total along the axis); rotating one map relative to the other map by 180 degrees; cross-correlating between different fields; and creating simulated maps with the same auto-power spectra as the observational data, but with random phases. For the cross-correlation between different fields, we cross-correlate the CFIS-$r$ LSB map of each field with a portion of SPIRE map in the HATLAS-NGP field at a random position, but with the same size as the LSB map. For the HATLAS-NGP field, we use the southern halves of the SPIRE maps, which do not overlap with the CFIS coverage of the field (see Fig.~\ref{fig_fields}) and thus were not used for our fiducial measurements of signals. 

\begin{figure*}
\includegraphics[width=0.99\linewidth]{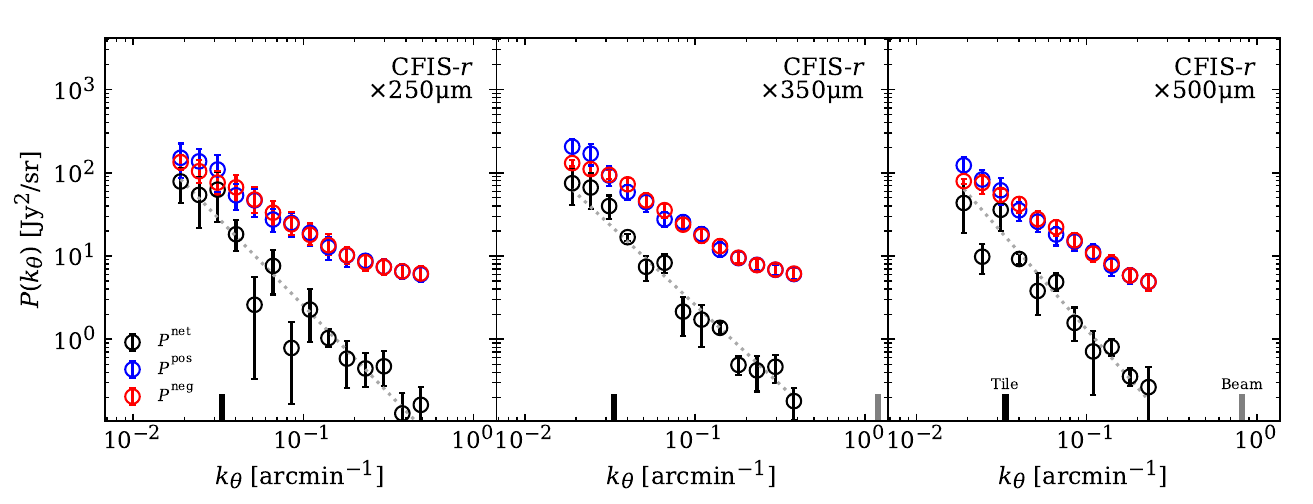}
\caption{Cross-power spectra between the CFIS $r$-band (where galaxies are masked) and SPIRE maps at 250, 350 and 500\,$\micron$, measured and averaged from the following four cases of null tests: translating one map relative to the other map by one-fourth the size of a field along the major axis (the axis along which the non-blank data are most contiguous) of the field each time (thus repeating three times in total along the axis); rotating one map relative to the other map by 180 degrees; cross-correlating between different fields; and creating simulated maps with the same auto-power spectra as the observational data, but with random phases. The black data points show the net power spectra as calculated using Eq.~\ref{eq_Xpower}, while the blue and red data points show the positive and negative power spectra as defined by Eq.~\ref{eq_Xpower_pos_neg}. As can be seen, the amplitudes of positive and negative power spectra are similar to each other, and the net power spectra are consistent with the noise level (dotted), which are all consistent with expectations as described in Sect.~\ref{ssec_res}. This indicates that our cross-correlation results are not biased by the method or by the interpretations using the net, positive, and negative cross-power spectra. }
\label{fig_null_test}
\end{figure*}

The averages from the four test cases are taken and shown in Fig.~\ref{fig_null_test}. We do not find significant differences among the results from the four cases. Here we present only the results based on the CFIS maps where galaxies are masked, while we find the same conclusions for those without masking. As explained earlier in Sect.~\ref{ssec_res}, in the case of no correlation, it is expected that the positive and negative cross-power spectra (Eq.~\ref{eq_Xpower_pos_neg}) will be similar to each other in amplitude within statistical uncertainties, although the net spectra (Eq.~\ref{eq_Xpower}) will still not be completely zero due to the limited number of realisations. As can be seen, these are all consistent with the null tests presented here, showing the utility of our interpretation of the net, positive, and negative cross-power spectra, as well as reassuring us that our measurements are not biased by our method. 

\section{Visual correlations between the SPIRE and CFIS images}
\label{sec_appC}
\begin{figure*}
\includegraphics[width=0.9\linewidth]{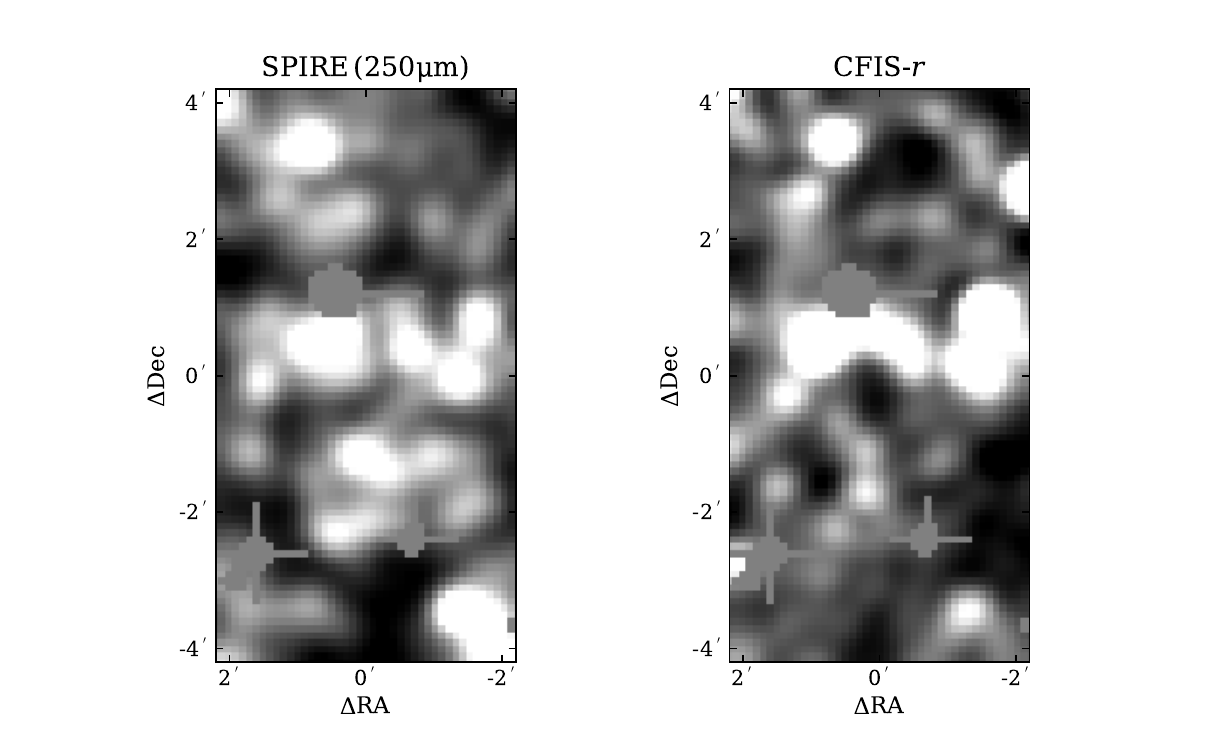}
\caption{Representative region from the EGS field, which shows a clear visual correlation between SPIRE and CFIS maps, with multiple sources (white areas) at almost the same positions. The grey areas here are masks for stars and artefacts. The real-space correlation at the relevant scales is about 25~per cent.  After removing all detected sources the correlation falls to about 5~per cent and would no longer be visually apparent.}
\label{fig_visualcorr_EGS}
\end{figure*}

\begin{figure*}
\includegraphics[width=0.98\linewidth]{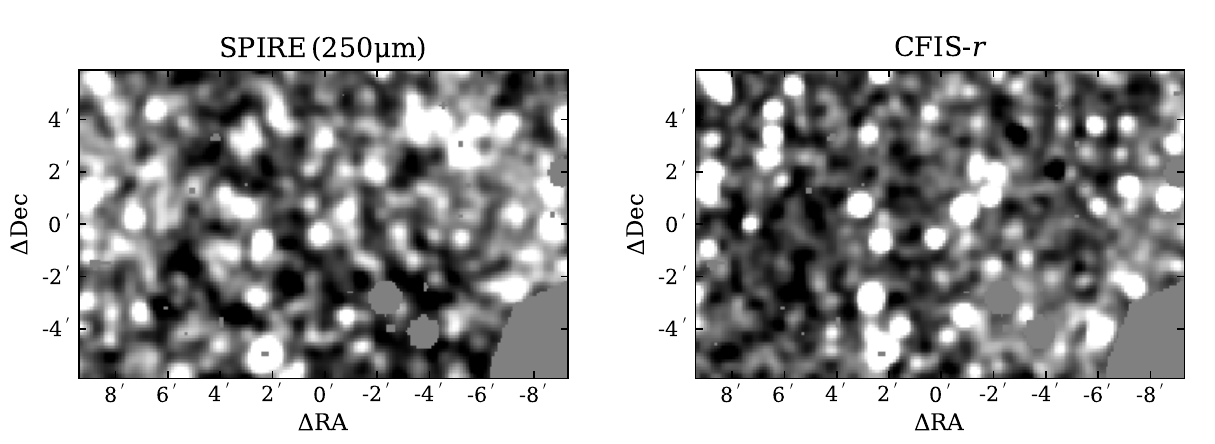}
\caption{Same as Fig.~\ref{fig_visualcorr_EGS} for part of the HATLAS-NGP field.}
\label{fig_visualcorr_NGP}
\end{figure*}

\begin{figure*}
\includegraphics[width=0.98\linewidth]{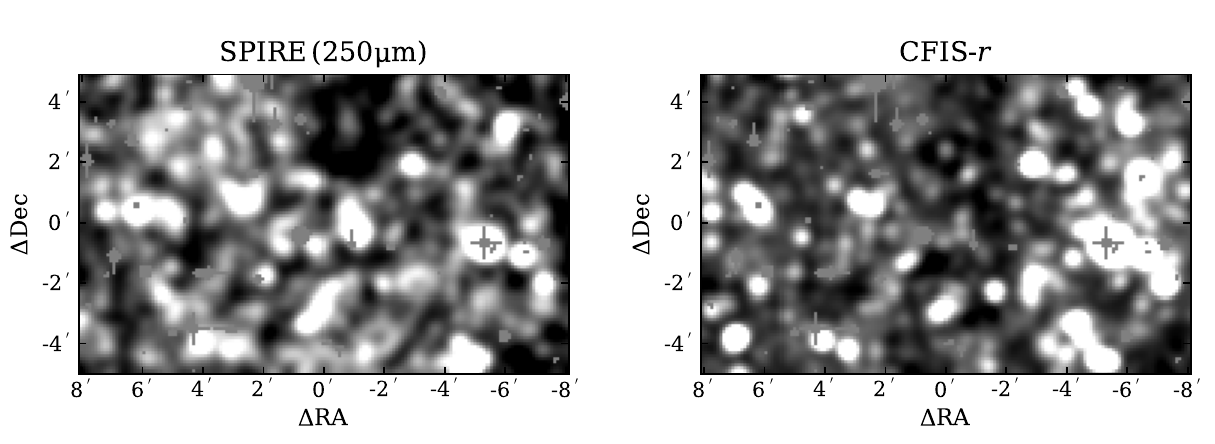}
\caption{Same as Fig.~\ref{fig_visualcorr_EGS} for part of the ELAIS-N1 field.}
\label{fig_visualcorr_N1}
\end{figure*}

\begin{figure*}
\includegraphics[width=0.98\linewidth]{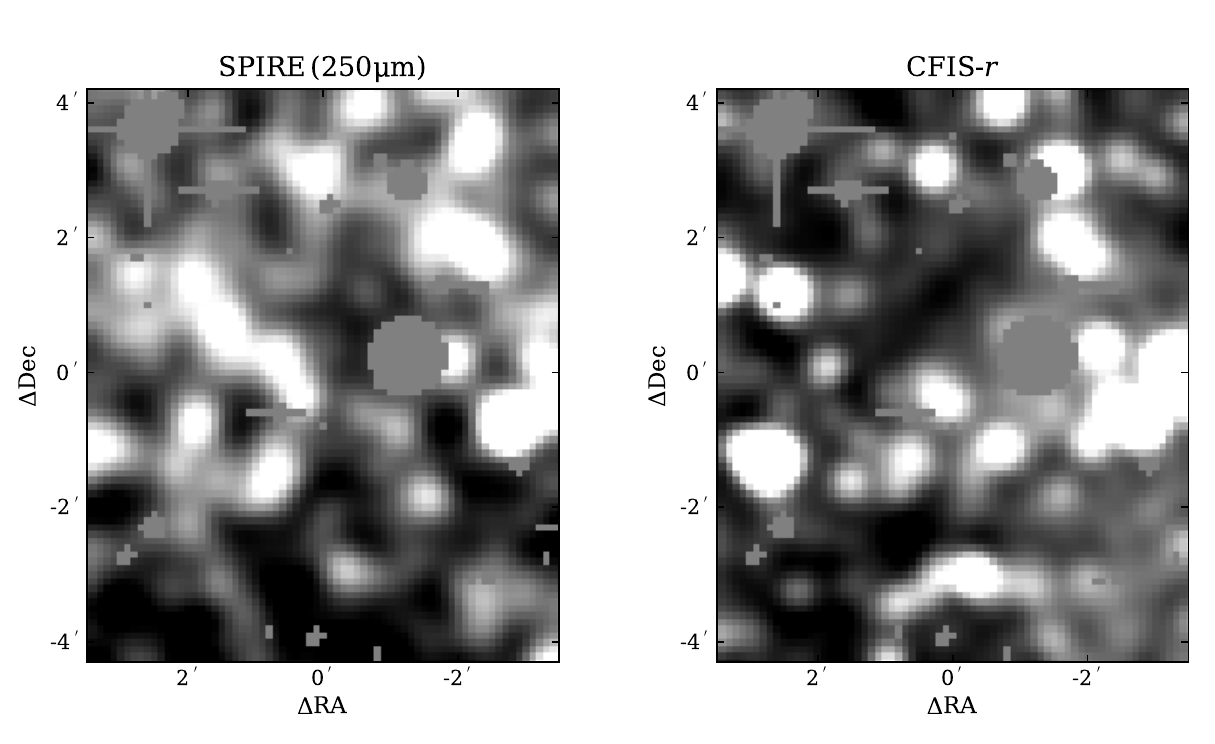}
\caption{Same as Fig.~\ref{fig_visualcorr_EGS} for part of the ELAIS-N2 field.}
\label{fig_visualcorr_N2}
\end{figure*}

Figures~\ref{fig_visualcorr_EGS}--\ref{fig_visualcorr_N2} show cut-out images about 10\,arcmin across for each field, from the SPIRE and CFIS maps where individually identified galaxies are not masked. As can be seen, there are clearly correlations confirmed by eye directly between the images, reassuring us that there is a strong correlation signal in the measurements of the cross-power spectra. 

However, if we remove individually-detected sources, the visual cross-correlation is not perhaps as striking as one might naively expect -- this is because the correlation is fairly weak, is strongest on relatively small scales, and only builds up high signal-to-noise when averaged over large areas. In fact, we find that it is much less obvious to see the correlations by eye between the maps where identified galaxies have been removed. This is because the correlations are much weaker for the background maps, as seen in Fig.~\ref{fig_Xpower_mask} in comparison to Fig.~\ref{fig_Xpower_nomask}. In fact the correlation coefficient between the background maps, calculated as $P_{\rm C\times S} / (P_{\rm C}\, P_{\rm S})^{0.5}$ averaged over the $k$-bins, is only about $0.05$, while the same coefficient between the full maps (the maps where galaxies are not masked) is $\simeq 0.25$. We also tested using simulations with similar power spectra and signal-to-noise to the data, and confirmed that the correlations between the simulated maps are only very weakly identifiable by eye.  Hence we do not show such examples here.

\label{lastpage}

\end{document}